\documentclass[reqno]{amsart}

\usepackage{tikz}
\usepackage{mathptmx}
\usepackage{xspace}
\usepackage{amsmath,amsfonts,amsthm,eucal}

\newtheorem{Thm}{Theorem}[section]
\newtheorem{Prop}[Thm]{Proposition}
\newtheorem{Lem}[Thm]{Lemma}
\newtheorem{Coro}[Thm]{Corollary}
\newtheorem{Def}{Definition}

\newcommand{\vac}[1]{\draw (#1) circle (1/5);
 	\fill[white] (#1) circle (1/5)}
\newcommand{\defect}[1]{\draw (#1) --++ (3,0)}
\newcommand{\bord}[2]{\draw[very thick] (#1,-1/2) --++ (0,#2)}
\newcommand{\link}[4]{\draw (#1) .. controls (#2) and (#3) .. (#4)}

\newcommand{\wave}[2]{\draw[dashed] (#1,#2) --++ (3,0);}

\newcommand{\dtl}[1]{\mathsf{dTL}_{#1}}
\newcommand{\odtl}[1]{\mathsf{odTL}_{#1}}
\newcommand{\edtl}[1]{\mathsf{edTL}_{#1}}
\newcommand{\tl}[1]{\mathsf{TL}_{#1}}

\newcommand{\ds}[1]{\mathsf S_{#1}}

\newcommand{\dl}[1]{\mathsf I_{#1}}
\newcommand{\dP}[1]{\mathsf P_{#1}}

\newcommand{\kc}{k_c}
\newcommand{\xf}{\times\hspace{-0.27em}_f}
\newcommand{\F}[1]{F_{#1}}
\newcommand{\U}[1]{U_{#1}}
\newcommand{\T}[2]{\mathsf{T}^{#1}_{#2}}
\newcommand{\B}[2]{\mathsf{B}^{#1}_{#2}}
\newcommand{\lc}{\ell}
\newcommand{\Hom}[1]{\text{\rm{Hom}}\big( #1 \big)}

\newcommand{\Res}[1]{#1{}\hspace{-0.18em} \downarrow{}\hskip -0.18em}
\newcommand{\Ind}[1]{#1{}\hspace{-0.18em} \uparrow{}\hskip -0.18em}

\DeclareMathOperator{\id}{id}
\DeclareMathOperator{\im}{im}

\DeclareMathOperator{\Coker}{Coker}

\definecolor{rougePompier}{rgb}{0.93,0.11,0.14}
\definecolor{vertForet}{rgb}{0.04,0.75,0.07}

\title[Fusion rules for Temperley-Lieb families.]{Fusion rules for the Temperley-Lieb algebra\\ and its dilute generalisation}
\author[J Bellet\^ete]{Jonathan Bellet\^ete}
\address[Jonathan Bellet\^ete]{
CRM and D\'epartement de physique\\ 
Universit\'e de Montr\'eal\\
Montr\'eal, QC, Canada, H3C 3J7}
\email{\tt jonathan.belletete@umontreal.ca}
\date{\today}
\begin{document}
  
\begin{abstract}

The Temperley-Lieb (TL) family of algebras is well known for its role in building integrable lattice models. Even though a proof is still missing, it is agreed that these models should go to conformal field theories in the thermodynamic limit and that the limiting vector space should carry a representation of the Virasoro algebra. The fusion rules are a notable feature of the Virasoro algebra. One would hope that there is an analogous construction for the TL family. Such a construction was proposed by Read and Saleur [Nucl. Phys. B 777, 316 (2007)] and partially computed by Gainutdinov and Vasseur [Nucl. Phys. B 868, 223-270 (2013)] using the bimodule structure over the Temperley-Lieb algebras and the quantum group Uq(sl2).
 
We use their definition for the dilute Temperley-Lieb (dTL) family, a generalisation of the original TL family. We develop a new way of computing fusion by using induction and show its power by obtaining fusion rules for both dTL and TL. We recover those computed by Gainutdivov and Vasseur and new ones that were beyond their scope. In particular, we identify a set of irreducible TL- or dTL-representations whose behavior under fusion is that of some irreducibles of the CFT minimal models.

\medskip
  
\end{abstract}
  
\maketitle  

\tableofcontents
	
\begin{section}{Introduction}
The Temperley-Lieb family of algebras $\lbrace \tl{n}(q) \rbrace_{n\in \mathbb{Z}_{>0}}$, introduced in \cite{TemperleyLieb}, is well-known for its use in building integrable lattice models that correspond to a large variety of different physical systems \cite{Martin,Nienhuis}, particularly to quantum spin chains. Many properties of these physical models can be interpreted in terms of the algebraic properties of the family, which can be obtained by studying the representation theory of these algebras. As such, it has received a lot of attention over the years. Since its introduction, many generalizations have been proposed: the periodic Temperley-Lieb algebras \cite{GrahamLehrer,GreenFan,Greenseul,ErdmannGreen}, the boundary or blob algebras \cite{GrimmMartin}, the multi-colored Temperley-Lieb algebras \cite{GrimmPearce}, etc... One such generalization which is of particular interest is the dilute Temperley-Lieb family $\lbrace \dtl{n}(q) \rbrace_{n\in \mathbb{Z}_{>0}}$ \cite{Grimm,BelSY}, which has been introduced to build dilute lattice models, i.e., ones where lattice sites can be empty.

It has been conjectured that the lattice models built from $\tl{n}(q)$ should correspond, in the continuum limit, to conformal field theories \cite{KooSaleur,PasSaleur,PeRasZu}. A consequence to these conjectures is that the Temperley-Lieb family should bear a structure of Virasoro-module when $n$ goes to infinity. In order to study these conjectures, or at least to give them credibility, there has been a lot of interest towards identifying similar algebraic structures between $\tl{n}$ and the Virasoro algebra, like module structure \cite{GoodmanWenzl,Westbury,RiSY,BelRiSy} and fusion rules \cite{ReadSaleur1,ReadSaleur2,GainVass,PeaRasm}.

Fusion rules, from a physical point of view, describe how fields interact at short distance. From a mathematical point of view however, it is a way of defining a product between modules over the algebras underlying the theory. For chiral algebras in CFT's, these rules have been widely studied, and while defining these rules in terms of functors is relatively simple, computing them explicitly as proven to be very challenging. The recursive algorithm described by Nahm \cite{Nahm} and developed by Gaberdiel and Kausch \cite{GabKa} remains the leading tool. 

On the Temperley-Lieb family, there has been two main suggestions on how to define and then compute such functors. The first, suggested by Pearce, Rasmussen \cite{PeaRasm}, is built in terms of the lattice models and rely on properties of their transfer matrices instead of relying directly on the algebras. The second, proposed by Read and Saleur \cite{ReadSaleur1,ReadSaleur2} and later studied by Gainutnidov and Vasseur \cite{GainVass}, is built around the following description. To compute the fusion product between two spin chains, one joins them together at one of their extremities and then one lets them evolve. While heuristic, they used this idea to build a purely categorical description of the fusion rules which, while motivated from spin chain analysis, rely entirely on algebraic properties of the algebras. This paper will focus on the latter definition.

Instead of computing these fusion rules directly, Gainutnidov and Vasseur opted to follow a route closer to how these rules are defined in the Virasoro algebra. There, fusion is defined by first pushing modules to modules over a quantum group, using the co-multiplication on Virasoro, and are then pulled back to modules over the Lie algebra. However, there is no co-multiplication on  $\tl{n}$, so they instead used the quantum Schur-Weyl duality between $\tl{n}$ and the quantum group $U_{q}(sl_{2})$ \cite{Jimbo, SWMartin, SWCorfWeston}. Modules over $\tl{n}$ are first pushed to modules over this quantum group, where the co-multiplication naturally defines a fusion product, and the result is then pulled back to $\tl{n}$. They then argued that the resulting construction was equivalent to Read and Saleur's original one. Using this argument, they were able to compute fusion rules for most of the main classes of Temperley-Lieb modules\cite{GainVass}.

We are interested in generalizing this construction for the other, more exotic Temperley-Lieb algebras, in particular, the dilute $\dtl{n}$. While generalizing Read and Saleur's construction is simple enough, generalizing Gainutnidov and Vasseur's argument is not, mainly because the duality between $\dtl{n}$ and $U_{q}(sl_{2})$ is not so clear. Our goal is thus to compute directly this fusion product, without using this duality. We instead rely purely on category theory and the representation theory of $\tl{n}$ and $\dtl{n}$.

The outline of the paper is as follows. In section \ref{sec:rappel_tl}, we present a quick overview of the the representation theory of the regular $\tl{n}$ and the dilute $\dtl{n}$ families. None of these results are proved here; the reader can consult \cite{RiSY,BelSY,BelRiSy} for their proofs. In section \ref{sec:deffusion}, we present the generalization of Read and Saleur's construction for general family of algebras and then for dilute and regular cases. A natural consequence of this construction is the existence of a dual product, the \emph{fusion quotient}. Studying this new operation is beyond the scope of this paper but some results are nevertheless presented in appendix \ref{sec:fusion_quotient}. The fusion of projective modules is studied in section \ref{sec:fusionproj}. These turn out to admit a representation in terms of Chebyshev polynomials of the second kind. In section \ref{sec:fusion.standard}, we study the fusion of standard modules, first with projective modules and then with other standard modules. Fusion rules for irreducible modules are first studied in sections \ref{sec:fusionIvsP} and \ref{sec:fusionIvsS1}. These show the appearance of two other classes of modules, the $B$s and the $T$s. The fusion rules for those are studied in sections \ref{sec:fusionBvsP}, \ref{sec:fusionBvsS} and \ref{sec:fusionTvsP}, \ref{sec:fusionTvsS}, respectively. Fusion rules for pairs of irreducible modules are finally computed in section \ref{sec:fusionIvsIfinal}. In particular, a subset of irreducible modules is shown to behave under fusion like primary fields in a minimal model of the Virasoro algebra.
\end{section}

\begin{section}{Temperley-Lieb algebras}\label{sec:rappel_tl}
\emph{The results of this section first appeared in \cite{GoodmanWenzl,Martin,Westbury}. The definitions and results presented here are based on \cite{RiSY,BelSY,BelRiSy}.}

The Temperley-Lieb algebras can be defined in terms of generators or in terms of diagrams. The later is presented here and will be used throughout the paper as it gives a more intuitive description of the fusion product. After introducing this definition, the classes of indecomposable modules are introduced in terms of extensions. Loewy diagrams are given and can be used as a quick way of assessing the various properties of these modules. Finally, the algebra's families are described in terms of the induction and restriction functors.

The basic objects, $n$-diagrams, are first introduced. Draw two vertical lines, each with $n$ points on it, $n$ being a positive integer. Choose first $2m$ points, $0\leq m \leq n$ an integer, and put a $\circ$ on each of them. A point with a $\circ$ will be called a \emph{vacancy}. Now connect the remaining points, pairwise, with non-intersecting strings. The resulting object is called a {\em $n$-diagram}. If the diagram contains no vacancy, it is said to be \emph{dense}, and is called \emph{dilute} otherwise. If the number of vacancies on the left side of a $n$-diagram is odd (even), it is called \emph{odd}, (\emph{even}). For example,

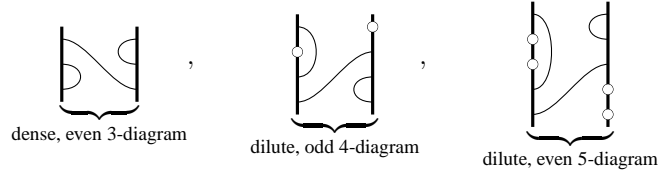
\begin{equation*}
	\underbrace{\begin{tikzpicture}[baseline={(current bounding box.center)},scale=1/3]
		\draw[very thick] (0,-1/2) -- (0,5/2);
		\draw[very thick] (3,-1/2) -- (3,5/2);
			\draw (0,0) .. controls (1,0) and (1,1) .. (0,1);
			\draw (0,2) .. controls (1,2) and (2,0) .. (3,0);
			\draw (3,1) .. controls (2,1) and (2,2) .. (3,2);
	\end{tikzpicture}}_{\text{dense, even $3$-diagram}}, \qquad
	\underbrace{\begin{tikzpicture}[baseline={(current bounding box.center)},scale=1/3]
		\draw[very thick] (0,-1/2) -- (0,7/2);
		\draw[very thick] (3,-1/2) -- (3,7/2);
			\draw (0,0) .. controls (1,0) and (2,2) .. (3,2);
			\draw (3,0) .. controls (2,0) and (2,1) .. (3,1);
			\draw (0,1) .. controls (1,1) and (1,3) .. (0,3);
				\vac{0,2};
				\vac{3,3};
	\end{tikzpicture}}_{\text{dilute, odd $4$-diagram}}, \qquad
	\underbrace{\begin{tikzpicture}[baseline={(current bounding box.center)},scale=1/3]
		\draw[very thick] (0,-1/2) -- (0,9/2);
		\draw[very thick] (3,-1/2) -- (3,9/2);
			\draw (0,0) .. controls (1,0) and (2,2) .. (3,2);
			\draw (0,1) .. controls (1,1) and (1,4) .. (0,4);
			\draw (3,3) .. controls (2,3) and (2,4) .. (3,4);
				\vac{0,2};
				\vac{0,3};
				\vac{3,0};
				\vac{3,1};
	\end{tikzpicture}}_{\text{dilute, even $5$-diagram}}
\end{equation*}

On the set of formal linear combinations of all $n$-diagrams a product is defined by extending linearly the product of two $n$-diagrams obtained as follows. The two diagrams are put side by side, the inner borders and the points on them are identified, then removed. A string which no longer ties two points is called a \emph{floating string}. A floating string that closes on itself is called a \emph{closed loop}. If all floating strings are closed loops, the result of the product of the two $n$-diagrams is then the diagram obtained by reading the vacancies on the left and right vertical lines and the strings between them multiplied by a factor of $\beta = q+q^{-1}$, $q$ a non-zero complex number, for each closed loop. Otherwise, the product is the zero element of the algebra. For example, 
\begin{equation*}
	\begin{tikzpicture}[baseline={(current bounding box.center)},scale=1/3]
		\draw[very thick] (0,-1/2) -- (0,5/2);
		\draw[very thick] (3,-1/2) -- (3,5/2);
			\draw (0,0) .. controls (1,0) and (1,1) .. (0,1);
			\draw (0,2) .. controls (1,2) and (2,0) .. (3,0);
			\draw (3,1) .. controls (2,1) and (2,2) .. (3,2);
	\end{tikzpicture} \times
	\begin{tikzpicture}[baseline={(current bounding box.center)},scale=1/3]
		\draw[very thick] (0,-1/2) -- (0,5/2);
		\draw[very thick] (3,-1/2) -- (3,5/2);
			\draw (0,0) .. controls (1,0) and (2,2) .. (3,2);
			\draw (0,2) .. controls (1,2) and (1,1) .. (0,1);
				\vac{3,0};
				\vac{3,1};
	\end{tikzpicture} = \beta
	\begin{tikzpicture}[baseline={(current bounding box.center)},scale=1/3]
		\draw[very thick] (0,-1/2) -- (0,5/2);
		\draw[very thick] (3,-1/2) -- (3,5/2);
			\draw (0,0) .. controls (1,0) and (1,1) .. (0,1);
			\draw (0,2) -- (3,2);
				\vac{3,0};
				\vac{3,1};
	\end{tikzpicture}  
\end{equation*}
If $q$ is a root of unity, the integer $\ell $ is defined as the smallest strictly positive integer such that $q^{2\ell}=1$. If $q$ is not a root of unity, $\ell $ is said to be infinite.

A dashed string represents the formal sum of two diagrams: one where the points are linked by a regular string, and one where the points are both vacancies. For example,
\begin{equation*}
\begin{tikzpicture}[baseline={(current bounding box.center)},scale=1/3]
	\bord{0}{3};
	\bord{3}{3};
	\link{0,0}{1,0}{1,1}{0,1};
	\wave{0}{2}
	\vac{3,0};
	\vac{3,1};
	\node at (4,1) {$=$};
\end{tikzpicture}
\begin{tikzpicture}[baseline={(current bounding box.center)},scale=1/3]
	\bord{0}{3};
	\bord{3}{3};
	\link{0,0}{1,0}{1,1}{0,1};
	\defect{0,2};
	\vac{3,0};
	\vac{3,1};
	\node at (4,1) {$+$};
	\bord{5}{3};
	\bord{8}{3};
	\link{5,0}{6,0}{6,1}{5,1};
	\vac{5,2};
	\vac{8,2};
	\vac{8,0};
	\vac{8,1};
\end{tikzpicture}\ \ .
\end{equation*}
Note that the diagram where each point is linked by a dashed line to the corresponding point on the opposite side acts as the identity on all $n$-diagrams and is a sum of $2^n$ $n$-diagrams.

Note finally that the product is clearly associative: the reading of how the left and right sides are connected in a product of three diagrams is blind to the order of glueing, and so is the number of closed loops. The set of $n$-diagrams with the formal sum with complex number coefficients and the product just introduced is the dilute Temperley-Lieb algebra $\dtl{n}=\dtl{n}(\beta)$. The subset spanned by only even (odd) diagrams is closed under the product and this subalgebra will be called the even (odd) dilute Temperley-Lieb subalgebra, denoted by $\edtl{n}$ ($\odtl{n}$).  Clearly any dilute $n$-diagram is either even or odd. Since the product of two diagrams of distinct parities is zero, it is clear that the even and odd subalgebras are two-sided ideals of $\dtl{n}$ and
$$\dtl n=\edtl n\oplus \odtl n.$$
A module on which every odd (even) diagram acts as zero is called even (odd). It follows that every module can be split into a direct sum of an even, and an odd modules. 

The regular Temperley-Lieb algebra $\tl{n} = \tl{n}(\beta)$ is obtained by considering only dense diagrams, that is, those containing no vacancies. As such, every non-zero $\tl{n}$-module is even. In the case $\beta =0 $ ($\ell = 2$), the structure of $\tl{n}$ will be slightly more complicated than for the other cases. It will thus be treated separately in many calculations and definitions.

%
\begin{subsection}{The indecomposable modules}\label{sec:rappel.indec}
%
Since the Temperley-Lieb algebras are finite dimensional associative algebras over the complex numbers, they have finitely many non-isomor\- phic, irreducible modules. In both algebras, these can be indexed by a single integer $0 \leq k \leq n$, which must be of the same parity as $n$ in $\tl{n}$, and are written $\dl{n,k}$. The only exception is when $\ell = 2$ in $\tl{n}$, where $\dl{n,0} \equiv 0$.

 These integers $k$ are first classified in orbits. If $\ell$ is a finite number, an integer $k \geq 0$ is said to be \emph{critical}, and is written  $\kc$ if $k+1 \equiv 0 \mod \ell$. If $\ell$ is not a finite number, every integer is said to be critical; this is also the case if $\ell = 1$. For a non-critical integer $k$, define $k_{1}$ to be the smallest non-critical integer strictly bigger than $k$  such that $(k_{1} + k )/2 +1 \equiv 0 \mod \ell$. Similarly, define $k_{-1}$ to be the biggest non-critical integer strictly smaller than $k$ such that $(k_{-1} + k)/2 +1 \equiv 0 \mod \ell$. Define inductively $(k_{i})_{j}= k_{i+j}$, so that for instance $(k_{1})_{-1} = k_{0} = k$. Two integers $r,k$ are then said to be in the same orbit if there exist $i \in \mathbb{Z}$ such that $r=k_{i}$; the modules $\dl{n,k},\dl{n,r}$ are also said to be on the same orbit. The irreducible modules $\dl{n,\kc}$ are each alone on their orbit. For instance, when $\ell = 3$, figure \ref{fig:orbits} shows the orbits between $-3$ and $16$.
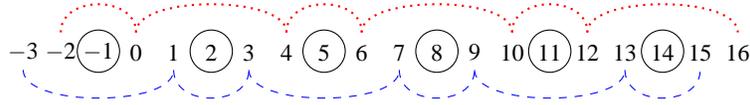
\begin{figure}[h]
\caption{Orbits when $\ell=3$: the critical numbers are circled, and the two other orbits are linked with dashed, and dotted lines respectively.}\label{fig:orbits}
\begin{tikzpicture}
	\def \n {16};
	\foreach \s in {-3,...,\n}
	{	
		\node[anchor = south] at (\s /2,0) {\small{$\s$}};
	}
	\foreach \s in {2,8,14}
	{
		\draw[dashed, blue] (\s/2 -1/2,0) .. controls (\s/2 -1/2,-1/2) and (\s/2 +1/2,-1/2) .. (\s/2+1/2,0);
		\draw[thick, dotted, red] (\s/2 -1,1/2) .. controls (\s/2 -1,1) and (\s/2 +1,1) .. (\s/2+1,1/2);
	}
	\foreach \s in {-1, 5, 11}
	{
		\draw[dashed,blue] (\s/2 -2/2,0) .. controls (\s/2 -2/2,-1/2) and (\s/2 +2/2,-1/2) .. (\s/2+2/2,0);
		\draw[thick, dotted, red] (\s/2 -1/2,1/2) .. controls (\s/2 -1/2,1) and (\s/2 +1/2,1) .. (\s/2+1/2,1/2);
	}
	\foreach \s in {-1,2,5,8,11,14}
	{
		\draw (\s/2,1/4) circle (8 pt); 
	}
\end{tikzpicture}
\end{figure}
\begin{Prop}\label{prop:extensions}
For $0 \leq r,k \leq n$,
\begin{equation}
\text{\rm{Ext}}\left(\dl{n,r},\dl{n,k}\right) \simeq \mathbb{C}\delta_{r,k_{\pm 1}}.
\end{equation}
\end{Prop} 

There is then a unique indecomposable module $\ds{n,k}$, up to isomorphism\footnote{Whenever we say that a module is unique, we will always mean ``up to isomorphism", but it will not always be mentioned.}, satisfying the short exact sequence
\begin{equation}
0 \longrightarrow \dl{n,k_{1}} \longrightarrow \ds{n,k} \longrightarrow \dl{n,k} \longrightarrow 0.
\end{equation} 
This defines the \emph{standard module} $\ds{n,k}$. In $\tl{n}$, when $\ell =2$, $\dl{n,0}=0$, so that $\ds{n,0}\equiv \dl{n,2}$. Note also that if $k_{1}>n$, the module $\dl{n,k_{1}}$ is simply not defined, in which case $\ds{n,k} \simeq \dl{n,k}$. It is generally consistent to set undefined irreducible modules to the zero module; we shall use this convention unless otherwise noted.

There is also a unique indecomposable module $\U{n,k}$, satisfying the short exact sequence
\begin{equation}
0 \longrightarrow \dl{n,k} \longrightarrow \U{n,k} \longrightarrow \dl{n,k_{1}} \longrightarrow 0.
\end{equation} 
This defines the \emph{dual standard module} $\U{n,k}$.

Let $\T{1}{n,k}= \ds{n,k}$, then $\T{2i}{n,k}$ is defined as the unique indecomposable extension of $\dl{n,k_{2i}}$ by $\T{2i-1}{n,k}$ and $\T{2i+1}{n,k}$ as the unique indecomposable extension of $\T{2i}{n,k}$ by $\dl{n,k_{2i+1}}$. Figure \ref{fig:LoewyofT} shows the Loewy diagrams of the smallest $\mathsf{T}$ modules.
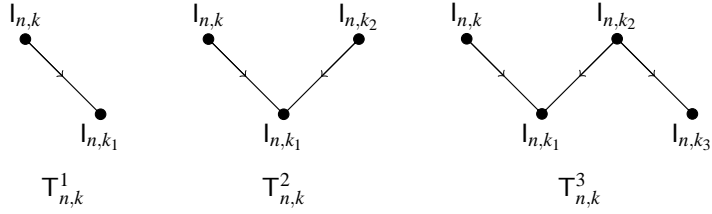
\begin{figure}[h]
\caption{Loewy diagrams of some $\mathsf{T}$ modules.}\label{fig:LoewyofT}
\begin{tikzpicture}[scale=1/3]
	\filldraw
			(0,0) circle (6pt) node[anchor = south] {$\dl{n,k}$}
		--	(3,-3) circle (6pt) node[anchor = north] {$\dl{n,k_{1}}$};
	\draw[->] (0,0) -- (1.5,-1.5);
	\node[anchor=north] at (1.5,-5) {$\T{1}{n,k}$};
\end{tikzpicture}\qquad
\begin{tikzpicture}[scale=1/3]
	\filldraw
			(0,0) circle (6pt) node[anchor = south] {$\dl{n,k}$}
		--	(3,-3) circle (6pt) node[anchor = north] {$\dl{n,k_{1}}$}
		--	(6,0) circle (6pt) node[anchor = south] {$\dl{n,k_{2}}$};
	\draw[->] (0,0) -- (1.5,-1.5);
	\draw[->] (6,0) -- (4.5,-1.5);
	\node[anchor=north] at (3,-5) {$\T{2}{n,k}$};
\end{tikzpicture}\qquad
\begin{tikzpicture}[scale=1/3]
	\filldraw
			(0,0) circle (6pt) node[anchor = south] {$\dl{n,k}$}
		--	(3,-3) circle (6pt) node[anchor = north] {$\dl{n,k_{1}}$}
		--	(6,0) circle (6pt) node[anchor = south] {$\dl{n,k_{2}}$}
		--	(9,-3) circle (6pt) node[anchor = north] {$\dl{n,k_{3}}$};
	\draw[->] (0,0) -- (1.5,-1.5);
	\draw[->] (6,0) -- (4.5,-1.5);
	\draw[->] (6,0) -- (7.5,-1.5);
	\node[anchor=north] at (4.5,-5) {$\T{3}{n,k}$};
\end{tikzpicture}
\end{figure}

Similarly, let $\B{1}{n,k} = \U{n,k}$ and define $\B{2i}{n,k}$ as the unique indecomposable extension of $\B{2i-1}{n,k}$ by $\dl{n,k_{2i}}$, and $\B{2i+1}{n,k}$ as the unique indecomposable extension of $\dl{n,k_{2i+1}}$ by $\B{2i}{n,k}$. Figure \ref{fig:LoewyofB} shows the Loewy diagrams of a few $\mathsf{B}$ modules.

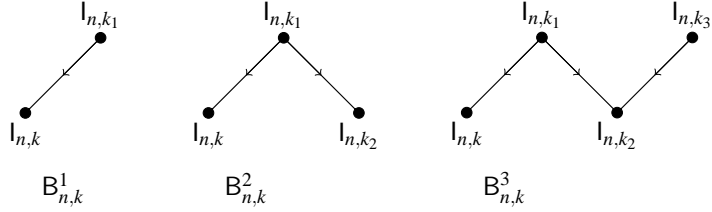
\begin{figure}[h]
\caption{Loewy diagrams of some $\mathsf{B}$ modules.}\label{fig:LoewyofB}
\begin{tikzpicture}[scale=1/3]
	\filldraw
			(0,0) circle (6pt) node[anchor = north] {$\dl{n,k}$}
		--	(3,3) circle (6pt) node[anchor = south] {$\dl{n,k_{1}}$};
	\draw[->] (3,3) -- (1.5,1.5);
	\node[anchor = north] at (1.5,-2) {$\B{1}{n,k}$}; 
\end{tikzpicture}\qquad
\begin{tikzpicture}[scale=1/3]
	\filldraw
			(0,0) circle (6pt) node[anchor = north] {$\dl{n,k}$}
		--	(3,3) circle (6pt) node[anchor = south] {$\dl{n,k_{1}}$}
		--	(6,0) circle (6pt) node[anchor = north] {$\dl{n,k_{2}}$};
	\draw[->] (3,3) -- (1.5,1.5);
	\draw[->] (3,3) -- (4.5,1.5);
	\node[anchor = north] at (1.5,-2) {$\B{2}{n,k}$}; 
\end{tikzpicture}\qquad
\begin{tikzpicture}[scale=1/3]
	\filldraw
			(0,0) circle (6pt) node[anchor = north] {$\dl{n,k}$}
		--	(3,3) circle (6pt) node[anchor = south] {$\dl{n,k_{1}}$}
		--	(6,0) circle (6pt) node[anchor = north] {$\dl{n,k_{2}}$}
		--	(9,3) circle (6pt) node[anchor = south] {$\dl{n,k_{3}}$};
	\draw[->] (3,3) -- (1.5,1.5);
	\draw[->] (3,3) -- (4.5,1.5);
	\draw[->] (9,3) -- (7.5,1.5);
	\node[anchor = north] at (1.5,-2) {$\B{3}{n,k}$}; 
\end{tikzpicture}
\end{figure}

The $\mathsf{P}$ modules are defined a bit differently. In the case $\lc =2$ of $\tl{n}$, $\dP{0}$ is the zero module. For all other cases, when $k $ is critical or smaller than $\ell-1$, $\dP{n,k}=\ds{n,k}$; otherwise, $\dP{n,k}$ is the unique indecomposable extension of $\ds{n,k}$ by $\ds{n,k_{-1}}$. Figure \ref{fig:LoewyofP} shows the Loewy diagrams of the $\mathsf{P}$ modules.
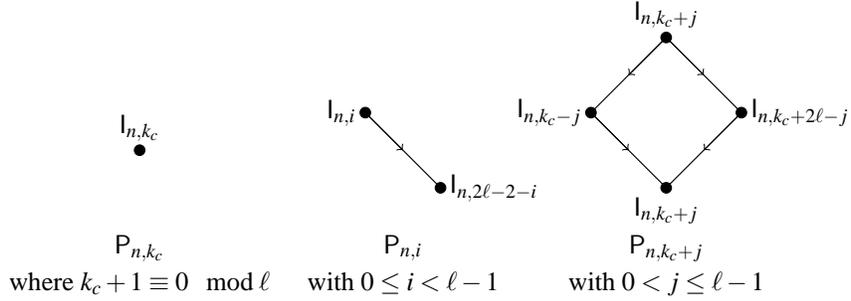
\begin{figure}[h]
	\caption{The Loewy diagrams of the $\mathsf{P}$ modules.}\label{fig:LoewyofP}
	\begin{tikzpicture}[scale=1/3]
	\filldraw
		(0,0) circle (6pt) node[anchor = east] {$\dl{n,i}$}
	--	(3,-3) circle (6pt) node[anchor = west] {$\dl{n,2\ell -2 -i}$};
	\draw[->] (0,0) -- (1.5,-1.5);
	\node[anchor = north] at (1.5,-4.5) {$\dP{n,i}$};
	\node[anchor = north] at (1.5,-6) {with $0\leq i<\ell -1$};
	\filldraw
		(-9,-1.5) circle (6pt) node[anchor = south] {$\dl{n,\kc}$};
	\node[anchor = north] at (-9,-4.5) {$\dP{n,\kc}$};
	\node[anchor = north] at (-9,-6) {where $\kc +1\equiv 0 \mod \ell$};
	\filldraw
	 	(9,0) circle (6pt) node[anchor = east] {$\dl{n,\kc -j}$}
	--	(12,-3) circle (6pt) node[anchor = north] {$\dl{n,\kc+j}$}
	--	(15,0) circle (6pt) node[anchor = west] {$\dl{n,\kc +2\ell -j}$}
	--	(12,3) circle (6pt) node[anchor = south] {$\dl{n,\kc +j}$}
	--	(9,0);
	\draw[->] (12,3) -- (10.5,1.5);
	\draw[->] (12,3) -- (13.5,1.5);
	\draw[->] (9,0) -- (10.5,-1.5);
	\draw[->] (15,0) -- (13.5,-1.5);
	\node[anchor = north] at (12,-4.5) {$\dP{n,\kc + j}$};
	\node[anchor = north] at (12,-6) {with $0 <j\leq\ell-1  $};
	\end{tikzpicture}	
\end{figure}

These modules satisfy several exact sequences which can all be read from their Loewy diagrams. For example, the short exact sequence
$$0 \longrightarrow \dl{n,k} \longrightarrow \B{3}{n,k} \longrightarrow \T{2}{n,k_{1}} \longrightarrow 0,$$
can be seen by noticing that in the Loewy diagram of $\B{3}{n,k}$, the part circled in a dashed line is precisely the Loewy diagram of $\T{2}{n,k_{1}}$:
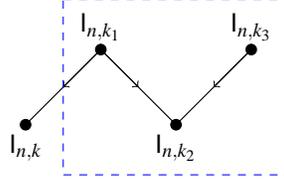
\begin{equation*}
\begin{tikzpicture}[scale=1/3]
	\filldraw
			(0,0) circle (6pt) node[anchor = north] {$\dl{n,k}$}
		--	(3,3) circle (6pt) node[anchor = south] {$\dl{n,k_{1}}$}
		--	(6,0) circle (6pt) node[anchor = north] {$\dl{n,k_{2}}$}
		--	(9,3) circle (6pt) node[anchor = south] {$\dl{n,k_{3}}$};
	\draw[->] (3,3) -- (1.5,1.5);
	\draw[->] (3,3) -- (4.5,1.5);
	\draw[->] (9,3) -- (7.5,1.5);
	\draw[dashed, blue] (1.5,5) -- (10.5,5) -- (10.5,-2) -- (1.5,-2) -- (1.5,5);
\end{tikzpicture}
\end{equation*}
The Hom spaces can similarly be read off their diagrams.
\begin{Prop}\label{prop:indec.prop}
\begin{itemize}
\item The $P$ modules are all projective; they form a complete set of non-isomorphic indecomposable projective modules.
\item $\dP{n,k}$ is the projective cover of $\dl{n,k}$.
\item If $f:\B{2i}{n,k} \to \overset{i}{\underset{j=0}{\bigoplus}} \dP{n,k_{2j}} $ is injective, then $\Coker f \simeq \B{2(i+1)}{n,k_{-1}}$. If $k_{-1}<0$, there are no such morphism.
\item If $g:\T{2i+1}{n,k_{-1}} \to \overset{i}{\underset{j=0}{\bigoplus}} \dP{n,k_{2j}}$ is injective, then $\Coker g \simeq \T{2i+1}{n,k}$.
\item The modules $\dP{n,k}$ are injective for all $k\geq \ell-1$, and the modules $\B{1}{n,k}$ are also for all $k<\ell -1$, except if $\ell =2$ in $\tl{n}$ in which case $\B{1}{n,0}$ is not injective. They form a complete set of non-isomorphic indecomposable injective modules.
\item The injective hull of $\dl{n,k}$ is $\B{1}{n,k}$ if $k<\ell-1$ and $\dP{n,k}$ otherwise.
\end{itemize}
\end{Prop}
\end{subsection}
\begin{subsection}{A basis of $\ds{n,k}$}\label{sec:rappel.base.standard}
Our computations will almost all be based on the short-exact sequences satisfied by the various modules and on their homological properties, they will therefore be completely independent of a choice of basis. However, a basis of the standard module $\ds{n,k}$ will be needed. The bases we present here are the usual ones used in the Temperley-Lieb algebras so the reader should feel free to skip this section if they are already familiar with them.

Start by defining the basic objects, the $n$-link diagrams, which are built in the following way. First, take a dilute $n$-diagram and remove its right (left) side as well as the points that were on it. An object, whether it is a string or a vacancy that no longer touches any point, is simply removed. The other floating strings are straightened out and called \emph{defects}. For example,
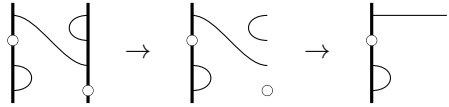
\begin{alignat*}{3}
\begin{tikzpicture}[scale = 1/3]
	\bord{0}{4};
	\bord{3}{4};	
	\link{0,3}{1,3}{2,1}{3,1};
	\link{0,0}{1,0}{1,1}{0,1};
	\link{3,2}{2,2}{2,3}{3,3};
	\vac{0,2};
	\vac{3,0};
	\node at (5,1.5) {$\rightarrow$     };
\end{tikzpicture}&\quad
\begin{tikzpicture}[scale=1/3]
	\bord{0}{4};
	\link{0,3}{1,3}{2,1}{3,1};
	\link{0,0}{1,0}{1,1}{0,1};
	\link{3,2}{2,2}{2,3}{3,3};
	\vac{0,2};
	\vac{3,0};
	\node at (5,1.5) {$\rightarrow$     };
\end{tikzpicture}& \quad
\begin{tikzpicture}[scale=1/3]
	\bord{0}{4};
	\defect{0,3};
	\link{0,0}{1,0}{1,1}{0,1};
	\vac{0,2};
	\node at (4,1.5) {\ };
\end{tikzpicture}
\end{alignat*}
The resulting diagram is called a {\em left $n$-link} ({\em right $n$-link}). It is seen that a dilute $n$-diagram induces a unique pair of one left and one right $n$-link diagrams and that, given such a pair, there can be at most one $n$-diagram, if any, that could have induced them. It will thus be useful to denote an $n$-diagram by its induced $n$-links, $b= |l r|$, where $l$ ($r$) is the left (right) link diagram induced from $b$.  This notation can also be used for linear combinations of $n$-diagrams as in $b=|(l+j)r| + |uv|$ where $l, j, u$ are left $n$-links and $r, v$ right ones. If $u$ is a left link, then $\bar u$ will denote its (right) mirror image.

A natural action can be defined of $n$-diagrams on left (and right) $n$-link diagrams. We start with the left action. Draw the $n$-diagram on the left side of the left $n$-link, identify the points on its right side with those on the link and remove them. Each floating string that is not connected to the remaining side is removed and yields a factor $\beta$ if it is closed and zero if it opened, or touches a vacancy. If a floating string starting on the remaining side is connected to a defect in the $n$-link diagram, it becomes a defect. Finally, remove any remaining vacancies on the right side. The remaining drawing is the resulting $n$-link diagram, weighted by factors of $\beta$, one for each closed floating strings. For example
\begin{align*}
\begin{tikzpicture}[scale = 1/3]
	\bord{0}{4};
	\bord{3}{4};
	\link{0,3}{1,3}{2,1}{3,1};
	\link{0,0}{1,0}{1,1}{0,1};
	\link{3,2}{2,2}{2,3}{3,3};
	\vac{0,2};
	\vac{3,0};
	\bord{4}{4};
	\defect{4,3};
	\link{4,2}{5,2}{5,1}{4,1};
	\vac{4,0};
	\node at (8,2) {$=$ };
\end{tikzpicture}&
\begin{tikzpicture}[scale = 1/3]
	\bord{0}{4};
	\link{0,3}{1,3}{2,1}{3,1};
	\link{0,0}{1,0}{1,1}{0,1};
	\link{3,2}{2,2}{2,3}{3,3};
	\vac{0,2};
	\vac{3,0};
	\defect{3,3};
	\link{3,2}{4,2}{4,1}{3,1};
	\vac{3,0};
	\node at (7,2) {$=$ };
\end{tikzpicture}
\begin{tikzpicture}[scale = 1/3]
	\bord{0}{4};
	\defect{0,3};
	\link{0,0}{1,0}{1,1}{0,1};
	\vac{0,2};
\end{tikzpicture}
\end{align*}
\begin{Prop}
Over $\dtl{n}$, the formal sums of all $n$-link diagrams having exactly $k$-defects, with the action defined above, defines a basis of $\ds{n,k}$.

Over $\tl{n}$, the formal sums of all $n$-link diagrams having exactly $k$-defects and no vacancies, with the action defined above, defines a basis of $\ds{n,k}$.
\end{Prop}
\end{subsection}
\begin{subsection}{The Temperley-Lieb families}\label{sec:rappel.induction}
%
There is a natural inclusion of the symmetric group $\mathsf{S}_{n}$ into $\mathsf{S}_{n+1}$. There are similar inclusion for the Temperley-Lieb algebras. Consider the following transformation: take a $n$-diagram and add a dashed line at its bottom. The result is an element of $\dtl{n+1}$. Similarly, taking a dense $n$-diagram and adding a straight line at its bottom yields a dense $(n+1)$-diagram which is an element of $\tl{n+1}$. Extending the first transformation linearly gives a subalgebra of $\dtl{n+1}$ isomorphic to $\dtl{n}$, while doing the same thing to the second yields a subalgebra of $\tl{n+1}$ isomorphic to $\tl{n}$. There are thus two ascending families of algebras
\begin{equation*}
\dtl{1} \subset \dtl{2} \subset \dtl{3} \subset \hdots, \qquad \text{and }\tl{1} \subset \tl{2} \subset \tl{3} \subset \hdots
\end{equation*}
The functor $\Ind{-}^{n+1}_{n}$ is the induction functor from $\dtl{n}$ to $\dtl{n+1}$, or from $\tl{n}$ to $\tl{n+1}$. While this really defines multiple functors, they will have similar properties so we  write them all $\Ind{-}$, unless it is not clear which one we are talking about from the context. The induction functor from a subalgebra $B$ to an algebra $A$ is always a right-exact linear functor defined on all $B$-module $U$ by 
\begin{equation*}
\Ind{U} =  A \otimes_{B} U,
\end{equation*}
where $A$ is seen as a left $A$-module and a right $B$-module, and the index $B$ next to the tensor product sign means that elements of $B$ can pass freely through it.

As the induction functors ``moves up" along the families, its adjoint, the restriction functor $\Res{-}$ ``moves down", taking $\dtl{n+1}$-modules to $\dtl{n}$-modules or $\tl{n+1}$-modules to $\tl{n}$-modules. The restriction functor from an algebra $A$ to a subalgebra $B$ is always an exact, linear functor defined on an $A$-module $V$ by 
\begin{equation*}
\Res{V} = \text{Hom}_{A} \left( A , V \right),
\end{equation*}
where $A$ is seen as a left $A$-module and a right $B$-module.

These functors have been computed before for all indecomposable modules over either family of Temperley-Lieb algebras in \cite{RiSY,BelSY,BelRiSy}. These results will be very important for computing the fusion rules and they will be stated where they are needed.
%
\end{subsection}
%

\end{section}
\begin{section}{The Fusion ring}\label{sec:deffusion}
Fusion is first defined for left modules over a general family of algebras. This definition is a straightforward generalization of the definition in \cite{ReadSaleur1,ReadSaleur2,GainVass}, which works for the regular Temperley-Lieb family. Some general results are then proven before studying fusion in the Temperley-Lieb families.
\begin{subsection}{The fusion product}
Consider $(A_i)_{i \in \mathbb{N}}$ a family of associative algebras over $\mathbb{C}$ such that for all positive integers $i,j$ the tensor algebra $A_{i} \otimes_{\mathbb{C}} A_{j}$ is isomorphic to a subalgebra of $A_{i+j}$. The tensor algebra  $A_{i} \otimes_{\mathbb{C}} A_{j}$ is defined such that $(a \otimes b)(c\otimes d) = ac \otimes bd$ for all $a,c \in A_{i}$ and all $b,d \in A_{j}$. Given $U$ a $A_{i}$-module and $V$ a $A_{j}$-module, the \emph{fusion} of $U$ and $V$ is defined as 
\begin{equation}
U \xf V = A_{i+j} \otimes_{A_{i}\otimes_{\mathbb{C}} A_{j}}\left( U \otimes_{\mathbb{C}} V \right)\text{.}
\end{equation}
Note that $U \otimes_{\mathbb{C}} V$ is naturally a $A_{i}\otimes_{\mathbb{C}} A_{j}$-module. The fusion can thus be seen as a simple induction from $A_{i}\otimes_{\mathbb{C}} A_{j}$ to $A_{i+j}$ and, hence, $U \xf V$ is an $A_{i+j}$-module. Note that to each induction functor corresponds an adjoint restriction functor. As such, there exists a construction adjoint to the fusion product which is called the \emph{fusion quotient}. This construction will only be used while computing the fusion product of irreducible modules, and the argument to obtain the needed fusion quotients are slightly different from those used to compute fusion products. These results will therefore be presented in appendix \ref{sec:fusion_quotient}.

 The following propositions follow readily from the properties of tensor products.
\begin{Prop}\label{prop:fusion_resproj}
For $U$,$V$ two $A_{i}$-modules and $W$,$Z$ two $A_{j}$-modules, $$(U\oplus V)\xf(W\oplus Z) \simeq (U\xf W) \oplus (U \xf Z) \oplus (V\xf W) \oplus (V\xf Z) \text{.}$$Furthermore, if $U$ and $W$ are projective then so is $U\xf W$.
\end{Prop}
\begin{proof}
The first result follows readily from the linearity of tensor products.

Suppose now that $U$ and $W$ are two projective $A_{i}$- and $A_{j}$-modules respectively. By definition, this means that there are two sets $\Lambda$ and $\Sigma$ and two projective modules $P$ and $Q$ such that $A^{\Lambda}_{i} \simeq U \oplus P$ and $A^{\Sigma}_{j}\simeq W \oplus Q$. Here $A^{\Lambda}_{i}$ is a direct sum of copies of $A_{i}$ indexed by the elements of $\Lambda$ and similarly for $A^{\Sigma}_{j}$. Using the first result,
\begin{equation}
A^{\Lambda}_{i} \xf A^{\Sigma}_{j} \simeq (U \xf W) \oplus (U\xf Q) \oplus (P\xf W) \oplus (P\xf Q) \simeq A^{\Gamma}_{i+j}, 
\end{equation}
were $\Gamma$ is a set whose elements are the pairs $(\lambda, \sigma)$ with $\lambda \in \Lambda$,$\sigma \in \Sigma$. The second equality is obtained by noting that the induction to an algebra $A$ of a subalgebra $B$ is always isomorphic to $A$. Since $A^{\Gamma}_{i+j} $ is a free module by definition, $U\xf W$ is projective.
\end{proof}
\begin{Prop}\label{prop:fusionexact}
If the sequence $$ 0 \longrightarrow U \overset{f}{\longrightarrow} V \overset{g}{\longrightarrow} W \longrightarrow 0$$ of $A_{i}$-modules is exact, the sequence of $A_{i+j}$-modules $$ U\xf S \overset{\bar{f}}{\longrightarrow} V\xf S \overset{\bar{g}}{\longrightarrow} W\xf S \longrightarrow 0$$ is also exact for all $A_{j}$-modules $S$.
\end{Prop}
\begin{proof}
Note that $\mathbb{C}$ is semi-simple, so that all $\mathbb{C}$-modules are flat. The sequence of $A_{i}\otimes_{\mathbb{C}} A_{j}$-modules $$0 \longrightarrow U \otimes_{\mathbb{C}} S \overset{f \otimes_{\mathbb{C}} \text{id}_{A_{j}}}{\longrightarrow} V\otimes_{\mathbb{C}} S \overset{g\otimes_{\mathbb{C}} \text{id}_{A_{j}}}{\longrightarrow} W\otimes_{\mathbb{C}} S \longrightarrow 0 ,$$ is therefore exact. The conclusion is obtained by using the fact that induction is right-exact.
\end{proof}
It should also be noted that for any $A_{i}$-module $U$,
\begin{equation}
U \xf A_{j} = A_{i+j} \otimes_{A_{i}\otimes_{\mathbb{C}} A_{j}} U \otimes_{\mathbb{C}} A_{j} \simeq A_{i+j} \otimes_{A_{i}} U \text{,}
\end{equation}
which is simply the induction functor from $A_{i}$ to $A_{i+j}$. Note also that just like the induction functor, it will depend on the actual embedding $A_{i} \to A_{i+j}$. 
\end{subsection}
\begin{subsection}{Fusion on the dilute Temperley-Lieb family}
Of the many ways of including $\dtl{n}$ as a subalgebra of $\dtl{n+p}$, we focus on two. The first is to insert $p$ dashed lines at the bottom of every diagram in $\dtl{n}$ and the other is to add them at the top. The simplest way to define the inclusion of $\dtl{n}\otimes\dtl{p}$ in $\dtl{n+p}$ is thus to draw the diagram $a\in \dtl{n}$ on top of $b\in\dtl{p}$. For example,
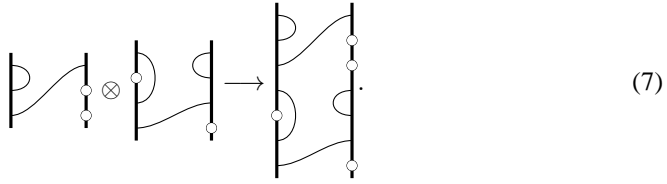
\begin{equation}
\begin{tikzpicture}[baseline={(current bounding box.center)},scale = 1/3]
\draw[very thick] (0,-3/2) -- (0,3/2);
\draw[very thick] (3,-3/2) -- (3,3/2);
	\draw (0,-1) .. controls (1,-1) and (2,1) .. (3,1);
	\draw (0,0) .. controls (1,0) and (1,1) .. (0,1);
	\vac{3,0};
	\vac{3,-1};
\node  at (4,0) {$\otimes$};
\draw[very thick] (5,-2) -- (5,2);
\draw[very thick] (8,-2) -- (8,2);
	\draw (5,-3/2) .. controls (6,-3/2) and (7,-1/2) .. (8,-1/2);
	\draw (5,-1/2) .. controls (6,-1/2) and (6,3/2) .. (5,3/2);
	\vac{5,1/2};
	\vac{8,-3/2};
	\draw (8,1/2) .. controls (7,1/2) and (7,3/2) .. (8,3/2);
\end{tikzpicture}
\longrightarrow
\begin{tikzpicture}[baseline={(current bounding box.center)},scale = 1/3]
\draw[very thick] (0,-7/2) -- (0,7/2);
\draw[very thick] (3,-7/2) -- (3,7/2);
	\draw (0,1) .. controls (1,1) and (2,3) .. (3,3);
	\draw (0,2) .. controls (1,2) and (1,3) .. (0,3);
	\vac{3,2};
	\vac{3,1};
	\draw (0,-3) .. controls (1,-3) and (2,-2) .. (3,-2);
	\draw (0,-2) .. controls (1,-2) and (1, 0) .. (0,0);
	\vac{0,-1};
	\vac{3,-3};
	\draw (3,-1) .. controls (2,-1) and (2,0) .. (3,0);
\end{tikzpicture}.
\end{equation}

Notice that we could have defined it the other way around, drawing $b$ on top of $a$. It can be shown that the two inclusions yield isomorphic bi-module structures on $\dtl{n+m}$. It follows that fusion is commutative on the dilute Temperley-Lieb family.
\begin{Prop}
For $U,V$, modules over $\dtl{n}$ and $\dtl{p}$, respectively, $$U\xf V \simeq V\xf U. $$ 
\end{Prop}
Note that the inclusion used is compatible with the parity of diagrams. Take $a,b$ two diagrams with well-defined parity in $\dtl{n}$ and $\dtl{m}$, respectively. If $a$ is odd but $b$ is even, $a \otimes b$ is odd  while if they are both odd or even, $a\otimes b$ is even. It follows that fusing two modules with the same parity yields an even module while if their parities are different it yields an odd one. Note also that fusing a module with $\dtl{1}$ gives the induction of this module as defined in \ref{sec:rappel.induction}. Since $\dtl{1}\simeq \dP{1,1} \oplus \dP{1,0}$ the following proposition is obtained.
\begin{Prop}\label{prop:fusion_parity}
For a $\dtl{n}$-module $V$ with a well-defined parity,
$$ V\xf \dtl{1} \simeq \Ind{V} \simeq V\xf \dP{1,0} \oplus V \xf \dP{1,1},$$
 $V\xf\dP{1,1}$ has the same parity as $V$, while $V\xf \dP{1,0}$ has a different parity.
\end{Prop}

Furthermore, tensor products are associative and it is easy to verify that the chosen inclusion process is also. It thus follows that the fusion algebra of the dilute Temperley-Lieb family is associative.
\begin{Prop}
For $U$ a  $\dtl{n}$-module, $V$ a $\dtl{m}$-module and $W$ a $\dtl{p}$-module,
\begin{equation}
\left(U \xf V\right)\xf W \simeq U\xf \left(V \xf W \right).
\end{equation}
\end{Prop}
\end{subsection}
\begin{subsection}{Fusion on the regular Temperley-Lieb algebra}
Fusion for the regular Temperley-Lieb family is very similar to that on the dilute family. Again inclusion of $\tl{n}$ in $\tl{n+p}$ can be obtained by adding straight lines below or above an $n$-diagram and inclusion of $\tl{n}\otimes\tl{p}$ in $\tl{n+p}$ by drawing $n$-diagrams atop $p$-diagrams. For example,
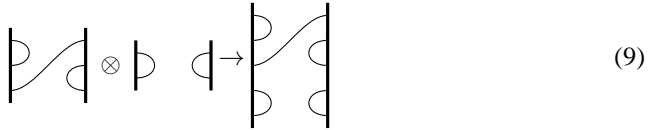
\begin{equation}
\begin{tikzpicture}[baseline={(current bounding box.center)},scale = 1/3]
\draw[very thick] (0,-3/2) -- (0,3/2);
\draw[very thick] (3,-3/2) -- (3,3/2);
	\draw (0,-1) .. controls (1,-1) and (2,1) .. (3,1);
	\draw (0,0) .. controls (1,0) and (1,1) .. (0,1);
	\draw (3,-1) .. controls (2,-1) and (2,0) .. (3,0);
\node  at (4,0) {$\otimes$};
\draw[very thick] (5,-1) -- (5,1);
\draw[very thick] (8,-1) -- (8,1);
	\draw (5,-1/2) .. controls (6,-1/2) and (6,1/2) .. (5,1/2);
	\draw (8,-1/2) .. controls (7,-1/2) and (7,1/2) .. (8,1/2);
\end{tikzpicture}
\to
\begin{tikzpicture}[baseline={(current bounding box.center)},scale = 1/3]
\draw[very thick] (0,-5/2) -- (0,5/2);
\draw[very thick] (3,-5/2) -- (3,5/2);
	\draw (0, -2) .. controls (1,-2) and (1,-1) .. (0,-1);
	\draw (0,0) .. controls (1,0) and (2,2) .. (3,2);
	\draw (0,1) .. controls (1,1) and (1,2) .. (0,2);
	\draw (3,-2) .. controls (2,-2) and (2,-1) .. (3,-1);
	\draw (3,0) .. controls (2,0) and (2,1) .. (3,1);
\end{tikzpicture}
\end{equation}

The definition mimics very closely that on the dilute family and the proofs of the various results will be nearly identical. In particular, the same arguments yields the following proposition.
\begin{Prop}\label{prop:fusion_regular}
For $U$ a $\tl{n}$-module, $V$ a $\tl{m}$-module and $W$ a $\tl{k}$-module,
\begin{equation}
U \xf \tl{1} \simeq U \xf \ds{1,1} \simeq \Ind{U},
\end{equation}
\begin{equation}
U \xf V \simeq V\xf U,
\end{equation}
\begin{equation}
\left(U \xf V\right)\xf W \simeq U\xf \left(V \xf W \right).
\end{equation}
\end{Prop}
\end{subsection}
\end{section}
\begin{section}{Fusion of projective modules}\label{sec:fusionproj}
It was proved in proposition \ref{prop:fusion_resproj} that the fusion of two projective modules always yields a projective module. Since the projective modules of the Temperley-Lieb algebras are all known, it is natural to start by computing their fusion rules. The projective indecomposables of $\dtl{n}$ and $\tl{n}$ falls in three different classes (see section \ref{sec:rappel.indec}), the standard modules $\ds{n,k}=\dP{n,k}$ with $k<\lc-1$, which we will often call the \emph{small projectives}, the standard modules $\ds{n,\kc}=\dP{n,\kc}$ where $\kc$ is critical and the projective indecomposable $\dP{n,\kc+i}$ for $0<i<\lc$. We use the same notation for the two families, but recall that in $\dtl{n}$, modules such as $\dP{n,k}$ are defined for all integer $k \in [0,n]$, while in $\tl{n}$, they are only defined when $k \equiv n \mod 2$.
Propositions \ref{prop:fusion_parity} and \ref{prop:fusion_regular} show that fusion is closely related to the process of induction; the following proposition gives the induction of projective modules \cite{BelSY,BelRiSy}.
\begin{Prop}
For all critical $\kc$, $0<i<\lc$, and $n-1\geq \kc +i$,
\begin{equation}
\Ind{\dP{n-1,i-1}} \simeq \begin{cases}
\dP{n,i-2} \oplus \dP{n,i-1} \oplus \dP{n,i} \text{,} & \text{ on } \dtl{n}\\
\dP{n,i-2} \oplus \dP{n,i} , & \text{ on } \tl{n}
\end{cases},
\end{equation}
\begin{equation}
\Ind{\dP{n-1,\kc}} \simeq \begin{cases}
 \dP{n,\kc} \oplus \dP{n,\kc +1} \text{,} & \text{ on } \dtl{n}\\
\dP{n,\kc +1} , & \text{ on } \tl{n}
 \end{cases},
\end{equation}
\begin{multline}
\Ind{\dP{n-1,\kc+i}} \simeq \left. \begin{cases}
\dP{n,\kc+i},& \text{ on } \dtl{n}\\
0, & \text{ on } \tl{n}
\end{cases} \right\}
 \oplus \left.\begin{cases}
\dP{n,\kc} \oplus \dP{n,\kc}, & \text{ if } i=1\\
\dP{n,\kc +i-1}, & \text{ otherwise}
\end{cases}\right\} \\
\oplus \left.\begin{cases}
\dP{n,\kc -\lc} \oplus \dP{n,\kc + \lc}, & \text{ if } i=\lc-1\\
 \dP{n,\kc+i+1}, & \text{ otherwise}
 \end{cases}\right\},
\end{multline}
where it is understood that $\dP{n,j} \simeq 0$ if $j<0$.
\end{Prop}
Proposition \ref{prop:fusion_parity} described how fusion behaves regarding parity of modules: the fusion of two odd or even modules yields an even module while the fusion of an odd and an even module yields an odd one. A projective module $\dP{n,k}$ is odd (even) if $n-k$ is odd (even); the following proposition is thus easily proven.
\begin{Prop}\label{prop_fusion_pvs01}
For all critical $\kc$, $0<i<\lc$, and $n-1\geq \kc +i$, on the dilute family
\begin{equation}
\dP{n-1,i-1} \xf \dP{1,0} \simeq \dP{n,i-1} \text{,} \qquad \dP{n-1,\kc} \xf \dP{1,0} \simeq \dP{n,\kc},
\end{equation}
\begin{equation}
 \dP{n-1,\kc +i} \xf\dP{1,0} \simeq \dP{n,\kc +i} \text{,}
\end{equation}
and in both families
\begin{equation}
\dP{n-1,i-1} \xf \dP{1,1} \simeq \dP{n,i-2} \oplus \dP{n,i}\text{,}\qquad \dP{n-1,\kc} \xf \dP{1,1} \simeq \dP{n,\kc +1} \text{,}
\end{equation}
\begin{equation}
\dP{n-1,\kc +i}\xf \dP{1,1} \simeq \begin{cases}
\dP{n,\kc} \oplus \dP{n,\kc} & \text{ if } i=1\\
\dP{n,\kc +i-1} & \text{ otherwise}
\end{cases} \oplus \begin{cases}
\dP{n,\kc -\lc} \oplus \dP{n,\kc + \lc} & \text{ if } i=\lc-1\\
 \dP{n,\kc+i+1} & \text{ otherwise}
 \end{cases}.
\end{equation}
\end{Prop}
\begin{proof}
It follows from the previous proposition together with the linearity of fusion, the breakdown according to parity and the fact that $\dtl{1} \simeq \dP{1,0} \oplus \dP{1,1}$ and $\tl{1} \simeq \dP{1,1}$.
\end{proof}
For all projective modules in the dilute family, fusion of projectives with $\dP{1,0}$ simply increases the parameter $n$ by one. Since fusion is associative, fusions can be computed using the smallest $n$ for which the modules make sense, and fuse the result with the appropriate number of $\dP{1,0}$ needed to reach the required $n$. For instance $$\dP{10,3} \xf \dP{6,4} \simeq \dP{1,0} \xf (\dP{9,3} \xf \dP{6,4}) \simeq \dP{2,0} \xf (\dP{8,3} \xf \dP{6,4}) \simeq \hdots \simeq \dP{9,0} \xf (\dP{3,3} \xf \dP{4,4}) .$$ In the regular family, this role is played by $\dP{2,0}$, when $\lc \neq 2$. Then
$$\dP{2,0}\xf \dP{n,p} \simeq \dP{n+2,p},  $$ for all $p$. The proof is much more involved and based on diagrammatic arguments; it is presented in appendix \ref{sec:magicp2}. When $\lc =2$, it will be proved as a corollary of proposition \ref{prop:fusionIvsP} that this role is played by $\dl{4,2} \simeq \ds{4,0}$. The results could therefore depend on the parity of $n/2$. Nevertheless, most of our proofs will be independent of $n$, so we will simply write $\dP{p} = \dP{n,p}$ and assume that $n$ is big enough for the module to exist. Proofs where $n$ is important will be dealt with separately.

\begin{subsection}{The fusion matrix}

For a projective module $\mathsf{P}$, define the \emph{fusion matrix} $\text{F}(\mathsf{P})$ by $$ \mathsf{P} \xf \dP{j} \simeq \bigoplus_{k} (\text{F}(\mathsf{P}))_{j}^{k} \dP{k}$$ where it is understood that a non-negative integer multiple of a module stands for that many copies of this module. To simplify the notation, $k$ is allowed to run over all non-negative integers, but it is assumed that $\dP{n,k}\simeq 0$ when $k >n$, or when $k \not\equiv n \mod 2$ in the regular family. Define also $X = \text{F}(\dP{1,1})$, $\F{i} = \text{F} (\dP{i})$ and write $(\F{i})^{k}_{j} = \F{i,j}^{k}$. This definition will reduce the computation of fusion rules to simple products of matrices. Note that since fusion is commutative, $\F{i,j}^{k} = \F{j,i}^{k}$.

Proposition \ref{prop_fusion_pvs01} already gives the fusion matrices of $\dP{0}$ and $\dP{1}$:
\begin{equation}
\F{0,i}^{j} = \delta_{i,j},
\end{equation}
\begin{equation}
\F{1,i}^{j} = \left. \begin{cases}
\delta_{j,i+1}, & \textrm{if } i=0 \text{ or } i+1 \equiv 0 \mod \lc\\
2\delta_{j,i-1} + \delta_{j,i+1}, & \textrm{if } i \equiv 0\mod \lc \text{ and } \lc \neq 2\\
\delta_{j,i-1} + \delta_{j,i+1} + \delta_{j, i-2\lc +1}, & \textrm{if } i>\lc-1 \text{ and } i+2 \equiv 0\mod \lc \text{ and } \lc \neq 2\\
2\delta_{j,i-1} + \delta_{j,i+1} + \delta_{j, i-2\lc +1}, & \textrm{if } i>\lc-1 \text{ and } i+2 \equiv 0\mod \lc \text{ and } \lc =2 \\
\delta_{j,i-1} + \delta_{j,i+1}, & \textrm{otherwise}
\end{cases}\right\}
\end{equation}
where $\delta_{i,j}$ is the Kronecker delta.

The following proposition shows that a finite projective module is uniquely determined by its fusion matrix.
\begin{Prop}\label{prop:fmatrixunique}
For $P$,$Q$ two finite projective $\dtl{n}$- or $\tl{n}$-modules,
\begin{equation*}
\text{F}(P) = \text{F}(Q) \iff P \simeq Q.
\end{equation*}
\end{Prop}
\begin{proof}

Every finite projective module is isomorphic to a direct sum of principal indecomposable modules. For a projective module $T$, define the set $\alpha(T)$ as the set of integers such that $$T \simeq \oplus_{i \in \alpha(T)} \dP{i}, $$ where each integer can occur more than once. Define $i(T)$ as the maximum of $\alpha(T)$, and $\# i(T)$ as the number of times this maximum appears. From proposition \ref{prop_fusion_pvs01}, it is clear that $ i(T\xf \dP{1,1}) = i(T) + 1$ and $ \# i(T\xf \dP{1,1}) = \# i(T)$.

Now, for $P,Q$ two projective $\dtl{n}$- or $\tl{n}$-modules, if $\text{F}(P) = \text{F}(Q)$, then in particular $ P \xf \dP{1,1} \simeq Q\xf \dP{1,1}$. Thus $i(P\xf \dP{1,1}) = i(Q\xf \dP{1,1})$ and $\# i(P\xf \dP{1,1}) = \# i(Q\xf \dP{1,1})$. Therefore $i(P)= i(Q)$, $\# i(P)=\# i(Q)$, and
\begin{equation}
P \simeq P' \oplus \# i(P)\dP{i(P)}, \qquad Q \simeq Q' \oplus \# i(P) \dP{i(P)},
\end{equation}
where $i(P') < i(P)$ and $i(Q')<i(Q)$. Since fusion is linear, $ P' \xf \dP{1,1} \simeq Q'\xf \dP{1,1}$. Proceeding by recursion on the cardinality of $i(P)$, the result is obtained.
\end{proof}

\end{subsection}
\begin{subsection}{Fusion matrices of small projectives}\label{sec:fmatrixsmall}
By using the formulas in proposition \ref{prop_fusion_pvs01}, for $0 \leq i<\lc-1$ and all $j$,
\begin{equation}
\dP{1} \xf ( \dP{i} \xf \dP{j}) \simeq (\dP{i-1} \oplus \dP{i+1}) \xf \dP{j} \text{.}
\end{equation}
In terms of the fusion matrices, this is simply
\begin{equation}
\sum_{m}\F{1,m}^{p} \F{i,j}^{m} = \F{i-1,j}^{p} + \F{i+1,j}^{p}
\end{equation}
and this gives the recurrence relation
\begin{equation}
X\F{i} = \F{i-1} + \F{i+1} \text{,}\qquad \F{0} = \text{id} \text{,  } \F{1} = X , 
\end{equation}
where $$ (X\F{i})^{p}_{j} = \sum_{m=0}^{n} X_{m}^{p}(\F{i})^{m}_{j}=\sum_{m=0}^{n} \F{1,m}^{p} \F{i,j}^{m}.$$ One should recognize the recurrence relation\footnote{Note that the Chebyshev solution to this recurrence relation is valid on $\mathbb{C}[ X ]$ even when $X$ is a matrix.} of Chebyshev polynomials of the second kind $\U{i}\left(\frac{X}{2}\right)$ and thus find $$\F{i} = \U{i}\left(\frac{X}{2}\right) \text{,}\qquad  0\leq i \leq l-1 \text{.}$$ Since the matrix $X$ is known, this can be used in principle to compute the fusion matrix of all small projectives. Note that this proof fails when $\lc =2$ on the regular family because in this case, there are no small projectives.
\end{subsection}
\begin{subsection}{Fusion matrices for the indecomposable projective $\dP{\kc+i}$}
Using again proposition \ref{prop_fusion_pvs01}, for $0\leq i \leq \lc-1$
\begin{equation}
\dP{1} \xf \dP{\kc+i}  \simeq 
\left. \begin{cases}
0, & \text{ if } i=0 \\
\dP{\kc} \oplus \dP{\kc} & \text{ if } i=1\\
\dP{\kc +i-1} & \text{ otherwise}
\end{cases}\right\} \oplus \left. \begin{cases}
\dP{\kc -\lc} \oplus \dP{\kc + \lc} & \text{ if } i=\lc-1\\
 \dP{\kc+i+1} & \text{ otherwise}
 \end{cases}\right\} \text{.}
\end{equation}
Expressing this in terms of fusion matrices gives the following recurrence relation
\begin{equation}
X \F{\kc} = \F{\kc +1},
\end{equation}
\begin{equation}\label{eq:fusionmatrix1}
X \F{\kc +1} = 2 \F{\kc} + \F{\kc +2},
\end{equation}
\begin{equation}
X \F{\kc +i} = \F{\kc +i-1} + \F{\kc +i +1} , \qquad \text{if } \kc + i\pm 1 \text{ are not critical} 
\end{equation}
\begin{equation}\label{eq:fusionmatrix2}
X \F{\kc + \lc -1} = \F{\kc + \lc -2} + \F{\kc + \lc} + \F{\kc - \lc},
\end{equation}
where it was implicitly assumed that $\lc \neq 2$. When $\lc =2$, equations \eqref{eq:fusionmatrix1} and \eqref{eq:fusionmatrix2}  becomes
\begin{equation}
X \F{\kc + \lc -1} = X \F{\kc +1} = 2\F{\kc } + \F{\kc + 2} + \F{\kc - 2}.
\end{equation}
Using the fact that $\F{\lc -1} = \U{\lc-1}\left(\frac{X}{2}\right)$, it can be checked directly that the solution to this system is
\begin{Prop}
For $0 \leq i \leq \ell -1$,
\begin{equation}
 \F{\kc + i} = \begin{cases}
 \U{\kc }\left(\frac{X}{2}\right), & \text{ if } i =0\\
 \U{\kc -i }\left(\frac{X}{2}\right) + \U{\kc + i}\left(\frac{X}{2}\right), & \text{ otherwise }
 \end{cases}.
\end{equation}
\end{Prop}
\end{subsection}
\begin{subsection}{A closed expression for $\dP{i} \xf \dP{j}$}
Using fusion matrices, computing fusion rules is reduced to evaluating a Chebyshev polynomial at a matrix $X $, but since this matrix is not diagonal, computing this polynomial may be far from trivial. However, since the projective indecomposable modules are all finite dimensional, proposition \ref{prop:fmatrixunique} implies that if
\begin{equation}
F_{i}F_{j} = F(G),
\end{equation}
where $F(G)$ is the fusion matrix of some finite-dimensional projective module $G$, then 
$$\dP{i} \xf \dP{j} \simeq G. $$
Computing fusion rules thus reduces to expressing a product of Chebyshev polynomials as a linear combination of other Chebyshev polynomials. Using this fact will greatly simplify the proof of the following explicit formulas. These are written in a particular way to express the fact that they are identical to those obtained by Gainutnidov and Vasseur \cite{GainVass}.
\begin{Prop}\label{prop:fusionproj}
If $k,r \geq 1$, $0<i,j<\lc$,
\begin{equation}
\dP{i} \xf \dP{j} \simeq \overset{i+j-\lc+1}{\underset{\sigma = (i+j+\lc+1)\bmod{2}}{\bigoplus}}\dP{\lc-1 +\sigma} \oplus \overset{\text{\rm{Min}}(i+j,2\lc - (i+j) -4)}{\underset{\sigma = |i-j|}{\bigoplus}}\dP{\sigma},
\end{equation}
\begin{equation}
\dP{i}\xf \dP{k \lc -1} \simeq \overset{i}{\underset{\sigma = i\bmod{2}}{\bigoplus}}\dP{k \lc -1 + \sigma},
\end{equation}
\begin{align}
\dP{k \lc -1} \xf \dP{r \lc -1} & \simeq \overset{k+r-1}{\underset{\rho = |k-r| +1}{\bigoplus}}\overset{\lc-1}{\underset{\sigma = (\lc +1)\bmod{2}}{\bigoplus}}\dP{\rho \lc -1 +\sigma}\\
 & \simeq \overset{k+r-1}{\underset{\rho = |k-r| +1}{\bigoplus}} \left(\dP{\rho \lc -1} \xf \dP{\lc -1} \right),
\end{align}
\begin{align}
\dP{j} \xf \dP{k \lc -1+i} &\simeq \overset{i+j-\lc}{\underset{\sigma = (i+j+\lc)\bmod{2}}{\bigoplus}}\left(\dP{(k-1) \lc -1 + \sigma} + \dP{(k+1)\lc-1 +\sigma}\right) \oplus 2 \overset{j-i}{\underset{\sigma = |i-j|\bmod{2}}{\bigoplus}}\dP{k \lc -1 +\sigma} \notag \\
& \oplus \overset{\text{\rm{Min}}(i+j, 2\lc - (i+j)-2)}{\underset{\sigma = \text{\rm{Max}}(i-j,j-i+2)}{\bigoplus}}\dP{k \lc -1 +\sigma},
\end{align}
\begin{align}
\dP{r \lc -1} \xf \dP{k \lc -1+i} &\simeq \overset{i-1}{\underset{\sigma = (i+1)\bmod{2}}{\bigoplus}}\left(\dP{|k-r| \lc -1 + \sigma} + \dP{(k+r)\lc-1 +\sigma}\right)\notag \\
&\oplus 2\overset{k+r -1}{\underset{\rho = |k-r|+1}{\bigoplus}}\overset{\lc-i-1}{\underset{\sigma = (i+\lc+1)\bmod{2}}{\bigoplus}}\dP{\rho \lc -1 +\sigma} \oplus 2\overset{k+r -2}{\underset{\rho = |k-r|+2}{\bigoplus}}\overset{i-1}{\underset{\sigma = (i-1)\bmod{2}}{\bigoplus}}\dP{\rho \lc -1 +\sigma}\\
&\simeq \left(\dP{|k-r| \lc -1} + \dP{(k+r)\lc-1 }\right) \xf \dP{i-1} \notag \\
& \oplus 2\overset{k+r -1}{\underset{\rho = |k-r|+1}{\bigoplus}}\left(\dP{\rho \lc -1}\xf \dP{\lc - i -1}\right) \oplus 2\overset{k+r -2}{\underset{\rho = |k-r|+2}{\bigoplus}}\left(\dP{\rho \lc -1}\xf \dP{i-1} \right)
\end{align}
\begin{multline}
\dP{k \lc -1+i} \xf \dP{r \lc- 1+j } \simeq \overset{k+r+1}{\underset{\rho = |k-r|-1}{\bigoplus} }\left( 4\phi_{\rho}  \overset{\lc - (l+i+j+1)\bmod{2}}{\underset{\sigma = 2\lc - (i+j)+1}{\bigoplus} } \dP{(\rho+1)\lc -1-\sigma}\right)\\
\oplus 2 \overset{k+r-1}{\underset{\rho = |k-r|+1}{\bigoplus} }\left( \overset{\text{\rm{Min}}(i+j-1,2\lc - i - j -1)}{\underset{\sigma = |i-j| +1}{\bigoplus} }\dP{(\rho + 1)\lc -1 -\sigma} \oplus 2\overset{\lc - (l+i+j+1)\bmod{2}}{\underset{\sigma = i+j +1}{\bigoplus} }\dP{(\rho + 1)\lc -1 -\sigma} \right)\\
\oplus 2 \overset{k+r}{\underset{\rho = |k-r|}{\bigoplus} }\psi_{\rho}\left( \overset{\text{\rm{Min}}(\lc -i+j-1,\lc + i - j -1)}{\underset{\sigma = |\lc-i-j| +1}{\bigoplus} }\dP{(\rho + 1)\lc -1 -\sigma} \oplus2\overset{\lc - \gamma_1}{\underset{\sigma = \text{\rm{Min}}(\lc - i +j +1, \lc+i-j+1)}{\bigoplus} }\dP{(\rho + 1)\lc -1 -\sigma} \right)\text{,}
\end{multline}

\noindent where $\gamma_1 = (i+j+1)\bmod{2}$,$\gamma_2 = (i+j+q+\lc)\bmod{2}$,$\phi_{\rho} = 1 - \frac{3}{4}\delta_{\rho,|k-r|-1} - \frac{1}{4}\delta_{\rho,|k-r|+1} - \frac{1}{4}\delta_{\rho,k+r-1} - \frac{3}{4}\delta_{\rho,k+r+1}$, $\psi_{\rho} = 1- \frac{1}{2} \delta_{\rho,|k-r|} - \frac{1}{2}\delta_{\rho,k+r}$ and all sums have ``step=2".
\end{Prop}
The proof of all these are done using the same argument. Start by using the identity 
$$\U{i}(y)\U{j}(y) = \sum_{\underset{step =2}{k=|i-j|}}^{i+j} \U{k}(y),$$ to write the product of the fusion matrices as a sum of Chebyshev polynomials then gather them in appropriate combinations to obtain a linear combination of fusion matrices. Using the fact that a fusion matrix uniquely determines a projective module and that fusion of two projective modules always yields a projective modules, the conclusion is obtained. Here are a few examples on how this is done. Since the argument of the polynomials involved will always be $\frac{X}{2}$, we will simply omit them and write $\U{i}$ instead of $\U{i}\left(\frac{X}{2}\right)$.

For $\lc=5$, here are some fusion of small projectives.
 \begin{equation}
 \F{3}\F{2}= \U{3}\U{2} = \U{1} + \U{3} + \U{5} = \F{1} + \F{5} \text{,}
 \end{equation}
 \begin{equation}
 \F{3}\F{4} = \U{1} + \U{3} +\U{5} + \U{7} = (\U{1} + \U{7}) + (\U{3} + \U{5}) = \F{5} + \F{7} \text{,}
 \end{equation}
 \begin{equation}
 \F{2}\F{4} = \U{2} + \U{4} + \U{6} = \U{4} + (\U{2} + \U{6}) = \F{4} + \F{6}\text{.}
 \end{equation}
For the fusion of a small projective and a projective indecomposable,
\begin{equation}
\begin{tabular}{r l}
$\F{4}\F{8} $&$= \U{4}(\U{0} + \U{8}) = 2 \U{4} + \U{6} + \U{8} + \U{10} + \U{12}$\\
&$ = 2\U{4} + (\U{8} + \U{10}) + (\U{6} + \U{12}) = 2\F{4} + \F{10} + \F{12}$
\end{tabular}
\end{equation}
giving the fusion rule
\begin{equation}
\dP{4} \xf \dP{8} \simeq 2\dP{4} \oplus \dP{10} \oplus \dP{12} \text{.}
\end{equation}
The fusion matrix of $\dP{11} \xf \dP{28}$ is
\begin{align}
\F{11}\F{28} & =  (\U{7} + \U{11})(\U{20} + \U{28}) = \sum_{\underset{step =2}{i = 13}}^{27}(\U{i}) + \sum_{\underset{step =2}{i = 9}}^{31}(\U{i})+\sum_{\underset{step =2}{i = 21}}^{35}(\U{i})+\sum_{\underset{step =2}{i = 17}}^{39}(\U{i})\notag \\
& = \U{9} + (\U{11} + \U{17}) + 2(\U{13} + \U{15}) + 3 \U{19} + 2 (\U{17} + \U{21}) + 4(\U{23} + \U{25})\notag\\
& + 2 (\U{21}+\U{27}) + 3\U{29} + 2(\U{27}+\U{31}) + 2(\U{33}+\U{35}) + (\U{31} + \U{37}) + \U{39}\notag\\
& = \F{9} + \F{17} + 2 \F{15} + 3\F{19}+ 2 \F{21} + 4 \F{25} + 2\F{27} + 3\F{29}\notag\\
& +2\F{31}+2\F{35} + \F{37} + \F{39}
\end{align}
giving the fusion rule
\begin{align}
\dP{11} \xf \dP{28} &  \simeq \dP{9}\oplus 2\dP{15}\oplus \dP{17} \oplus 3\dP{19} \oplus 2\dP{21} \oplus 4\dP{25} \oplus 2\dP{27} \oplus 3\dP{29} \notag\\
& \oplus 2 \dP{31} \oplus 2 \dP{35} \oplus \dP{37} \oplus \dP{39}.
\end{align}
Note that we used the same notation in this proposition than in \cite{GainVass}, where they compute the fusion rules in $\tl{n}$. This makes it obvious that the two fusion rules are identical.

\end{subsection}
\begin{subsection}{The semi-simple case}
When $q$ is not a root of unity different from $\pm 1$, the algebras $\tl{n}$ and $\dtl{n}$ are semi-simple and the standard modules $\ds{n,i}$ are all irreducible and projective. They satisfy the induction rules $$ \Ind{\ds{n,i}} \simeq \ds{n+1,i-1}\oplus \ds{n+1,i} \oplus \ds{n+1,i+1}, $$ where it is understood that $\ds{n,i+1} = 0$ if $n \neq i+1 \mod 2$ in the regular family. Using arguments identical to those in section \ref{sec:fmatrixsmall} yields
\begin{equation*}
\ds{n,i} \xf \ds{1,1} \simeq \ds{n+1,i-1} \oplus \ds{n+1,i+1}, \qquad \ds{n,i} \xf \ds{1,0} \simeq \ds{n+1,i},
\end{equation*}
where the second rule is replaced by $$\ds{n,i} \xf \ds{2,0} \simeq \ds{n+2,i}, $$ in the regular family. This gives the following recurrence relation for the fusion matrices
\begin{equation}
X\F{i} = \F{i+1} + \F{i-1}, \qquad \F{0} = \text{id}, \qquad \F{1} = X,
\end{equation}
where $X$ is simply $(X)_{i}^{j} = \delta_{i}^{j+1} + \delta_{i}^{j-1}.$ Using the same argument as in section \ref{sec:fmatrixsmall} then gives the following fusion rules.
\begin{Thm} 
If $q$ is not a root of unity different from $\pm 1$, then for $0\leq i \leq n$, $0 \leq j \leq m$,
$$\ds{n,i} \xf \ds{m,j} \simeq \bigoplus_{\underset{step=2}{k = |i-j|}}^{i+j} \ds{n+m,k}.$$
\end{Thm}
\end{subsection}
\end{section}
\begin{section}{Fusion of standard modules}\label{sec:fusion.standard}
It was noted in section \ref{sec:deffusion} that fusion is closely related with induction, we thus start by giving the behaviour of the non-projective standard modules under the induction functor \cite{RiSY,BelSY,BelRiSy}.
\begin{Prop}
If $i$ with $0\leq i \leq n-1$ is not critical, \begin{equation}
\Ind{\ds{n-1,i}} \simeq \begin{cases}
\ds{n,i-1} \oplus \ds{n,i} \oplus \ds{n,i+1}, & \text{ in the dilute family},\\
\ds{n,i-1} \oplus \ds{n,i+1}, & \text{ in the regular family},
\end{cases}
\end{equation} 
where it is understood that $\ds{n,-1} = 0$.
\end{Prop}
Using the same arguments as in proposition \ref{prop_fusion_pvs01}, this gives the following fusion rules.
\begin{Prop}\label{prop_fusion_svs01}
If $i$ with $0\leq i \leq n-1$ is not critical, in the dilute family
\begin{equation}
\ds{n-1,i}\xf \dP{1,0} \simeq \ds{n,i} \text{,}
\end{equation}
while in both families
\begin{equation}\label{eq:fusion_svs1}
\ds{n-1,i}\xf \dP{1,1} \simeq \ds{n,i-1} \oplus \ds{n,i+1}\text{,}
\end{equation}
where $\ds{n,j}\simeq\dP{n,j}$ if $j$ is critical.
\end{Prop}
Using the same argument as in the projective case with the first fusion rule, and proposition \ref{prop:magickSP2} in the regular case, $$\ds{n,i} \xf \ds{m,j} \simeq \dP{n-i + m-j,0} \xf(\ds{i,i} \xf \ds{j,j}) \text{.}$$ We will therefore always omit the parameter $n$, writing $\ds{n,i} = \ds{i}$, and assume that $n$ is big enough and of the right parity, in the regular case, for the module to exists. Note that in the regular case when $\lc =2$, the module $\ds{n,0}$ is very particular because $\ds{n,0} \simeq \dl{n,2}$. This module will therefore be treated in section \ref{sec:tl_l2}.

 Once a formula for the fusion of $\ds{k \ell}$, $k \in \mathbb{N}$, with some module $M$ is obtained, the second fusion rules \eqref{eq:fusion_svs1} will be used to obtain a formula for the fusion of $M$ with the other standard modules by simple induction. We start by studying the fusion of a standard module with a projective module then consider the fusion of two standard modules. Finally, we give a simple rule that can be used to quickly compute the fusion of standard modules. 
\begin{subsection}{Fusion of a standard and a projective module}
The general formula that will be obtained is quite complex and the inductive proof is very technical. The argument is thus split in four propositions that will be simpler to prove. Each one will be preceded by an example with $\lc=5$ before moving to the general case. The proof for general $\lc$ is very straightforward once these examples are understood so we highly suggest that the reader works them out carefully.

Consider the case $\lc=5$ and the standard module $\ds{n,25} = \ds{25}$ which is not projective. Proposition \ref{prop_fusion_svs01} then gives 
\begin{equation}\label{eq_s25vsp1}
\ds{25}\xf\dP{1} \simeq \dP{24} \oplus \ds{26} \text{.}
\end{equation}
 Note that $\ds{24}\simeq\dP{24}$ is projective. Fusing the left side of this isomorphism with $\dP{1}$ and using the associativity of fusion with proposition \ref{prop_fusion_svs01} and \ref{prop_fusion_pvs01} gives
 \begin{equation}
 	\ds{25} \xf \left(\dP{1}\xf \dP{1}\right)  \simeq \ds{25} \xf \dP{0} \oplus \ds{25} \xf \dP{2} \simeq \ds{25} \oplus \ds{25} \xf \dP{2},
 \end{equation}
 while fusing its right side with $\dP{1}$ and using the same propositions gives
 \begin{equation}
 	\left(\dP{24} \oplus \ds{26} \right)\xf \dP{1} \simeq \dP{24} \xf \dP{1} \oplus \dP{26} \xf \dP{1} \simeq \dP{25} \oplus \ds{25} \oplus \ds{27}.
 \end{equation}
 Comparing the two results yields $$\ds{25} \xf \dP{2} \simeq \dP{25} \oplus \ds{27} \text{.}$$ Repeating the same arguments gives the fusion rules
\begin{equation}
\ds{25} \xf \dP{3} \simeq \dP{24} \oplus \dP{26} \oplus \ds{28},
\end{equation}
\begin{equation}\label{eq_s25vsp4}
\ds{25} \xf \dP{4} \simeq  \dP{25} \oplus \dP{27} \oplus \ds{29}
\end{equation}
where $\ds{29}=\dP{29}$ is projective. A pattern can be identified here: for all $i<5$, $$ \ds{25} \xf \dP{i} \simeq \dP{24} \xf \dP{i-1} \oplus \ds{25+i} \text{.}$$
\begin{Prop}\label{prop_fusion_sklvspi}
For $i < \lc$, $\lc>2$ in $\tl{n}$, and $k \in \mathbb{N}$,
\begin{equation}\label{eq_sklvspi}
\ds{k \lc} \xf \dP{i} \simeq \dP{k \lc -1}\xf \dP{i-1} \oplus \ds{ k \lc +i} \text{.}
\end{equation}
\end{Prop}
\begin{proof}
The proof proceeds by induction on $i$. Proposition \ref{prop_fusion_svs01} already gives the case $i=0$ and $i=1$. Suppose therefore the result for $i<\lc-1$ and $i-1$. Applying propositions \ref{prop_fusion_svs01} and \ref{prop_fusion_pvs01} on the left side of equation \eqref{eq_sklvspi} gives
\begin{equation}
\ds{ k \lc} \xf \dP{i} \xf \dP{1} \simeq \ds{ k \lc} \xf \left(\dP{i} \xf \dP{1}\right) \simeq \ds{k \lc} \xf \dP{i-1} \oplus \ds{ k \lc } \xf \dP{i+1}.
\end{equation}
Using the same proposition on the right side of \eqref{eq_sklvspi} yields
\begin{equation}
\dP{i-1} \xf \dP{k \lc -1} \xf \dP{1} \oplus \ds{ k \lc +i}\xf \dP{1} \simeq \dP{k \lc -1} \xf \left(\dP{i-2} \oplus \dP{i}\right) \oplus \ds{k \lc +i-1} \oplus \ds{k \lc +i+1}.
\end{equation}
Comparing the two results and using the induction hypothesis for $i-1$ gives the conclusion. Note that we implicitly assumed that $\lc \neq 2$. In this case, there is only $i=0$ and $i=1$, which are both covered by proposition \ref{prop_fusion_svs01}.
\end{proof}
Let us return to the preceding $\lc=5$ example. Using again the associativity and commutativity of fusion with proposition \ref{prop_fusion_pvs01} if equation \eqref{eq_s25vsp4} is fused with $\dP{1}$, the left side gives
\begin{equation}
\ds{25}\xf \left(\dP{4} \xf \dP{1}\right) \simeq \ds{25} \xf \dP{5}, 
\end{equation}
while the right one becomes
\begin{equation}
\left(\dP{24}\xf \dP{1}\right) \xf \dP{3}  \oplus \underbrace{\ds{29}}_{\simeq \dP{29}} \xf \dP{1} \simeq \dP{25}\xf \dP{3} \oplus \dP{30}.  
\end{equation}
Comparing the two gives
\begin{equation}
	 \ds{25} \xf \dP{5} \simeq \dP{25}\xf \dP{3} \oplus \dP{30}.
\end{equation}
Repeating this operation yields
\begin{equation}
\ds{25} \xf \dP{6} \simeq \dP{26} \xf \dP{3}  \oplus \dP{31}\text{,}
\end{equation}
\begin{equation}
\ds{25} \xf \dP{7} \simeq \dP{27} \xf \dP{3} \oplus \dP{32}\text{,}
\end{equation}
\begin{equation}
\ds{25} \xf \dP{8} \simeq \dP{28} \xf \dP{3} \oplus \dP{33}\text{.}
\end{equation}
Fusing $\dP{1}$ again on the last rule\footnote{Note that $9,14,19,24,29$ and $34$ are critical.}, the left side becomes
\begin{equation}
	\ds{25} \xf \left(\dP{8} \xf \dP{1}\right) \simeq \ds{25} \xf \left( \dP{7} \oplus \dP{9}\right),
\end{equation}
while the right one becomes
\begin{align}
	\left(\dP{28} \xf \dP{1} \right)\xf \dP{3} \oplus \dP{33}\xf \dP{1} &\simeq \left(\dP{19} \oplus \dP{27} \oplus \dP{29} \right)\xf \dP{3} \oplus \dP{24} \oplus \dP{32} \oplus \dP{34}\\
	& \simeq \underbrace{\left(\dP{27} \xf \dP{3} \oplus \dP{32} \right)}_{\simeq \ds{25} \xf \dP{7}} \oplus \left(\dP{19}\oplus \dP{29}\right)\xf \dP{3} \oplus \dP{24} \oplus \dP{34}.
\end{align}
This is simply
\begin{equation}
\ds{25} \xf \dP{9} \simeq (\dP{19} \oplus \dP{29}) \xf \dP{3} \oplus \dP{24} \oplus \dP{34} \text{.}
\end{equation}
We can then proceed with the general case.
\begin{Prop}\label{prop_fusion_sqlvspkl}
For $0\leq i<\lc$, $k,s \in \mathbb{Z}_{>0}$,
\begin{equation}\label{eq_sqlvspkl}
\ds{k \lc } \xf \dP{s \lc -1+i} \simeq \overset{k+s-1}{\underset{\underset{step =2}{r = |k-s| + 1}}{\bigoplus}}\left( \dP{r \lc -1+i} \xf \dP{\lc-2}\right) \oplus \overset{k+s}{\underset{\underset{step =2}{r = |s-(k+1)| + 1}}{\bigoplus}} \dP{r \lc -1+i} \text{.}
\end{equation}
In the case $\lc =2$ in $\tl{n}$, the fusion with $\dP{\lc-2}$ must be removed.
\end{Prop}
\begin{proof}
The  proof proceeds by induction on $s$ and $i$. Let us start by proving that for a given $k$, if the result stands for $i=0$ then it will also stands for all $i \leq \lc$. Note that the case $i=1$ follows directly from the case $i=0$ since for all $p\geq 1$, $\dP{p \lc} \simeq \dP{1} \xf \dP{p \lc -1}$. Suppose therefore that the result stands for $i-1, i<\lc$. Fusing \eqref{eq_sqlvspkl} with $\dP{1}$ and using proposition \ref{prop_fusion_pvs01} with the associativity and commutativity of fusion then gives, on the left side
\begin{equation}
	\ds{k \lc} \xf \left(\dP{s \lc -1 +i}\xf \dP{1} \right) \simeq (1+\delta_{i,1})\ds{k \lc} \xf \dP{s \lc -2 +i} \oplus \ds{k \lc} \xf \left(\dP{s \lc +i} \oplus \delta_{i,\lc-1} \dP{(s-1)\lc -1}\right),
\end{equation}
and on the right side
\begin{align*}
	\overset{k+s-1}{\underset{\underset{step =2}{r = |k-s| + 1}}{\bigoplus}}&\left( \left(\dP{r \lc -1+i}\xf \dP{1} \right) \xf \dP{\lc-2}\right) \oplus \overset{k+s}{\underset{\underset{step =2}{r = |s-(k+1)| + 1}}{\bigoplus}}\left( \dP{r \lc -1+i} \xf \dP{1} \right)\\
	& \simeq \overset{k+s-1}{\underset{\underset{step =2}{r = |k-s| + 1}}{\bigoplus}}\left( \left((1+\delta_{i,1})\dP{r \lc -2+i} \oplus \dP{r \lc + i}  \oplus \delta_{i,\lc -1}\dP{(r-1)\lc -1}\right) \xf \dP{\lc-2}\right)\\
	&\oplus \overset{k+s}{\underset{\underset{step =2}{r = |s-(k+1)| + 1}}{\bigoplus}}\left((1+\delta_{i,1}) \dP{r \lc -2+i} \oplus \dP{r\lc +i} \oplus \delta_{i,\lc-1} \dP{(r-1)\lc -1}\right)\\
	& \simeq (1+\delta_{i,1})\left(\overset{k+s-1}{\underset{\underset{step =2}{r = |k-s| + 1}}{\bigoplus}}\left(\dP{r \lc -2+i} \xf \dP{\lc -2} \right) \oplus \overset{k+s}{\underset{\underset{step =2}{r = |s-(k+1)| + 1}}{\bigoplus}}\dP{r \lc -2+i}\right)\\
	&\oplus \delta_{i,\lc-1} \left(\overset{k+s-1}{\underset{\underset{step =2}{r = |k-s| + 1}}{\bigoplus}}\left(\dP{(r-1)\lc -1}\xf \dP{\lc-2}\right) \oplus \overset{k+s}{\underset{\underset{step =2}{r = |s-(k+1)| + 1}}{\bigoplus}}  \dP{(r-1)\lc -1} \right)\\
	& \oplus \left(\overset{k+s-1}{\underset{\underset{step =2}{r = |k-s| + 1}}{\bigoplus}}\left(\dP{r \lc -1+i+1} \xf \dP{\lc -2} \right) \oplus \overset{k+s}{\underset{\underset{step =2}{r = |s-(k+1)| + 1}}{\bigoplus}}\dP{r \lc -1+i+1}\right).
\end{align*}
If $i \neq \lc -1 $, collecting the relevant terms, comparing the two sides and applying the induction hypothesis then gives the result for $i+1$. If $i =\lc-1 $, there is a slight subtlety involved. In the preceding expression, collect the terms being fused with $\dP{\lc -2}$ and note that
\begin{align*}
	\overset{k+s-1}{\underset{\underset{step =2}{r = |k-s| + 1}}{\bigoplus}}\left(\dP{(r-1)\lc -1} \oplus \dP{r \lc -1+\lc } \right)& \simeq \overset{k+s-2}{\underset{\underset{step =2}{r' = |k-s|}}{\bigoplus}} \dP{r'\lc -1} \oplus \overset{k+s}{\underset{\underset{step =2}{r' = |k-s|+2}}{\bigoplus}} \dP{r'\lc -1}\\
	& \simeq \overset{k+(s-1)-1}{\underset{\underset{step =2}{r = |k-(s-1)| + 1}}{\bigoplus}} \dP{r \lc -1} \oplus \overset{k+s+1-1}{\underset{\underset{step =2}{r = |k-(s+1)| + 1}}{\bigoplus}} \dP{r \lc -1},
\end{align*}
where we rearranged the terms between the two sums and used the fact that $\dP{-1} \equiv 0$. The exact rearranging required depends on the value of $k-s$. Doing the same rearranging on the other terms gives
\begin{align*}
	\ds{k \lc} \xf \left(\dP{(s+1) \lc -1} \oplus  \dP{(s-1)\lc -1}\right) & \simeq \overset{k+(s-1)-1}{\underset{\underset{step =2}{r = |k-(s-1)| + 1}}{\bigoplus}}\left( \dP{r \lc -1} \xf \dP{\lc-2}\right) \oplus \overset{k+s-1}{\underset{\underset{step =2}{r = |s-1-(k+1)| + 1}}{\bigoplus}} \dP{r \lc -1} \\
	& \oplus \overset{k+s+1-1}{\underset{\underset{step =2}{r = |k-(s+1)| + 1}}{\bigoplus}}\left( \dP{r \lc -1} \xf \dP{\lc-2}\right) \oplus \overset{k+s+1}{\underset{\underset{step =2}{r = |s+1-(k+1)| + 1}}{\bigoplus}} \dP{r \lc -1}.  
\end{align*}
It follows that if the statement holds for $(s-1,i=0)$, $(s,i=0)$, it will also stand for $(s,i)$ for all $i \leq \ell -1$ and $(s+1,i =0)$. 

The only remaining step is to prove that the result stands for $k=1$, $i=0$. This is precisely proposition \ref{prop_fusion_sklvspi}. In the case $\lc =2$ of $\tl{n}$, the result and its proof are slightly different, because then $1 = \lc -1$, so that
\begin{equation*}
 \dP{s \lc} \xf \dP{1} \simeq 2\dP{s \lc -1} \oplus \dP{(s+1)\lc -1}\oplus \dP{(s-1)\lc -1}.
\end{equation*}
However, the same arguments can be used to induce on $s$ and on $i$.
\end{proof}
Now that the expression for the fusion of $\ds{k \lc}$ is known, proposition \ref{prop_fusion_svs01} can be used to compute the fusion of the other standard modules with the projective. We return to the $\lc=5$ example. It was previously found that
\begin{equation*}
\ds{25} \xf \dP{8} \simeq \dP{28}\xf \dP{3} \oplus \dP{33}.
\end{equation*}
Fusing the left side with $\dP{1}$ gives,
\begin{equation}\label{eq:exampleS26P8.1}
	\ds{25} \xf \dP{8} \xf \dP{1} \simeq \left(\ds{25} \xf \dP{1}\right) \xf \dP{8} \simeq \left( \dP{24} \oplus \ds{26} \right) \xf \dP{8},
\end{equation}
while fusing the right side with $\dP{1}$ yields
\begin{align}\label{eq:exampleS26P8.2}
	\dP{28} \xf \left(\dP{3}\xf \dP{1}\right) \oplus \dP{33}\xf \dP{1} & \simeq \dP{28} \xf \left(\dP{2} \oplus \dP{4}  \right) \oplus \dP{33} \xf \dP{1} \notag \\
	& \simeq \left( \dP{28} \xf \dP{2} \oplus \dP{33} \xf \dP{1} \right) \oplus \dP{28} \xf \dP{4}.
\end{align}
Using proposition \ref{prop:fusionproj}, notice that
\begin{align*}
	\dP{24} \xf \dP{8} \simeq \left(\dP{19} \oplus \dP{29}\right)\xf \dP{3} \oplus 2 \dP{24}\simeq \dP{4} \xf \dP{28}.
\end{align*}
Comparing \eqref{eq:exampleS26P8.1} with \eqref{eq:exampleS26P8.2} then gives the fusion rule
\begin{equation}
	\ds{26} \xf \dP{8} \simeq  \dP{28} \xf \dP{2} \oplus \dP{33} \xf \dP{1}.
\end{equation}
Repeating the same steps gives
\begin{equation}
	\ds{27} \xf \dP{8} \simeq  \dP{28} \xf \dP{1} \oplus \dP{33} \xf \dP{2},
\end{equation}
\begin{equation}
	\ds{28} \xf \dP{8} \simeq  \dP{28} \xf \dP{0} \oplus \dP{33} \xf \dP{3}.
\end{equation}
\begin{Thm}\label{prop:fusion.standvsproj}
For $0<i<\lc$, $0 \leq j <\lc$ and $k,s\in\mathbb{Z}_{>0}$,
\begin{equation}\label{eq:fusionSP}
\ds{k \lc -1 +i} \xf \dP{s \lc -1+j} \simeq \overset{k+s-1}{\underset{\underset{step =2}{r = |k-s| + 1}}{\bigoplus}}\left( \dP{r \lc -1+j} \xf \dP{\lc -1-i}\right) \oplus \overset{k+s}{\underset{\underset{step =2}{r = |s-(k+1)| + 1}}{\bigoplus}} \left(\dP{r \lc -1+j} \xf \dP{i-1} \right)\text{.}
\end{equation}
\end{Thm}
\begin{proof}
In this case the proof is a simple induction on $i$. The case $i = 1$ is covered by proposition \ref{prop_fusion_sqlvspkl}. Fusing the left side of \eqref{eq:fusionSP} ($i=1$) with $\dP{1}$ and using proposition \ref{prop_fusion_svs01} gives
\begin{equation}\label{eq:fusionSP.G}
 \left( \ds{k \lc} \xf \dP{1} \right) \xf \dP{s \lc -1 +j} \simeq \left(\dP{k \lc -1} \oplus \ds{k \lc -1 + 2} \right) \xf \dP{s \lc -1 +j},
\end{equation}
while fusing the right side of the same equation with $\dP{1}$ and using proposition \ref{prop_fusion_pvs01} yields
\begin{align}\label{eq:fusionSP.D}
\overset{k+s-1}{\underset{\underset{step =2}{r = |k-s| + 1}}{\bigoplus}}&\left( \dP{r \lc -1+j} \xf \left( \dP{\lc -2} \xf \dP{1} \right)\right)  \oplus \overset{k+s}{\underset{\underset{step =2}{r = |s-(k+1)| + 1}}{\bigoplus}} \left(\dP{r \lc -1+j} \xf \dP{1} \right) \notag\\
& \simeq \overset{k+s-1}{\underset{\underset{step =2}{r = |k-s| + 1}}{\bigoplus}}\left( \dP{r \lc -1+j} \xf \left( \dP{\lc -1} \oplus \dP{\lc -3} \right)\right) \oplus \overset{k+s}{\underset{\underset{step =2}{r = |s-(k+1)| + 1}}{\bigoplus}} \left(\dP{r \lc -1+j} \xf \dP{1} \right) \notag\\
& \simeq \Big( \overset{k+s-1}{\underset{\underset{step =2}{r = |k-s| + 1}}{\bigoplus}}\left( \dP{r \lc -1+j} \xf  \dP{\lc -3} \right)\oplus \overset{k+s}{\underset{\underset{step =2}{r = |s-(k+1)| + 1}}{\bigoplus}} \left(\dP{r \lc -1+j} \xf \dP{1} \right)\Big) \notag\\
&\oplus \overset{k+s-1}{\underset{\underset{step =2}{r = |k-s| + 1}}{\bigoplus}}\left( \dP{r \lc -1+j} \xf  \dP{\lc -1} \right).
\end{align}
However, proposition \ref{prop:fusionproj} gives
\begin{align}
\dP{k\lc -1} \xf \dP{s \lc -1+j} & \simeq \left(\dP{|s-k| \lc -1} + \dP{(s+k)\lc-1 }\right) \xf \dP{j-1} \notag \\
& \oplus 2\overset{k+s -1}{\underset{\rho = |k-s|+1}{\bigoplus}}\left(\dP{\rho \lc -1}\xf \dP{\lc - j -1}\right) \oplus 2\overset{k+s -2}{\underset{\rho = |k-s|+2}{\bigoplus}}\left(\dP{\rho \lc -1}\xf \dP{j-1} \right),
\end{align}
and
\begin{align}
\overset{k+s-1}{\underset{\underset{step =2}{r = |k-s| + 1}}{\bigoplus}}\left( \dP{r \lc -1+j} \xf  \dP{\lc -1} \right) &\simeq \overset{k+s-1}{\underset{\underset{step =2}{r = |k-s| + 1}}{\bigoplus}}\left( \left(\dP{(r-1)\lc -1} \oplus \dP{(r+1)\lc -1} \right)\xf \dP{j-1} \oplus 2 \dP{r \lc -1} \xf \dP{\lc - (j+1)}\right) \notag\\
& \simeq 2 \overset{k+s-1}{\underset{\underset{step =2}{r = |k-s| + 1}}{\bigoplus}}\left(\dP{r \lc -1} \xf \dP{\lc - (j+1)}\right) \oplus \overset{k+s-2}{\underset{\underset{step =2}{r' = |k-s|}}{\bigoplus}}\left(\dP{r'\lc -1}\xf \dP{j-1} \right) \notag\\
& \oplus \overset{k+s}{\underset{\underset{step =2}{r' = |k-s| + 2}}{\bigoplus}}\left(\dP{r'\lc -1}\xf \dP{j-1} \right).
\end{align}
Collecting identical terms in the last two sums then gives the identity
\begin{equation}
\dP{k\lc -1} \xf \dP{s \lc -1+j} \simeq \overset{k+s-1}{\underset{\underset{step =2}{r = |k-s| + 1}}{\bigoplus}}\left( \dP{r \lc -1+j} \xf  \dP{\lc -1} \right).
\end{equation}
Comparing equations \eqref{eq:fusionSP.G}, and \eqref{eq:fusionSP.D} and using this identity then gives the result for $i=2$. 

Suppose now that the result stands for $i-1,i$ with $1<i<\lc$. Fusing the left side of equation \eqref{eq:fusionSP} with $\dP{1}$ and using proposition \ref{prop_fusion_svs01} gives
\begin{equation}
\left( \ds{k \lc -1 +i} \xf \dP{1} \right) \xf \dP{s \lc -1+j} \simeq \left( \ds{k \lc -1 +(i-1)} \oplus \ds{k \lc -1 +(i+1)}  \right) \xf \dP{s \lc -1+j},
\end{equation}
while fusing the right side of equation \eqref{eq:fusionSP} with $\dP{1}$ and using proposition \ref{prop_fusion_pvs01} yields
\begin{align}
\overset{k+s-1}{\underset{\underset{step =2}{r = |k-s| + 1}}{\bigoplus}}&\left( \dP{r \lc -1+j} \xf \left(\dP{\lc -1-i}\xf \dP{1}\right)\right) \oplus \overset{k+s}{\underset{\underset{step =2}{r = |s-(k+1)| + 1}}{\bigoplus}} \left(\dP{r \lc -1+j} \xf \left(\dP{i-1} \xf \dP{1}\right) \right) \notag \\
& \simeq \overset{k+s-1}{\underset{\underset{step =2}{r = |k-s| + 1}}{\bigoplus}}\left( \dP{r \lc -1+j} \xf \left(\dP{\lc -1-(i-1)}\oplus \dP{\lc -1-(i+1)}\right)\right) \notag \\
& \oplus \overset{k+s}{\underset{\underset{step =2}{r = |s-(k+1)| + 1}}{\bigoplus}} \left(\dP{r \lc -1+j} \xf \left(\dP{i-2} \oplus \dP{i}\right) \right).
\end{align}
Comparing these two results and using the induction hypothesis then gives the result for $i+1$. 

Note that it was implicitly assumed that $\lc \neq 2$, because this case is covered by proposition \ref{prop_fusion_sqlvspkl}.
\end{proof}
\end{subsection}
\begin{subsection}{Fusion of two standard modules}
The action of $\dP{1}$ has played a central role so far in the proofs. Projective modules can all be expressed as ``polynomials" in $\dP{1}$ and even the standard modules $\ds{k \lc +i}$ could be obtained by fusing $\ds{k l}$ with it. However, fusing $\ds{k \lc}$ repeatedly with $\dP{1}$ produced a sum of projective module, so that $\ds{(k+1)\lc}$ cannot be obtained from $\ds{k \lc}$. Another argument will thus be needed to ``cross" the critical lines without obtaining projective modules. It will eventually be proved that this is done by fusing with $\ds{\lc}$. The proofs are identical for the dilute and the regular family, except when $\lc =2$. The proof of proposition \ref{prop:fusionstandvssl} below is then very different. The result still stands in this case, but the proof will be presented in section \ref{sec:tl_l2}.

The first step is to compute the dimension of $ \ds{k, k} \xf \ds{r,r}$ as it will make the proof of proposition \ref{prop:fusionstandvssl} much easier. Note that the parameter $n$ in $\ds{n,k}$ is now important as the dimension of the modules depends on it. The general case is very simple but somewhat long. 
 We compute the dimension of $ \ds{3,3} \xf \ds{3,3}$. Define (see section \ref{sec:rappel.base.standard})
$$z = \begin{tikzpicture}[baseline={(current bounding box.center)},scale = 1/3]
	\draw[very thick] (0,-3/2) -- (0,3/2);
	\draw (0,-1) -- (3,-1);
	\draw (0,0) -- (3,0);
	\draw (0,1) -- (3,1);
\end{tikzpicture}, $$
which is such that $\ds{3,3} = A_{3}z$, where $A_n = \tl{n}$ or $\dtl{n}$. Then $$\ds{3,3} \xf \ds{3,3} \simeq A_{6} \left(\id_{A_6} \otimes_{A_{3} \otimes A_{3}} ( z \otimes_{\mathbb{C}} z)\right) .$$ Furthermore, notice that the only diagram in $A_{3}$ which does not act as zero on $z$ is the identity. It follows that the only diagrams of $A_{6}$ which do not act as zero on $\id_{A_6} \otimes_{A_{3} \otimes A_{3}} ( z \otimes_{\mathbb{C}} z)$ are those of the following form
\begin{equation}
\begin{tikzpicture}[baseline={(current bounding box.center)},scale = 1/3]
	\draw[very thick] (0,-3) -- (0,3);
	\draw[very thick] (3,-3) -- (3,3);
	\draw (3,-5/2) -- (0,-5/2);
	\draw (3,-3/2) -- (0,-3/2);
	\draw (3,-1/2) -- (0,-1/2);
	\draw (3,1/2) -- (0,1/2);
	\draw (3,3/2) -- (0,3/2);
	\draw (3,5/2) -- (0,5/2);
\end{tikzpicture},\qquad
\begin{tikzpicture}[baseline={(current bounding box.center)},scale = 1/3]
	\draw[very thick] (0,-3) -- (0,3);
	\draw[very thick] (3,-3) -- (3,3);
	\draw (3,-5/2) -- (2,-5/2);
	\draw (3,-3/2) -- (2,-3/2);
	\draw (3,-1/2) .. controls (2,-1/2) and (2,1/2) .. (3,1/2);
	\draw (3,3/2) -- (2,3/2);
	\draw (3,5/2) -- (2,5/2);
	\node at (1,0) {$x_1$};
\end{tikzpicture}, \qquad
\begin{tikzpicture}[baseline={(current bounding box.center)},scale = 1/3]
	\draw[very thick] (0,-3) -- (0,3);
	\draw[very thick] (3,-3) -- (3,3);
	\draw (3,-5/2) -- (2,-5/2);
	\draw (3,-3/2) .. controls (3/2,-3/2) and (3/2,3/2) ..  (3,3/2);
	\draw (3,-1/2) .. controls (2,-1/2) and (2,1/2) .. (3,1/2);
	\draw (3,5/2) -- (2,5/2);
	\node at (1,0) {$x_2$};
\end{tikzpicture}, \qquad
\begin{tikzpicture}[baseline={(current bounding box.center)},scale = 1/3]
	\draw[very thick] (0,-3) -- (0,3);
	\draw[very thick] (3,-3) -- (3,3);
	\draw (3,-5/2) .. controls (3/2,-5/2) and (3/2,5/2) .. (3,5/2);
	\draw (3,-3/2) .. controls (2,-3/2) and (2,3/2) ..  (3,3/2);
	\draw (3,-1/2) .. controls (5/2,-1/2) and (5/2,1/2) .. (3,1/2);
	\node at (1,0) {$x_3$};
\end{tikzpicture},
\end{equation}
where $x_{i}$ is a link diagram in $\ds{6,6-2i}$. It also follows that for $b \in A_{3} \otimes A_{3}$, $ b ( z \otimes_{\mathbb{C}} z) = 0$, unless $b$ can be expressed as $b = (\id\otimes \id) + c,$ for some $c\in\dtl{3}\otimes_{\mathbb{C}}\dtl{3}$. We thus conclude that these diagrams form a basis of $\ds{3,3}\xf \ds{3,3}$ and thus that $$\dim \ds{3,3}\xf\ds{3,3} = \dim \ds{6,6} + \dim \ds{6,4} + \dim \ds{6,2} + \dim \ds{6,0}.$$
The general case is obtained by a straightforward generalisation of this argument.
\begin{Lem}\label{lem:magickdim}
For all $k,r \in \mathbb{N}$,
\begin{equation}
\dim \left( \ds{k,k} \xf \ds{r,r} \right) = \sum_{i=0}^{\min\left(k,r \right)} \dim \ds{k+r, k+r-2i}.
\end{equation}
\end{Lem}

The proof of the general case $\ds{k \lc +i} \xf \ds{r \lc +j}$ will be done by induction on $k$,$r$,$i$ and $j$. Fusion with $\dP{1}$ will be used to induce from $i$ to $i+1$, and from $j$ to $j+1$, while fusion with $\ds{\lc}$ will be used to induce from $k$ to $k+1$,and $r$ to $r+1$. The inductive proof is split into numerous lemmas so that the various steps are clearer. Each lemma will be accompanied by an example to illustrate the result.

Use again the particular case $\lc=5$, and recall (see section \ref{sec:rappel.indec}) that the projective module $\dP{5,5}$ satisfies the short exact sequence $$ 0 \longrightarrow \dP{5,3} \longrightarrow \dP{5,5} \longrightarrow \ds{5,5} \longrightarrow 0 \text{,}$$ and using the right-exactness of fusion, proposition \ref{prop:fusionexact}, this implies the exact sequence
\begin{equation}
\dP{5,3}\xf \ds{5,5} \overset{f}{\longrightarrow} \dP{5,5}\xf \ds{5,5}  \longrightarrow \ds{5,5}\xf \ds{5,5}  \longrightarrow 0\text{.}
\end{equation}
Using the previously obtained fusion rules,  note that 
\begin{equation}
(\dP{5, 5}\xf \ds{5, 5})/(\dP{5, 3}\xf \ds{5, 5}) \simeq \frac{2 \dP{10,4} \oplus 2\dP{10,6} \oplus \dP{10,8} \oplus \dP{10,10}}{\dP{10,4} \oplus \dP{10,6} \oplus \ds{10,8}} \simeq \dP{10,4} \oplus \dP{10,6} \oplus \dP{10,8} \oplus \ds{10,10}
\end{equation}
by using the fact that $\dP{10,10}/\ds{10,8} \simeq \ds{10,10}$. 
However, lemma \ref{lem:magickdim} gives $$\dim{\ds{5,5}\xf\ds{5,5}} = \dim{\dP{10,4} \oplus \dP{10,6} \oplus \dP{10,8} \oplus \ds{10,10}},$$ so it follows that $f$ must be injective and thus 
$$\ds{5,5} \xf \ds{5,5} \simeq \dP{6,4}\xf \dP{4,4} \oplus \ds{10,10}. $$
Fusing the left side of this result with $\dP{1,1}$ and using proposition \ref{prop_fusion_svs01} gives
\begin{equation}\label{eq:exampleSvsS.1}
	\ds{5,5} \xf \left(\ds{5,5} \xf \dP{1,1} \right) \simeq \ds{5,5} \xf \left(\dP{6,4} \oplus \ds{6,6}\right),
\end{equation}
while fusing its right side with $\dP{1,1} $ and using propositions \ref{prop_fusion_svs01}, and \ref{prop_fusion_pvs01} yields
\begin{equation}\label{eq:exampleSvsS.2}
	\dP{6,4}\xf \left( \dP{4,4} \xf \dP{1,1}\right) \oplus \left(\ds{10,10}\xf \dP{1,1} \right) \simeq \dP{6,4} \xf \dP{5,5} \oplus \dP{11,9} \oplus \ds{11,11}.
\end{equation}
Using proposition \ref{prop:fusionproj}, and \ref{prop_fusion_sqlvspkl}, note that
\begin{equation*}
	\dP{6,4} \xf \dP{5,5} \simeq 2 \dP{6,4} \xf \dP{5,3} \oplus \dP{11,9} \simeq \dP{6,4} \xf \dP{5,3} \oplus \ds{5,5}\xf \dP{6,4}. 
\end{equation*}
Comparing equations \eqref{eq:exampleSvsS.1}, and \eqref{eq:exampleSvsS.2} and using this observation give the fusion rule
\begin{equation}
	\ds{6,6}\xf \ds{5,5} \simeq \dP{6,4} \xf \dP{5,3} \oplus \dP{11,9} \oplus \ds{11,11}
\end{equation} 
Repeating these arguments yields
\begin{equation}
\ds{7,7} \xf \ds{5,5} \simeq \dP{6,4} \xf \dP{6,2} \oplus \dP{11,9}\xf \dP{1,1} \oplus \ds{12,12}\text{,}
\end{equation}
\begin{equation}
\ds{8,8} \xf \ds{5,5} \simeq \dP{6,4} \xf \dP{7,1} \oplus \dP{11,9}\xf \dP{2,2} \oplus \ds{13,13}\text{,}
\end{equation}
\begin{equation}
\ds{9,9} \xf \ds{5,5} \simeq \dP{6,4} \xf \dP{8,0} \oplus \dP{11,9}\xf \dP{3,3} \oplus \ds{14,14}\text{.}
\end{equation}
Note that since $\ds{9,9} \simeq \dP{9,9}$ is projective, the last one could be obtained from proposition \ref{prop_fusion_sqlvspkl}. 
\begin{Prop}\label{prop:fusionstandvssl}
For $0 < i < \lc$, $k \in \mathbb{Z}_{>0}$ and in the regular family  $\lc > 2$,
\begin{equation}\label{eq:fusionstandvssl}
\ds{k \lc -1 +i} \xf \ds{\lc} \simeq \left(\dP{k \lc -1} \xf \dP{\lc -i}\right) \oplus \left(\dP{(k+1)\lc -1} \xf \dP{i - 2}\right) \oplus \ds{(k+1)\lc -1+i}.
\end{equation}
\end{Prop}
\begin{proof}
The proof proceeds by induction on $k$ and $i$. Let us start by proving that for a given $k$, if the result stands for $i=1$, it will also stand for all $i \leq \lc -1 $. 

Suppose that the result stands for $i=1$. Fusing the left side of equation \eqref{eq:fusionstandvssl} with $\dP{1}$ and using proposition \ref{prop_fusion_svs01} gives
\begin{equation}\label{eq:fusionSvsS.1}
	\left(\ds{k \lc} \xf \dP{1}\right) \xf \ds{\lc} \simeq \left( \dP{k \lc -1} \oplus \ds{k \lc +1} \right) \xf \ds{\lc}
\end{equation}
while fusing its right side and using propositions \ref{prop_fusion_svs01}, and \ref{prop_fusion_pvs01} yields
\begin{equation}\label{eq:fusionSvsS.2}
	\left(\dP{k \lc -1} \xf\left( \dP{\lc -1} \xf \dP{1} \right)\right) \oplus \ds{(k+1)\lc} \xf \dP{1} \simeq \left( \dP{k \lc -1} \xf \dP{\lc} \right) \oplus \dP{(k+1)\lc -1} \oplus \ds{(k+1)\lc -1+2}.
\end{equation}
However, proposition \ref{prop:fusionproj} gives 
\begin{equation*}
	\dP{k \lc -1} \xf \dP{\lc}  \simeq 2\dP{k \lc -1} \xf \dP{\lc -2} \oplus \dP{(k-1) \lc -1} \oplus \dP{(k+1)\lc -1},
\end{equation*}
and proposition \ref{prop:fusion.standvsproj},
\begin{equation*}
	\dP{k \lc -1} \xf \ds{\lc} \simeq \dP{k \lc -1} \xf \dP{\lc -2} \oplus \dP{(k-1) \lc -1} \oplus \dP{(k+1)\lc -1}.
\end{equation*}
Comparing equations \eqref{eq:fusionSvsS.1}, and \eqref{eq:fusionSvsS.2}, and using these two results gives the result for $i =2$.
Suppose then that the result stands for $i-1, i$, with $2 \leq i< \lc -1$. Fusing the left side of \eqref{eq:fusionstandvssl} with $\dP{1}$ and using proposition \ref{prop_fusion_svs01} gives
\begin{equation*}
	\left(\ds{k \lc -1 +i}\xf \dP{1}\right) \xf \ds{\lc} \simeq \left(\ds{k \lc -1 + i-1} \oplus \ds{k \lc -1 + i +1} \right) \xf \ds{\lc}, 
\end{equation*}
while fusing its right side with $\dP{1}$ and using propositions \ref{prop_fusion_svs01}, and \ref{prop_fusion_pvs01} yields
\begin{align}
	\left(\dP{k \lc -1} \right. & \left. \xf \left(\dP{\lc -i}\xf \dP{1}\right)\right) \oplus \left(\dP{(k+1)\lc -1} \xf \left(\dP{i - 2} \xf \dP{1}\right)\right) \oplus \ds{(k+1)\lc -1+i}\xf \dP{1} \notag\\
	& \simeq  \left(\dP{k \lc -1} \xf \left(\dP{\lc -(i-1)} \oplus \dP{\lc - (i+1)}\right)\right) \oplus \left(\dP{(k+1)\lc -1} \xf \left(\dP{i - 3} \oplus \dP{i-1}\right)\right) \notag\\
	& \qquad \oplus \ds{(k+1)\lc -1+i-1}\oplus \ds{(k+1)\lc -1 + i+1}\notag\\
	& \simeq  \left(\dP{k \lc -1} \xf \dP{\lc -(i-1)}\right) \oplus \left(\dP{(k+1)\lc -1} \xf \dP{(i-1) - 2}\right) \oplus \ds{(k+1)\lc -1+(i-1)} \notag\\
	& \qquad \oplus  \left(\dP{k \lc -1} \xf \dP{\lc -(i+1)}\right) \oplus \left(\dP{(k+1)\lc -1} \xf \dP{(i+1) - 2}\right) \oplus \ds{(k+1)\lc -1+(i+1)}
\end{align}
 Comparing the two and using the induction hypothesis yields the result for $i+1$.

We must now do the induction on $k$. Note that when $k=0$, $\ds{0 \lc -1+i} \simeq \dP{i-1}$ and is thus projective. Proposition \ref{prop_fusion_sklvspi} then gives the result when $k=0$. Suppose now that the result holds for $k$ and $i=\lc-1$. There is a short-exact sequence
\begin{equation}
0 \longrightarrow \ds{(k+1) \lc, k \lc -1 + (\lc-1)} \longrightarrow \dP{(k+1 )\lc,(k+1) \lc} \longrightarrow \ds{(k+1) \lc, (k+1) \lc} \longrightarrow 0.
\end{equation} 
Note that the $n = (k+1) \lc$ is important in this case so it is written explicitly. Fusing this sequence with $\ds{\lc,\lc}$ gives the exact sequence
\begin{equation}
\ds{(k+1) \lc, k \lc -1 + (\lc-1)}\xf \ds{\lc,\lc} \overset{f}{\longrightarrow} \dP{(k+1)\lc, (k+1) \lc}\xf \ds{\lc,\lc} \longrightarrow \ds{(k+1) \lc, (k+1) \lc}\xf \ds{\lc,\lc} \longrightarrow 0.
\end{equation}
We thus have the following inequality
\begin{equation}
\begin{tabular}{r l}
$\dim{\ds{(k+1) \lc,(k+1) \lc} \xf \ds{\lc,\lc}} $&$\leq \dim{\dP{(k+1)\lc,(k+1)\lc} \xf \ds{\lc,\lc}} - \dim{\ds{(k+1) \lc, k \lc -1 + (\lc-1)}\xf \ds{\lc,\lc} }$\\
& $= \dim{\dP{(k+1)\lc-1, (k+1)\lc -1} \xf \dP{\lc+1,\lc -1} \oplus \ds{(k+2)\lc,(k+2)\lc}}$, 
\end{tabular}
\end{equation}
where equality stands if and only if $\ker f = 0$, and the second line is obtained by using proposition \ref{prop_fusion_sqlvspkl} and the induction hypothesis with the structure of the projective modules (see section \ref{sec:rappel.indec}). However, lemma \ref{lem:magickdim} gives
$$\dim \ds{(k+1)\lc, (k+1) \lc} \xf \ds{\lc,\lc} = \dim{\dP{(k+1)\lc+1, (k+1)\lc -1} \xf \dP{\lc-1, \lc -1} \oplus \ds{(k+2)\lc,(k+2)\lc}}. $$
It follows that $\ker f =0$, and thus that
\begin{equation}
\begin{tabular}{r l}
$\ds{(k+1) \lc,(k+1) \lc} \xf \ds{\lc,\lc} $&$\simeq \left(\dP{(k+1)\lc,(k+1)\lc} \xf \ds{\lc,\lc}\right)/ \left(\ds{(k+1) \lc, k \lc -1 + (\lc-1)}\xf \ds{\lc,\lc} \right)$\\
& $\simeq \dP{(k+1)\lc-1, (k+1)\lc -1} \xf \dP{\lc+1, \lc -1} \oplus \ds{(k+2)\lc,(k+2)\lc}$, 
\end{tabular}
\end{equation}
where the second equality is obtained by using proposition \ref{prop_fusion_sqlvspkl} and the induction hypothesis with the structure of the projective modules. Note that once the result stands for $n= (k+1)\lc$, fusing it repeatedly with $\dP{2,0}$ will give the result for all $n\geq (k+1)\lc$. It follows that if the result stands for $k$ and $i=\lc-1$, it stands for $k+1$ and $i=1$. Using the first part of the proof, the conclusion is obtained.
\end{proof}

Fusion with $\ds{\lc}$ can thus be used to ``cross" the critical lines. The following continuation of the $\lc=5$ example illustrate how the argument works. Proposition \ref{prop:fusionstandvssl} gives
\begin{equation}
\ds{10} \xf \ds{5} \simeq \dP{9} \xf \dP{4} \oplus \ds{15}\text{.}
\end{equation}
Fusing the left side of this equation with $\ds{5}$ and using propositions \ref{prop:fusionstandvssl}, and \ref{prop:fusion.standvsproj} produces
\begin{align}
	\ds{10} \xf \left(\ds{5} \xf \ds{5}\right) &\simeq \ds{10} \xf \left(\left(\dP{4} \xf \dP{4}\right)  \oplus \ds{10} \right) \notag\\
	& \simeq \left(\dP{9} \xf \dP{3} \oplus \dP{14}\right) \xf \dP{4} \oplus \ds{10}\xf \ds{10},
\end{align}
while fusing its right side with $\ds{5}$ and using the same propositions gives
\begin{align}
	\left(\dP{9}\xf \ds{5} \right) \xf \dP{4} \oplus \ds{15}\xf \ds{5} &\simeq \left( \dP{9} \xf \dP{3}
\oplus \dP{4} \oplus \dP{14} \right) \xf \dP{4} \oplus \left(\dP{14} \xf \dP{4} \oplus \ds{20} \right) \notag\\
	& \simeq \left(  \dP{9} \xf \dP{3} \oplus \dP{14}\right) \xf \dP{4} \oplus \left(\dP{4} \oplus \dP{14} \right) \xf \dP{4} \oplus \ds{20}.
\end{align}
Comparing the two gives the fusion rule $$\ds{10} \xf \ds{10} \simeq (\dP{4} \oplus \dP{14})\xf \dP{4} \oplus \ds{20}.$$ 
\begin{Prop}\label{prop:fusionSqlvsSkl}
For $q$,$k \in \mathbb{Z}_{>0}$,
\begin{equation}\label{eq:propSqlvsSkl}
\ds{q \lc} \xf \ds{ k \lc} \simeq \overset{q+k-1}{\underset{\underset{step =2}{r = |q-k|+1}}{\bigoplus}}\left(\dP{r \lc -1} \xf \dP{\lc -1} \right) \oplus \ds{(q+k)\lc}\text{.}
\end{equation}
\end{Prop}
\begin{proof}
Since fusion is commutative, suppose without loss of generality that $k \leq q$. The proof then proceeds by induction on $k$. For $k=1$, proposition \ref{prop:fusionstandvssl} gives the result for all $q$. Suppose then that the result holds for some $k < q$. Fusing the left side of equation \eqref{eq:propSqlvsSkl} with $\ds{\lc}$ and using propositions \ref{prop:fusionstandvssl}, and \ref{prop_fusion_sqlvspkl} gives
\begin{align}
\ds{q \lc} & \xf \left(\ds{ k \lc} \xf \ds{\lc}\right)  \simeq \ds{q \lc} \xf \left( \dP{k \lc -1} \xf \dP{\lc -1} \oplus \ds{(k+1) \lc}\right)\notag\\
& \simeq \Big(  \overset{q+k-1}{\underset{\underset{step =2}{r = |q-k| + 1}}{\bigoplus}}\left( \dP{r \lc -1} \xf \dP{\lc-2}\right) \oplus \overset{k+q}{\underset{\underset{step =2}{r = |k-(q+1)| + 1}}{\bigoplus}} \dP{r \lc -1} \Big) \xf \dP{\lc -1} \oplus \ds{q \lc} \xf \ds{(k+1)\lc}.
\end{align}
Fusing its right side with $\ds{\lc}$ and using the same propositions yields
\begin{align}
\overset{q+k-1}{\underset{\underset{step =2}{r = q-k+1}}{\bigoplus}}&\left(\dP{r \lc -1} \xf \left(\dP{\lc -1}\xf \ds{\lc} \right) \right) \oplus \ds{(q+k)\lc}\xf \ds{\lc} \notag\\
&\simeq \overset{q+k-1}{\underset{\underset{step =2}{r = q-k+1}}{\bigoplus}}\left(\dP{r \lc -1} \xf \left(\dP{\lc -1}\xf \dP{\lc-2} \oplus \dP{2 \lc -1} \right) \right) \oplus \dP{(q+k)\lc-1}\xf \dP{\lc-1} \oplus \ds{(q+k+1)\lc -1} \notag\\
& \simeq \overset{q+k-1}{\underset{\underset{step =2}{r = q-k+1}}{\bigoplus}}\left(\dP{r \lc -1} \xf \dP{\lc -1}\xf \dP{\lc-2} \oplus \left(\dP{(r-1 )\lc -1}\oplus\dP{(r+1)\lc -1}\right) \xf\dP{\lc -1}  \right) \notag\\
&\qquad \oplus \dP{(q+k)\lc-1}\xf \dP{\lc-1} \oplus \ds{(q+k+1)\lc -1}.
\end{align}
Comparing these two equations gives
\begin{align*}
\overset{k+q}{\underset{\underset{step =2}{r = q+2-k}}{\bigoplus}} &\dP{r \lc -1} \xf \dP{\lc -1} \oplus \ds{q \lc} \xf \ds{(k+1)\lc} \\
&\simeq \overset{q+k-1}{\underset{\underset{step =2}{r = q-k+1}}{\bigoplus}}\left(\dP{(r-1 )\lc -1}\oplus\dP{(r+1)\lc -1}\right) \xf\dP{\lc -1} \oplus \dP{(q+k)\lc-1}\xf \dP{\lc-1} \oplus \ds{(q+k+1)\lc -1} \notag\\
& \simeq \overset{q+k}{\underset{\underset{step =2}{r = q-k+2}}{\bigoplus}}\dP{r\lc -1} \xf \dP{\lc -1} \oplus \overset{q+k}{\underset{\underset{step =2}{r = q-(k+1) +1 }}{\bigoplus}}\dP{r\lc -1} \xf \dP{\lc -1} \oplus \ds{(q+k+1)\lc -1},
\end{align*}
where the second equality is obtained by rearranging the terms in the sum. Comparing the two sides of this equation gives the conclusion.
\end{proof}
Knowing the fusion $\ds{q \lc} \xf \ds{k \lc}$, the fusion $\ds{q\lc +i} \xf \ds{k \lc +j}$ can be computed by using the fusion of $\ds{k \lc +j'}$ with $\dP{1}$.
\begin{Prop}
For $q,k\in \mathbb{Z}_{>0}$, $0 \leq i,j<\lc$,
\begin{multline}\label{eq:fusionStand}
\ds{q \lc-1 +i} \xf \ds{k \lc -1 +j} \simeq \overset{q+k-1}{\underset{\underset{step =2}{r = |q-k|+1}}{\bigoplus}}\left(\dP{r \lc -1} \xf \dP{\lc-|i-j|-1} \right) \oplus \overset{q+k}{\underset{\underset{step =2}{r = |k-q-\text{sign}(i-j)| + 1}}{\bigoplus}}\left(\dP{r \lc -1} \xf \dP{|i-j|-1} \right) \\
\oplus  \overset{\lc -|\lc-(i+j)|-1}{\underset{\underset{step =2}{s = |i-j|+1}}{\bigoplus}}\left(\ds{(q+k) \lc -1 + s} \right) \oplus \dP{(k+q+1)\lc -1} \xf \dP{i+j - \lc -1}.
\end{multline}
\end{Prop}
\begin{proof}
The proof proceeds by induction on $i, j$ and involves many different particular cases: $i<j$, $i=j$ or $i>j$ with $i+j < \lc$ or $i+j>\lc$.

Without loss of generality, suppose $q \geq k$. Proposition \ref{prop:fusionSqlvsSkl} gives the case $i=j=1$, proposition \ref{prop:fusion.standvsproj} gives the case $j=0$ for all $i>0$, and $i=0$ for all $j>0$, while proposition \ref{prop:fusionproj} gives the case $i=j=0$. Suppose now $j \geq 1$, fusing the left side of equation \eqref{eq:fusionStand} with $\dP{1}$ and using propositions \ref{prop_fusion_svs01} and \ref{prop:fusion.standvsproj} gives
\begin{align}\label{eq:fusionSvsS.3}
	\ds{q \lc-1 +i} \xf \left(\ds{k \lc -1 +j} \xf \dP{1} \right) \simeq \ds{q \lc -1+i} \xf \ds{k \lc -1 +j-1} \oplus \ds{q \lc -1 +i} \xf \ds{k \lc -1 +(j+1)},
\end{align}
while fusing the right side of this equation with $\dP{1}$ yields
\begin{align}
	\overbrace{\overset{q+k-1}{\underset{\underset{step =2}{r = |q-k|+1}}{\bigoplus}}\left(\dP{r \lc -1} \xf \left(\dP{\lc-|i-j|-1}\xf \dP{1} \right)\right)}^{a} \oplus \overbrace{\overset{q+k}{\underset{\underset{step =2}{r = |k-q-\text{sign}(i-j)| + 1}}{\bigoplus}}\left(\dP{r \lc -1} \xf \left(\dP{|i-j|-1} \xf \dP{1} \right) \right)}^{b} \notag \\
\oplus  \underbrace{\overset{\lc -|\lc-(i+j)|-1}{\underset{\underset{step =2}{s = |i-j|+1}}{\bigoplus}}\left(\ds{(q+k) \lc -1 + s}\xf \dP{1} \right)}_{c} \oplus \underbrace{\dP{(k+q+1)\lc -1} \xf \left(\dP{i+j - \lc -1}\xf \dP{1} \right)}_{d}.
\end{align}

The terms in $a$ can be written
\begin{align}
\overset{q+k-1}{\underset{\underset{step =2}{r = |q-k|+1}}{\bigoplus}}&\left(\dP{r \lc -1} \xf \left(\dP{\lc-|i-j|-1}\xf \dP{1} \right)\right) \notag\\
& \overset{1}{\simeq} \overset{q+k-1}{\underset{\underset{step =2}{r = |q-k|+1}}{\bigoplus}} \dP{r \lc -1} \xf \left. \begin{cases}
 \dP{\lc}, & \text{ if } i=j \\
\dP{\lc -|i-j|-1 -1}\oplus \dP{\lc -|i-j|-1 +1}, & \text{otherwise}
\end{cases} \right\rbrace \notag\\
& \overset{2}{\simeq} \overset{q+k-1}{\underset{\underset{step =2}{r = |q-k|+1}}{\bigoplus}} \left. \begin{cases}
2\dP{r \lc -1} \xf \dP{\lc -2} \oplus \dP{(r-1)\lc -1} \oplus \dP{(r+1)\lc -1}, & \text{ if } i=j,\\
\dP{r \lc -1} \xf \left(\dP{\lc -|i-j-1| -1}\oplus \dP{\lc -|i-j+1| - 1} \right) , & \text{otherwise}
\end{cases}\right\rbrace \notag\\
& \overset{3}{\simeq} \overset{q+k-1}{\underset{\underset{step =2}{r = |q-k|+1}}{\bigoplus}} \left(\dP{r \lc -1} \xf \left(\dP{\lc -|i-j-1| -1}\oplus \dP{\lc -|i-j+1| - 1} \right) \right) \notag\\
& \qquad \oplus \overset{q+k-1}{\underset{\underset{step =2}{r = |q-k|+1}}{\bigoplus}} \left. \begin{cases}
\dP{(r-1)\lc -1} \oplus \dP{(r+1)\lc -1}, & \text{ if } i=j,\\
0, & \text{otherwise}
\end{cases}\right\rbrace \notag\\
& \overset{4}{\simeq} \overset{q+k-1}{\underset{\underset{step =2}{r = |q-k|+1}}{\bigoplus}} \left(\dP{r \lc -1} \xf \left(\dP{\lc -|i-j-1| -1}\oplus \dP{\lc -|i-j+1| - 1} \right) \right) \notag\\
 & \qquad \oplus \left. 
 \begin{cases}
\overset{q+k}{\underset{\underset{step =2}{r = |k-q-\text{sign}(i-(j-1))| + 1}}{\bigoplus}}\Big(\dP{r \lc -1} \xf \dP{\underbrace{|i-(j-1)|-1}_{= 0}}\Big) & \\
\oplus \overset{q+k-2}{\underset{\underset{step =2}{r = |k-q-\text{sign}(i-(j+1))| + 1}}{\bigoplus}}\Big(\dP{r \lc -1} \xf \dP{\underbrace{|i-(j+1)|-1}_{= 0}}\Big) & \text{ if } i=j,\\
0 & \text{otherwise}.
\end{cases}\right\rbrace
\end{align}
The first equality is obtained by using proposition \ref{prop_fusion_pvs01}, the second by using proposition \ref{prop:fusionproj} and the third and fourth are obtained by noting that $|i-j+1| = |i-j-1| = 1$ when $i=j$ and rearranging the terms in the sums, respectively.

The terms in $b$ can be written
\begin{align}
	\overset{q+k}{\underset{\underset{step =2}{r = |k-q-\text{sign}(i-j)| + 1}}{\bigoplus}}&\left(\dP{r \lc -1} \xf \left(\dP{|i-j|-1} \xf \dP{1} \right) \right) \notag\\
	&\simeq \overset{q+k}{\underset{\underset{step =2}{r = |k-q-\text{sign}(i-j)| + 1}}{\bigoplus}}\left.
	\begin{cases}	
	0, & \text{ if } i=j,\\
	\dP{r \lc -1} \xf \left(\dP{|i-j|-2} \oplus \dP{|i-j|}\right), & \text{ otherwise}
	\end{cases}\right\rbrace \notag \\
	& \simeq \overset{q+k}{\underset{\underset{step =2}{r = |k-q-\text{sign}(i-j)| + 1}}{\bigoplus}}\left.
	\begin{cases}	
	0, & \text{ if } i=j,\\
	\dP{r \lc -1} \xf \left(\dP{|i-j-1|-1} \oplus \dP{|i-j+1|-1}\right), & \text{ otherwise}
	\end{cases}\right\rbrace \notag\\
	& \simeq \left. \begin{cases}
	0, & \text{ if } i=j,\\
	 \overset{q+k}{\underset{\underset{step =2}{r = |k-q-\text{sign}(i-j-1)| + 1}}{\bigoplus}}\left(\dP{r \lc -1} \xf \dP{|i-j-1|-1} \right) &\\
	  \oplus \overset{q+k}{\underset{\underset{step =2}{r = |k-q-\text{sign}(i-j+1)| + 1}}{\bigoplus}}\left(\dP{r \lc -1} \xf \dP{|i-j+1|-1} \right),& \text{ otherwise}
	 \end{cases}\right\rbrace .
\end{align}
The first equality is obtained by using proposition \ref{prop_fusion_pvs01} and the fact that $\dP{-1} \equiv 0$, while the second one is obtained by noting that if $i>j$, $|i-j| -2 = |i - (j+1)| -1$, $|i-j| = |i-(j-1)|-1$ while if $i<j$, $|i-j|-2 = |i-(j-1)|-1$, $|i-j| = |i - (j+1)|-1$. The third one is obtained by noting that if $i \neq j$ and $\text{sign}(i-j) \neq \text{sign}(i-j \pm 1)$, then $|i-j \pm 1| -1 < 0$, and thus $\dP{|i-j \pm 1| -1} \equiv 0$.

The terms in $c$ can be written
\begin{align}
	\overset{\lc -|\lc-(i+j)|-1}{\underset{\underset{step =2}{s = |i-j|+1}}{\bigoplus}}&\left(\ds{(q+k) \lc -1 + s}\xf \dP{1} \right) \notag\\
	& \simeq \overset{\lc -|\lc-(i+j)|-1}{\underset{\underset{step =2}{s = |i-j|+1}}{\bigoplus}}\left(
	\ds{(q+k)\lc -1 + s -1} \oplus \ds{(q+k)\lc -1 + s+1}\right) \notag\\
	& \simeq \overset{\lc -|\lc-(i+j)|-2}{\underset{\underset{step =2}{s = |i-j|}}{\bigoplus}}\ds{(q+k) \lc -1 + s} \oplus  \overset{\lc -|\lc-(i+j)|}{\underset{\underset{step =2}{s = |i-j|+2}}{\bigoplus}}\ds{(q+k) \lc -1 + s} \notag\\
	& \simeq \overset{\lc -|\lc-(i+j+1)|-1}{\underset{\underset{step =2}{s = |i-j-1|+1}}{\bigoplus}}\ds{(q+k) \lc -1 + s} \oplus \overset{\lc -|\lc-(i+j-1)|-1}{\underset{\underset{step =2}{s = |i-j+1|+1}}{\bigoplus}}\ds{(q+k) \lc -1 + s} \notag\\
	& \oplus \delta_{0,|i-j|} \dP{(k+q)\lc -1} \oplus \delta_{i+j, \lc} \dP{(k+q+1)\lc -1}.
\end{align}
The first equality is obtained by applying proposition \ref{prop_fusion_svs01}, the second by splitting the sum in two and renaming the indices while the third is obtained by considering the different possibilities for the absolute values and rearranging the two sums accordingly.

The terms in $d$ can be written
\begin{align}
	\dP{(k+q+1)\lc -1} \xf & \left(\dP{i+j - \lc -1}\xf \dP{1} \right) \notag\\
	& \simeq \dP{(k+q+1)\lc -1} \xf \left. \begin{cases}
	0,& \text{ if } i+j < \lc +1,\\
	(\dP{i+j-1 - \lc -1} \oplus \dP{i+j+1 - \lc -1}	), & \text{ otherwise}
	\end{cases}\right\rbrace
\end{align}
by simply using proposition \ref{prop_fusion_pvs01} and the fact that $\dP{t} \equiv 0$ when $t <0$.

Putting all of these together, grouping the terms in the appropriate manner and comparing the result with equation \eqref{eq:fusionSvsS.3} yields the conclusion for $i, j+1$, provided that it stands for $i,j$, $i,j-1$. The induction to $i+1,j$ from $i-1,j$ is done using the same arguments, except that in equation \eqref{eq:fusionSvsS.3}, $\dP{1}$ is fused with $\ds{q \lc-1 +i}$ instead of $\ds{k \lc-1 +j}$, and the rearranging used to reorder the sums in the different terms is slightly different.
\end{proof}
\end{subsection}
\begin{subsection}{The case $\lc =2$ in $\tl{n}$}\label{sec:tl_l2}
We treat here the regular Temperley-Lieb family when $\lc =2$. Recall that in this case the module $\dP{0}$ which was used to remove the dependence on $n$ is trivial, so the proof of proposition \ref{prop:fusionstandvssl} does not work. The method used here is more tedious than that of the previous section but it will ultimately give the same results.

\begin{Prop}\label{prop:fusion_ds2vsds2}
When $\lc =2$, in the regular family,
\begin{equation}
\ds{n,2}\xf \ds{m,2} \simeq \dP{n+m,2} \oplus \ds{n+m,4}.
\end{equation}
If $n \geq 4$,
\begin{equation}
\dl{n,2} \xf \ds{m,2} \simeq \ds{n+m,2}.
\end{equation}
\end{Prop}
\begin{proof}
The case $n=m=2$ is particular and it must be computed by hand. Using the same arguments as in lemma \ref{lem:magickdim}  the following set is a basis of $\ds{2,2} \xf \ds{2,2}$ :
\begin{equation}
\begin{tikzpicture}[baseline={(current bounding box.center)},scale = 1/3]
	\draw[very thick] (0,-2) -- (0,2);
	\draw[very thick] (3,-2) -- (3,2);
	\draw (3,-3/2) -- (0,-3/2);
	\draw (3,-1/2) -- (0,-1/2);
	\draw (3,1/2) -- (0,1/2);
	\draw (3,3/2) -- (0,3/2);
\end{tikzpicture}
\otimes \begin{tikzpicture}[baseline={(current bounding box.center)},scale = 1/3]
	\draw[very thick] (0,-5/2) -- (0,-1/2);
	\draw[very thick] (0,1/2) -- (0,5/2);
	\draw (0,-2) -- (2,-2);
	\draw (0,-1) -- (2,-1);
	\draw (0,1) -- (2,1);
	\draw (0,2) -- (2,2);
\end{tikzpicture} ,\qquad\underbrace{
\begin{tikzpicture}[baseline={(current bounding box.center)},scale = 1/3]
	\draw[very thick] (0,-2) -- (0,2);
	\draw[very thick] (3,-2) -- (3,2);
	\draw (3,-3/2) -- (2,-3/2);
	\draw (3,-1/2) .. controls (2,-1/2) and (2,1/2) .. (3,1/2);
	\draw (3,3/2) -- (2,3/2);
	\node at (1,0) {$x_1$};
\end{tikzpicture}\otimes \begin{tikzpicture}[baseline={(current bounding box.center)},scale = 1/3]
	\draw[very thick] (0,-5/2) -- (0,-1/2);
	\draw[very thick] (0,1/2) -- (0,5/2);
	\draw (0,-2) -- (2,-2);
	\draw (0,-1) -- (2,-1);
	\draw (0,1) -- (2,1);
	\draw (0,2) -- (2,2);
\end{tikzpicture}, \qquad
\begin{tikzpicture}[baseline={(current bounding box.center)},scale = 1/3]
	\draw[very thick] (0,-2) -- (0,2);
	\draw[very thick] (3,-2) -- (3,2);
	\draw (3,-3/2) .. controls (3/2,-3/2) and (3/2,3/2) ..  (3,3/2);
	\draw (3,-1/2) .. controls (2,-1/2) and (2,1/2) .. (3,1/2);
	\node at (1,0) {$x_2$};
\end{tikzpicture}\otimes \begin{tikzpicture}[baseline={(current bounding box.center)},scale = 1/3]
	\draw[very thick] (0,-5/2) -- (0,-1/2);
	\draw[very thick] (0,1/2) -- (0,5/2);
	\draw (0,-2) -- (2,-2);
	\draw (0,-1) -- (2,-1);
	\draw (0,1) -- (2,1);
	\draw (0,2) -- (2,2);
\end{tikzpicture}}_{V},
\end{equation}
where $x_{i}$ are the link diagrams in $\ds{4,4-2i}$ and it can be seen directly that the elements of $V$ spans a submodule of $\ds{2,2} \xf \ds{2,2} $. However when $\lc =2$, $\dP{4,2}\simeq\dP{3,1}\xf \dP{1,1} $ is spanned by
\begin{equation}
\begin{tikzpicture}[baseline={(current bounding box.center)},scale = 1/3]
	\draw[very thick] (0,-2) -- (0,2);
	\draw[very thick] (3,-2) -- (3,2);
	\node at (1,0) {$x_1$};
	\draw (2,-3/2) -- (3,-3/2);
	\draw (2,-1/2) -- (3,-1/2);
	\draw (3,1/2) .. controls (2,1/2) and (2,3/2) .. (3,3/2);
\end{tikzpicture} \otimes
\begin{tikzpicture}[baseline={(current bounding box.center)},scale = 1/3]
	\draw[very thick] (0,-1) -- (0,2);
	\draw (0,3/2) -- (3,3/2);
	\draw (0,-1/2) .. controls (1,-1/2) and (1,1/2) .. (0,1/2);
	\draw[very thick] (0,-5/2) -- (0,-3/2);
	\draw (0,-2) -- (3,-2);
\end{tikzpicture} , \qquad
\begin{tikzpicture}[baseline={(current bounding box.center)},scale = 1/3]
	\draw[very thick] (0,-2) -- (0,2);
	\draw[very thick] (3,-2) -- (3,2);
	\node at (1,0) {$x_2$};
	\draw (3,-3/2) ..controls (2,-3/2) and (2,-1/2) .. (3,-1/2);
	\draw (3,1/2) .. controls (2,1/2) and (2,3/2) .. (3,3/2);
\end{tikzpicture} \otimes
\begin{tikzpicture}[baseline={(current bounding box.center)},scale = 1/3]
	\draw[very thick] (0,-1) -- (0,2);
	\draw (0,3/2) -- (3,3/2);
	\draw (0,-1/2) .. controls (1,-1/2) and (1,1/2) .. (0,1/2);
	\draw[very thick] (0,-5/2) -- (0,-3/2);
	\draw (0,-2) -- (3,-2);
\end{tikzpicture}, 
\end{equation}
where  $x_{i}$ are the link diagrams in $\ds{4,4-2i}$. A simple verification shows that $\text{span}\{V\}\simeq \dP{4,2}$, and that $(\ds{2,2} \xf \ds{2,2})/\text{Span}\{V\} \simeq \ds{4,4}$. Using the fact that $\dP{4,2}$ is injective (see section \ref{sec:rappel.indec}) yields the conclusion.

Suppose that $n\geq m$, $n \geq 4$  and start with the exact sequence $$ 0 \longrightarrow \dl{n,2} \longrightarrow \dP{n,2} \longrightarrow \ds{n,2} \longrightarrow 0,$$ which becomes
\begin{equation}
\dl{n,2}\xf \ds{m,2} \overset{f}{\longrightarrow} \dP{n+m,2} \oplus \dP{n+m,4} \longrightarrow \ds{n,2} \xf \ds{m,2} \longrightarrow 0,
\end{equation}
by using the right-exactness of fusion with the fusion rules \ref{prop_fusion_sqlvspkl}. To find $\dl{n,2}\xf \ds{m,2}$, fuse the sequence
$$\dP{m,2} \longrightarrow \dP{m,2} \longrightarrow \ds{m,2} \longrightarrow 0, $$
with $\dl{n,2}$ to obtain,, 
\begin{equation}
\dP{n+m,2} \longrightarrow \dP{n+m,2} \longrightarrow \ds{m,2} \xf \dl{n,2} \longrightarrow 0,
\end{equation}
where proposition \ref{prop:fusionIvsP} was used. Note that the proof of this proposition is independent of this one so it can safely be used. It follows that there are three possibilities
\begin{equation*}
\ds{m,2} \xf \dl{n,2} \simeq \begin{cases}
\dP{n+m,2}, & \\
\ds{n+m,2}, & \\
0
\end{cases}.
\end{equation*}
But, proposition \ref{prop:fusionIvsP} gives $\ds{m,2}\xf \dl{n,2} \xf \dP{1,1} \simeq \ds{m,2}\xf \dP{n+1,1} \simeq \dP{n+m+1,1} \oplus \dP{n+m+1,3}$. Since $\dP{n+m,2}\xf \dP{1,1} \simeq 2\dP{n+m+1,1} \oplus \dP{n+m+1,3}$, it follows that $$\ds{m,2} \xf \dl{n,2} \simeq \ds{n+m,2}. $$
Now, the morphisms from $\ds{n+m,2}$ to $ \dP{n+m,2} \oplus \dP{n+m,4}$ are known (see their Loewy diagrams) and the cokernel of $f$ must be one of the following modules
\begin{equation*}
 \dP{n+m,2} \oplus \dP{n+m,4}, \qquad \ds{n+m,2} \oplus \dP{n+m,4}, \qquad \dP{n+m,2} \oplus \ds{n+m,4}.
\end{equation*}
Using propositions \ref{prop_fusion_pvs01} and \ref{prop_fusion_svs01},
\begin{equation*}
\left( \dP{n+m,2} \oplus \dP{n+m,4}\right)\xf \dP{1,1} \simeq 3\dP{n+m+1,1} \oplus 3\dP{n+m+1,3} \oplus \dP{n+m+1,5},
\end{equation*}
\begin{equation*}
(\ds{n+m,2} \oplus \dP{n+m,4})\xf \dP{1,1} \simeq 2\dP{n+m+1,1} \oplus 3\dP{n+m+1,3} \oplus \dP{n+m+1,5},
\end{equation*}
\begin{equation*}
(\dP{n+m,2} \oplus \ds{n+m,4})\xf \dP{1,1} \simeq 2\dP{n+m+1,1} \oplus 2\dP{n+m+1,3} \oplus \dP{n+m+1,5},
\end{equation*}
while $$\ds{n,2} \xf \ds{m,2} \xf \dP{1,1} \simeq \ds{n,2} \xf (\dP{m+1,1} \oplus \dP{m+1,3}) \simeq 2\dP{n+m+1,1} \oplus 2\dP{n+m+1,3} \oplus \dP{n+m+1,5}.$$
It thus follows that $\ds{n,2}\xf \ds{m,2} \simeq \dP{n+m,2} \oplus \ds{n+m,4}$, as long as one of $n$ or $m$ is bigger or equal to $4$. 
\end{proof}
Now that the fusion of $\ds{n,2}$ with itself is known, it can be used to compute the fusion of the other standard modules. Note that the fusion of $\ds{2,2}$ with standard modules other than $\ds{2,2}$ can be obtained by the same arguments as in proposition \ref{prop:fusionstandvssl}, so we will only give the proof for $\ds{n,2} = \ds{2}$ with $n\geq 4$. We present a few examples before proving the general case. There is an exact sequence $$0 \longrightarrow \ds{2} \longrightarrow \dP{4} \longrightarrow \ds{4} \longrightarrow 0, $$
which becomes
\begin{equation}
\ds{2} \overset{f}{\longrightarrow} \dP{4} \longrightarrow \dl{2} \xf \ds{4} \longrightarrow 0,
\end{equation}
by fusing it with $\dl{2}$ and using the preceding proposition with proposition \ref{prop:fusionIvsP}. Note that $\dl{2} \xf \ds{4} \xf \dP{1} \simeq \ds{4}\xf \dP{1} \simeq \dP{3} \oplus \dP{5}$. Since the cokernel of $f$ is either $\dP{4}$ or $\ds{4}$, it follows that $\dl{2} \xf \ds{4} \simeq \ds{4}$. Now, the exact sequence $$0 \longrightarrow \dl{2} \longrightarrow \dP{2} \longrightarrow \ds{2} \longrightarrow 0, $$
when fused with $\ds{4}$, yield the exact sequence
$$\ds{4} \longrightarrow \dP{4} \oplus \dP{6} \longrightarrow \ds{2} \xf \ds{4} \longrightarrow 0, $$ by using proposition \ref{prop:fusion.standvsproj}. There are thus three possibilities
$$\ds{2} \xf \ds{4} \simeq \dP{4} \oplus \dP{6}, \qquad \B{2}{2} \oplus \dP{6} \qquad \text{ or } \dP{4} \oplus \ds{6}. $$ 
But, using proposition \ref{prop:fusion.standvsproj}, we get
$$\ds{2} \xf \ds{4} \xf \dP{1} \simeq (\dP{1} \oplus \dP{3})\xf \ds{4} \simeq \dP{1} \oplus 2\dP{3} \oplus 2\dP{5} \oplus \dP{7}.$$
Then, we verify which of the three possibilities satisfies this rule:
\begin{equation*}
(\dP{4} \oplus \dP{6}) \xf \dP{1} \simeq \dP{1} \oplus 3\dP{3} \oplus 3\dP{5} \oplus \dP{7},
\end{equation*}
\begin{equation*}
(\B{2}{2} \oplus \dP{6}) \xf \dP{1} \simeq \dP{1} \oplus 2\dP{3} \oplus 3 \dP{5} \oplus \dP{7},
\end{equation*}
\begin{equation*}
(\dP{4} \oplus \ds{6}) \xf \dP{1} \simeq \dP{1} \oplus 2\dP{3} \oplus 2 \dP{5} \oplus \dP{7},
\end{equation*}
where propositions \ref{prop_fusion_pvs01}, \ref{prop_fusion_svs01}, and \ref{prop:fusionBvsP} were used. We are allowed to do so because the proofs of these propositions are independent of the fusion rules for standard modules. Comparing these fusion with $\ds{2} \xf \ds{4} \xf \dP{1} $, it follows that $$\ds{2} \xf \ds{4} \simeq \dP{4} \oplus \ds{6}.$$ 
The proof of the general result that follows is obtained by induction and repeats the preceding arguments.
\begin{Prop}
For $n \geq 4$, $m/2 \geq k \geq 1$,
$$\ds{n,2} \xf \ds{m,2k} \simeq \dP{n+m,2k} \oplus \ds{n+m, 2(k+1)},$$
and
$$\dl{n,2} \xf \ds{m,2k} \simeq \ds{n+m,2k}.$$
\end{Prop}
Note that a simple corollary of this proposition is that $\dl{n,2}\simeq \ds{n,0}$ when $\lc =2 $, plays the role of $\dP{n,0}$ when $\lc \neq 2$, except that in this case $n\geq 4$ instead of $n\geq 2$.
\end{subsection}

\begin{subsection}{A simple rule for fusion}
The fusion rules for standard modules and projective modules can be hard to apply in practice because of the numerous direct sum and fusions involved; we thus present a simple ``rule of thumb" to quickly compute fusion of standard modules.
\begin{Prop}
To a standard module $\ds{i}$ ($i$ can be critical), associate the Chebyshev polynomial of the second kind $U_{i}(\frac{x}{2})$ where $x$ is a formal parameter. To a projective module $\dP{\kc +j}$, $\lc>j>0$, associate the sum of Chebyshev polynomials $U_{\kc-j}(\frac{x}{2}) + U_{\kc +j}(\frac{x}{2})$. Call this association the \emph{polynomial representation} of the modules. Furthermore, since the polynomials all have the same argument, it will simply be omitted. To take the fusion of two modules, multiply their polynomial representations and split the result by using the product rule $$ U_{i}U_{j} = \overset{i+j}{\underset{step =2}{\underset{k=|i-j|}{\Sigma}}} U_{k}.$$ Collect the terms in this sum to form the polynomial representation of projective modules, starting with the smallest $k$. Remaining terms are then identified with the corresponding standard modules.
\end{Prop}
It is straightforward but tedious to prove that all fusion rules obtained so far respect this simple rule. 
\end{subsection}
\end{section}
\begin{section}{Fusion of quotients}\label{sec:fusionquotients}
We are now trying to compute the fusion of two irreducible modules. We begin by explaining the general idea which we will use to compute them. Suppose there are two modules $U,V$ and two resolutions $$U_{2} \longrightarrow U_{1} \longrightarrow U \longrightarrow 0, \qquad V_{2} \longrightarrow V_{1} \longrightarrow V \longrightarrow 0,  $$ by modules $U_{i},V_{i}$. It is a standard exercise in diagram chasing to obtain the exact sequence 
\begin{equation}
U_{2}\xf V_{1} \oplus U_{1}\xf V_{2} \longrightarrow U_{1} \xf V_{1} \overset{\phi}{\longrightarrow} U \xf V \longrightarrow 0.
\end{equation}
If $\phi$ can be computed somehow, the knowledge of the fusion rules for $U_{1} \xf V_{1}$, $U_{2} \xf V_{1}$ and $U_{1} \xf V_{2}$ will give $U \xf V$. If $U_{1}$, $V_{1}$ are ``close" to $U$ and $V$, the kernel of $\phi$ will be small, and its image will be much easier to compute. The idea is therefore to find the $U_{1}$, $V_{1}$ that are the ``closest" to $U$ and $V$ but such that their fusion can be computed. Of course the ``closest" module to an irreducible $\dl{n,i}$ is $\dl{n,i}$ itself, the second closest would be the standard module $\ds{n,i}$ and the third would be the projective module $\dP{n,i}$. The goal is thus to find the fusion of irreducible modules with  projective ones, which will then be used to compute the fusion of irreducible modules with standard modules. This is where the modules $\B{2i}{k}$s appear. We will then compute the fusion rules for these modules, introducing yet another class of modules, the $\T{2i+1}{k}$s. Computing the fusion of these modules with projective and standard modules will be the last step before arriving at the fusion of two irreducibles. Note that the same arguments will be used over and over again so we will not detail the proofs as much as in the preceding sections.
\begin{subsection}{Fusion of irreducible and projective modules}\label{sec:fusionIvsP}
We start by giving the rules for the induction of $\dl{n,k}$ \cite{BelSY}.
\begin{Prop}
If $n\geq k \lc- 1+i$, $0<i<\lc$,
\begin{multline}
\Ind{\dl{n,k \lc -1 -i}} \simeq \left. \begin{cases}
\dl{n+1,k \lc -1 -i}& \text{ in } \dtl{n} \\
0 & \text{ in } \tl{n}
\end{cases}\right\} \oplus \begin{cases}
\dP{n+1, (k-1) \lc -1} & \text{if } i =\lc-1\\
\dl{n+1,k\lc -1 -i-1} & \text{otherwise}
\end{cases}\\
\oplus \begin{cases}
0 & \text{if } i=1\\
\dl{n+1,k \lc -1 - i +1} & \text{ otherwise}\\
\end{cases}.
\end{multline}
\end{Prop}
\noindent The condition on $n$ ensures that the module under study is not a standard module. Using proposition \ref{prop:fusion_parity} with the parity of the irreducibles gives the following fusion rules.
\begin{Prop}\label{prop:fusionIvs01}
If $n\geq k \lc- 1+i$, $0 < i < \lc$,then in the dilute Temperley-Lieb family
\begin{equation}
\dl{n, k \lc -1 -i} \xf \dP{1,0} \simeq \dl{n+1, k \lc -1 -i},
\end{equation}
while in both families
\begin{equation}
\dl{n, k \lc -1 -i} \xf \dP{1,1} \simeq \begin{cases}
\dP{n+1, (k-1) \lc -1} & \text{if } i =\lc-1\\
\dl{n+1,k\lc -1 -i-1} & \text{otherwise}
\end{cases}
\oplus \begin{cases}
0 & \text{if } i=1\\
\dl{n+1,k \lc -1 - i +1} & \text{ otherwise}\\
\end{cases}.
\end{equation}
\end{Prop}
In the standard family, when $\lc \neq 2$, $$\dl{n,i} \xf \dP{2,0} \simeq \dl{n+2,i}, $$ which is proven in proposition \ref{prop:magickOthersP}. The proofs in this section will be independent of $n$ as long as it is big enough for the irreducible modules to be distinct from the standard modules; we will therefore simply omit the $n$. Note now that
\begin{equation*}
	\dl{k \lc -2} \xf \dP{1} \simeq \dl{k \lc -3}.
\end{equation*}
Fusing the left side of this equation with $\dP{1}$ and using proposition \ref{prop_fusion_pvs01} gives
\begin{equation}
	\dl{k \lc -2} \xf \left(\dP{1} \xf \dP{1} \right) \simeq \dl{k \lc -2} \xf \left(\dP{0} \oplus \dP{2}\right),
\end{equation}
while fusing its right side with $\dP{1}$ and using proposition \ref{prop:fusionIvs01} gives
\begin{equation}
\dl{k \lc -3} \xf \dP{1} \simeq \dl{k\lc -2} \oplus \dl{k \lc -4}.
\end{equation}
Comparing the two results then yields the fusion rule
\begin{equation*}
	\dl{k \lc -2} \xf \dP{2} \simeq \dl{k \lc -4}.
\end{equation*}
The following proposition is then obtained by simply repeating these arguments.
\begin{Prop}\label{prop:fusionIvssmallP}
For all $0\leq i<\lc-1$,
\begin{equation}
\dl{k \lc -2} \xf \dP{i} \simeq \dl{k \lc -2-i}.
\end{equation}
\end{Prop}
Once the fusion rules for $\dl{k\lc -2}$ are known, this proposition will be used to quickly compute the fusion of the other irreducible modules, since for all $0 < i < \lc$ and any module $M$,
\begin{equation*}
	\dl{k \lc -1 -i} \xf M \simeq \left(\dl{k \lc -2} \xf M \right) \xf \dP{i-1}.
\end{equation*}

 For $k>1$, $i=\lc-1$, the same arguments give
\begin{equation*}
\dl{k\lc -2} \xf \dP{\lc-1} \simeq \dP{(k-1)\lc -1}.
\end{equation*}
Fusing this repeatedly with $\dP{1}$ then yields
\begin{equation*}
\dl{k \lc -2} \xf \dP{\lc-1} \xf \dP{1} \simeq \dl{k\lc -2} \xf \dP{\lc} \simeq \dP{(k-1)\lc},
\end{equation*}
\begin{equation*}
\dl{k \lc -2} \xf \dP{\lc+1} \simeq \dP{(k-1)\lc +1},
\end{equation*}
\begin{equation*}
\dl{k \lc -2} \xf \dP{\lc+2} \simeq \dP{(k-1)\lc +2}.
\end{equation*}
Continuing in this manner eventually yields
\begin{equation*}
\dl{k\lc -2} \xf \dP{\lc  + \lc -1} \simeq \dP{(k-2)\lc -1} \oplus \dP{k \lc -1}.
\end{equation*}
Note that if $k=2$, $\dP{(k-2)\lc -1} \simeq \dP{ -1}\simeq 0$. The following proposition gives the general formula.
\begin{Prop}\label{prop:fusionIvsP}
For all $k>1$,$r\geq 1 $, $0\leq i < \lc-1$, $0\leq j<\lc$,
\begin{equation}\label{eq:fusionIvsP}
\dl{k \lc -2- i} \xf \dP{r \lc -1 +j} \simeq \overset{k+r-2}{\underset{\underset{step=2}{p = \max\left(k-r, r-k +2 \right)}}{\bigoplus}} \dP{p \lc -1 +j} \xf \dP{i}.
\end{equation}
\end{Prop}
\begin{proof}
The proof proceeds by induction on $r$ and $j$. The cases $r=1$ (for all $j$) and $r =2$, $j=0$ were proved in the preceding discussion, so  suppose that the result stands for some $r$ and $j =0$. Fusing the left side of equation \eqref{eq:fusionIvsP} with $\dP{1}$ and using proposition \ref{prop_fusion_pvs01} then gives
\begin{equation*}
	\dl{k \lc -2- i} \xf \left( \dP{r \lc -1} \xf \dP{1}\right) \simeq \dl{k \lc -2 -i} \xf \dP{r \lc} ,
\end{equation*}
while fusing its right side with $\dP{1}$ and using the same proposition yields
\begin{equation*}
	\overset{k+r-2}{\underset{\underset{step=2}{p = \max\left(k-r, r-k +2 \right)}}{\bigoplus}}\left( \dP{p \lc -1 } \xf \dP{1} \right)\xf \dP{i} \simeq \overset{k+r-2}{\underset{\underset{step=2}{p = \max\left(k-r, r-k +2 \right)}}{\bigoplus}} \dP{p \lc } \xf \dP{i}.
\end{equation*}
The case $j=1$ is then obtained by simply comparing the two results. Now, assume that the result stands for this $q$ and $j-1,j$, $1 \leq j < \lc -1$. Fusing the left side of equation \eqref{eq:fusionIvsP} and using proposition \ref{prop_fusion_pvs01} gives
\begin{equation*}
	\dl{k \lc -2- i} \xf \left(\dP{r \lc -1 +j} \xf \dP{1} \right) \simeq \dl{k \lc -2 -i} \xf \left((1+\delta_{j,1})\dP{r \lc -1 + (j-1)}  \oplus \dP{r \lc -1 +j+1} \right),
\end{equation*}
while fusing its right side and using the same proposition yields
\begin{align*}
	&\overset{k+r-2}{\underset{\underset{step=2}{p = \max\left(k-r, r-k +2 \right)}}{\bigoplus}}\left( \dP{p \lc -1 +j} \xf \dP{1}\right) \xf \dP{i} \notag\\
	&\qquad \simeq \overset{k+r-2}{\underset{\underset{step=2}{p = \max\left(k-r, r-k +2 \right)}}{\bigoplus}}\left( (1 + \delta_{j,1})\dP{p \lc -1 +j-1} \oplus \dP{p \lc -1 +j+1}\right) \xf \dP{i} \notag\\
	& \qquad \simeq (1 + \delta_{j,1})\overset{k+r-2}{\underset{\underset{step=2}{p = \max\left(k-r, r-k +2 \right)}}{\bigoplus}}\left( \dP{p \lc -1 +j-1} \xf \dP{i}\right)  \oplus \overset{k+r-2}{\underset{\underset{step=2}{p = \max\left(k-r, r-k +2 \right)}}{\bigoplus}}\left( \dP{p \lc -1 +j+1} \xf \dP{i}\right) .
\end{align*}
Comparing these two results and using the induction hypothesis then yields the conclusion for $j+1$. Note that doing the same thing for the case $j =\lc -1$ gives, for the left side
\begin{equation*}
	\dl{k \lc -2- i} \xf \left(\dP{(r+1) \lc -2} \xf \dP{1} \right) \simeq \dl{k \lc -2 -i} \xf \left((1+\delta_{\lc, 2})\dP{r \lc -1 + (\lc -2)}  \oplus \dP{(r+1) \lc -1 } \oplus \dP{(r-1)\lc -1} \right),
\end{equation*}
and for the right side
\begin{align*}
	&\overset{k+r-2}{\underset{\underset{step=2}{p = \max\left(k-r, r-k +2 \right)}}{\bigoplus}}\left( \dP{p \lc -1 +\lc - 1} \xf \dP{1}\right) \xf \dP{i} \\
	&\qquad \simeq (1 + \delta_{\ell,2})\overset{k+r-2}{\underset{\underset{step=2}{p = \max\left(k-r, r-k +2 \right)}}{\bigoplus}}\left( \dP{p \lc -1 +\ell -2} \xf \dP{i}\right) \oplus \overset{k+r-2}{\underset{\underset{step=2}{p = \max\left\{k-r, r-k +2 \right\}}}{\bigoplus}}\left( \dP{(p+1) \lc -1} \xf \dP{i}\right) \\
	& \qquad \qquad  \oplus \overset{k+r-2}{\underset{\underset{step=2}{p = \max\left\{k-r, r-k +2 \right\}}}{\bigoplus}}\left( \dP{(p-1) \lc -1} \xf \dP{i}\right)\\
	& \qquad \simeq (1 + \delta_{\ell ,2})\overset{k+r-2}{\underset{\underset{step=2}{p = \max\left(k-r, r-k +2 \right)}}{\bigoplus}}\left( \dP{p \lc -1 +\ell -2} \xf \dP{i}\right) \oplus \overset{k+r+1-2}{\underset{\underset{step=2}{p = \max\left(k-(r+1), r+1-k +2 \right)}}{\bigoplus}}\left( \dP{p \lc -1} \xf \dP{i}\right) \\
	& \qquad \qquad  \oplus \overset{k+r-1-2}{\underset{\underset{step=2}{p = \max\left\{k-(r-1), r-1-k +2 \right\}}}{\bigoplus}}\left( \dP{p \lc -1} \xf \dP{i}\right),
\end{align*}
where the last equality is obtained by rearranging the terms between the sums and considering the different values of $r-k$. Comparing the two sides, it follows that if the conclusion holds for $r-1, j=0$, $r,j=\lc-1, \lc-2 $, it will also hold for $r+1,j=0$.
\end{proof}
Note that if $k=1$, repeating the arguments leading to proposition \ref{prop:fusionIvssmallP} gives
\begin{equation*}
\dl{\lc-2} \xf \dP{\lc-3} \simeq \dl{1},
\end{equation*}
\begin{equation*}
\dl{\lc-2} \xf \dP{\lc-2} \simeq \dl{0},
\end{equation*}
\begin{equation*}
\dl{\lc-2} \xf \dP{\lc-1} \simeq 0.
\end{equation*}
This implies of course that $\dl{\lc-2} \xf \dP{i}\simeq 0$ for all $i\geq \lc-1$.
\begin{Prop}\label{prop:fusionIWeird}
For all $i\geq \lc-1$, $j < \ell -1$,
\begin{equation}
\dl{j} \xf \dP{i} \simeq 0.
\end{equation}
\end{Prop}
Note also that since fusion is right-exact, fusing $\dl{\lc-2}$ with any quotient of $\dP{k}$ will always yield $0$. This include the standard non-projective modules as well as all irreducibles $\dl{k}$ with $k>\lc-1$.
\end{subsection}
\begin{subsection}{Fusion of irreducible and standard modules, first part}\label{sec:fusionIvsS1}
Proposition \ref{prop:fusionstandvssl} can be used to obtain the non-projective standard modules by repeatedly fusing $\ds{n,\lc}$ with itself and small projectives. The first step to obtain the fusion of irreducible modules with standard modules is thus to compute $\dl{k \lc -2} \xf \ds{\lc}$, for $k>1$. There is a short exact sequence 
\begin{equation*}
0 \longrightarrow \dP{\lc-2} \longrightarrow \dP{\lc} \longrightarrow \ds{\lc} \longrightarrow 0.
\end{equation*}
Using the right-exactness of fusion together with known fusion rules, this yields the exact sequence
\begin{equation*}
\dl{(k-1)\lc} \overset{f}{\longrightarrow} \dP{(k-1)\lc} \longrightarrow \dl{k\lc-2} \xf \ds{\lc} \longrightarrow 0.
\end{equation*}
Since $\dl{(k-1)\lc}$ is irreducible, $f$ is either zero or injective. If it is injective, then $\dl{k\lc-2} \xf \ds{\ell} \simeq \B{2}{(k-1)\lc-2}$ by proposition \ref{prop:indec.prop} while if $f=0$, $\dl{k\lc-2} \xf \ds{\lc} \simeq \dP{(k-1)\lc}$. However, note that by propositions \ref{prop:fusionIvsP} and \ref{prop_fusion_sklvspi}
\begin{equation*}
\dl{k \lc -2} \xf \ds{\lc} \xf \dP{\lc-1} \simeq \dP{(k-1)\lc-1} \xf \ds{\lc},
\end{equation*}
while by proposition \ref{prop:fusionproj}
\begin{equation*}
\dP{(k-1)\lc} \xf \dP{\lc-1} \simeq \dP{(k-1)\lc-1} \xf \ds{\lc} \oplus \dP{(k-1)\lc-1} \xf \dP{\lc-2}. 
\end{equation*}
It follows that $f$ cannot be zero, and thus that $\dl{k\lc-2} \xf \ds{l} \simeq \B{2}{(k-1)\lc-2}$.

Note that the case $\lc=2$ in the regular family cannot be obtained from this discussion, since in this case the exact sequence satisfied by $\dP{2}$ is instead
\begin{equation}\label{eq:sesP2l2}
 0 \longrightarrow \dl{2} \longrightarrow \dP{2} \longrightarrow \ds{2} \longrightarrow 0. 
\end{equation}
In this case, proposition \ref{prop:fusion_ds2vsds2} gives
$$\dl{2} \xf \ds{2k} \simeq \ds{2 k}.$$ Instead, use the exact sequence 
$$ \ds{2(k+1)} \longrightarrow \ds{2k} \longrightarrow \dl{2k} \longrightarrow 0, $$ which becomes 
\begin{equation}
\ds{2(k+1)} \longrightarrow \ds{2k} \longrightarrow \dl{2k} \xf \dl{2} \longrightarrow 0,
\end{equation}
by fusing it with $\dl{2}$. Since $\dl{2k}\xf\dl{2} \xf \dP{1} \simeq \dl{2k} \xf \dP{1} \simeq \dP{2k-1}, $ it follows that $$\dl{2k} \xf \dl{2} \simeq \dl{2k}. $$ Using this fact with the exact sequence \eqref{eq:sesP2l2} gives
\begin{equation}
\dl{2k} \longrightarrow \dP{2k} \longrightarrow \dl{2k} \xf \ds{2} \longrightarrow 0.
\end{equation}
Now, since $\dl{2k}\xf\ds{2} \xf \dP{1} \simeq \dP{2k -1 } \xf \ds{2} \simeq \dP{2(k-1) -1} \oplus \dP{2(k +1)-1} $, while $\dP{2k} \xf \dP{1} \simeq \dP{2k-1} \oplus \dP{2(k+1) -1} \oplus 2 \dP{2(k-1)-1}$, it follows that $$\dl{2k} \xf \ds{2} \simeq \B{2}{2(k-1)}.$$
\begin{Prop}\label{prop:fusionIvsSl}
For all $k>1$, and $\ell \geq 2$
\begin{equation}
\dl{k \lc -2} \xf \ds{\lc} \simeq \B{2}{(k-1)\lc -2}.
\end{equation}
\end{Prop}
To proceed and compute the fusion of the irreducibles with the other standard modules, we therefore need the fusion of $\B{2}{(k-1)\lc-2}$ with $\ds{\lc}$, which requires the fusion of $\B{2}{(k-1)\lc-2}$ with projective modules. This is our next step.
\end{subsection}
\begin{subsection}{Fusion of $\B{2i}{n,k}$ and projective modules}\label{sec:fusionBvsP}
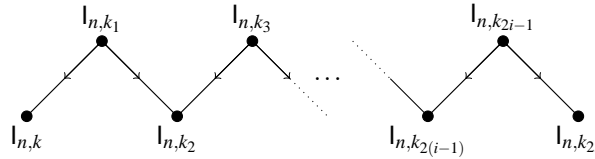
\begin{figure}[b]
\caption{The Loewy diagram of $\B{2i}{n,k}$, where $1 \leq i$, and $0\leq k<k_{2i}\leq n$.}\label{fig:Loewy.B2i}
\begin{tikzpicture}[scale=1/3]
\filldraw	
	   (0,0) circle (6 pt) node[anchor = north] {$\dl{n,k}$}
 	-- (3,3) circle (6 pt) node[anchor = south] {$\dl{n,k_{1}}$}
 	-- (6,0) circle (6 pt) node[anchor = north] {$\dl{n,k_{2}}$}
	-- (9,3) circle (6 pt) node[anchor = south] {$\dl{n,k_{3}}$};
\draw (9,3) -- (10.5,1.5);
\draw[dotted] (10.5,1.5) -- (12,0);
\node at (12,1.5) {$\hdots$};
\draw[dotted] (13,3) -- (14.5,1.5);
\draw (14.5,1.5) -- (16,0);
\filldraw
	   (16,0) circle (6 pt) node[anchor = north] {$\dl{n,k_{2(i-1)}}$}
	-- (19,3) circle (6 pt) node[anchor = south] {$\dl{n,k_{2i -1}}$}
	-- (22,0) circle (6 pt) node[anchor = north] {$\dl{n,k_{2i}}$};
\draw[->] (3,3) -- (1.5,1.5);
\draw[->] (3,3) -- (4.5,1.5);
\draw[->] (9,3) -- (7.5,1.5);
\draw[->] (9,3) -- (10.5,1.5);
\draw[->] (19,3) -- (17.5,1.5);
\draw[->] (19,3) -- (20.5,1.5);
\end{tikzpicture}
\end{figure}
The rules for the induction of these modules are \cite{BelRiSy}
\begin{multline}
	\Ind{\B{2i}{n,k}} \simeq  \left. \begin{cases}
	\B{2i}{n+1,k}, & \text{ in } \dtl{n}\\
	0, & \text{ in } \tl{n}
	\end{cases}\right\} \oplus \begin{cases}
	\overset{i}{\underset{p = 0}{\bigoplus}}\dP{n+1,k+2p\lc -1} & \text{ if } k = 0 \mod \lc \\
	 \B{2i}{n+1,k-1} & \text{otherwise}
	 \end{cases}\\
	 \oplus \begin{cases}
	 \overset{i-1}{\underset{p = 0}{\bigoplus}}\dP{n+1,k+2p\lc +1} & \text{ if } k+2 = 0 \mod \lc\\
	 \B{2i}{n+1,k+1} & \text{otherwise}
	 \end{cases}.
\end{multline}
The usual argument on the parity of the modules gives the following fusion rules.
\begin{Prop}\label{prop:fusionBvs01}
In the dilute family
\begin{equation}
\B{2i}{n,k} \xf \dP{1,0} \simeq \B{2i}{n+1,k},
\end{equation}
while in both families
\begin{multline}
\B{2i}{n,k} \xf \dP{1,1} \simeq \begin{cases}
	\overset{i}{\underset{p = 0}{\bigoplus}}\dP{n+1,k+2p\lc -1} & \text{ if } k = 0 \mod \lc \\
	 \B{2i}{n+1,k-1} & \text{otherwise}
	 \end{cases}\\
	 \oplus \begin{cases}
	 \overset{i-1}{\underset{p = 0}{\bigoplus}}\dP{n+1,k+2p\lc +1} & \text{ if } k+2 = 0 \mod \lc\\
	 \B{2i}{n+1,k+1} & \text{otherwise}
	 \end{cases}.
\end{multline}
\end{Prop}
The first formula shows that the parameter $n$ can be adjusted by simply fusing the module with $\dP{1,0}$. In the regular family,  proposition \ref{prop:magickOthersP} gives $$ \B{2i}{n,k} \xf \dP{2,0} \simeq \B{2i}{n+2,k}. $$ Like for the standard modules, we therefore omit this parameter and simply assume $n$ to be big enough for the modules to exist. 

We start by studying the fusion of $\B{2i}{23}$ in the $\lc=5$ case. The preceding proposition gives
\begin{equation}
	\B{2i}{23} \xf \dP{0} \simeq \B{2i}{23},
\end{equation}
\begin{equation}
	\B{2i}{23} \xf \dP{1} \simeq \overset{i-1}{\underset{p = 0}{\bigoplus}}\dP{(5+2p) 5 - 1} \oplus \B{2i}{22}.
\end{equation}
Fusing the last equation with $\dP{1}$ yields
\begin{equation}
	\B{2i}{23} \xf (\dP{0} \oplus \dP{2}) \simeq \overset{i-1}{\underset{p = 0}{\bigoplus}}\dP{(5+2p)5-1} \xf \dP{1} \oplus \B{2i}{21} \oplus \B{2i}{23}.
\end{equation}
Comparing this result with proposition \ref{prop:fusionBvs01}, it follows that
\begin{equation}
	\B{2i}{23} \xf \dP{2} \simeq \overset{i-1}{\underset{p = 0}{\bigoplus}}\dP{(5+2p)5-1} \xf \dP{1} \oplus \B{2i}{21}.
\end{equation}
Repeating the argument gives
\begin{equation}
	\B{2i}{23} \xf \dP{3} \simeq \overset{i-1}{\underset{p = 0}{\bigoplus}}\dP{(5+2p)5-1} \xf \dP{2} \oplus \B{2i}{20}.
\end{equation}
\begin{Prop}\label{prop:fusionBvssmallP}
For all $0 \leq j <\lc-1$, $\lc \geq 2$, $k > 1$
\begin{equation}
	\B{2i}{k\lc -2} \xf \dP{j} \simeq \overset{i-1}{\underset{p = 0}{\bigoplus}}\dP{(k+2p)\lc-1} \xf \dP{j-1} \oplus \B{2i}{k\lc-2-j},
\end{equation}
\begin{equation}
	\B{2i}{k\lc -2} \xf \dP{\lc-1} \simeq \overset{i-1}{\underset{p = 0}{\bigoplus}}\left(\ds{(k+2p)\lc} \xf \dP{\lc-1}\right) \oplus \dP{(k-1)\lc -1}.
\end{equation}
\end{Prop}
\begin{proof}
If $j=1,0$, proposition \ref{prop:fusionBvs01} already gives the conclusion. Suppose that the result stands for $j-1,j$ with $ j <\lc-2$. Then 
\begin{align}
\B{2i}{k\lc -2} \xf \dP{j} \xf \dP{1} & \simeq \B{2i}{k\lc -2} \xf \left(\dP{j-1} \oplus \dP{j+1} \right)\\
& \simeq \overset{i-1}{\underset{p = 0}{\bigoplus}}\left(\dP{(k+2p)\lc-1} \xf (\dP{j} \oplus \dP{j-2})\right) \oplus \B{2i}{k\lc-2-j-1} \oplus \B{2i}{k \lc -2 - j+1}.
\end{align}
Comparing the two lines and using the induction hypothesis yields the conclusion for $j+1$. In particular, this yields
\begin{equation}
	\B{2i}{k\lc -2} \xf \dP{\lc-2} \simeq \overset{i-1}{\underset{p = 0}{\bigoplus}}\dP{(k+2p)\lc-1} \xf \dP{\lc-3} \oplus \B{2i}{(k-1)\lc}.
\end{equation}
Fusing this result with $\dP{1}$ gives
\begin{align}
\B{2i}{k\lc -2} \xf \left(\dP{\lc-1}\oplus \dP{\lc-3}\right) \simeq & \overset{i-1}{\underset{p = 0}{\bigoplus}}\left(\dP{(k+2p)\lc-1} \xf \left(\dP{\lc-4}\oplus \dP{\lc-2}\right)\right) \oplus \B{2i}{(k-1)\lc+1}\\ 
& \oplus \overset{i}{\underset{p = 0}{\bigoplus}}\dP{(k-1+2p)\lc-1}.
\end{align}
Comparing the two sides, using the result of the first part and rearranging the terms gives
\begin{align}
\B{2i}{k\lc -2} \xf  \dP{\lc-1}  & \simeq \overset{i-1}{\underset{p = 0}{\bigoplus}}\left(\dP{(k+2p)\lc-1} \xf \dP{\lc-2} \oplus \dP{(k+2p +1)\lc-1}\right) \oplus \dP{(k-1)\lc-1}\\
&\simeq \overset{i-1}{\underset{p = 0}{\bigoplus}}\left(\ds{(k+2p)\lc} \xf \dP{\lc-1}\right) \oplus \dP{(k-1)\lc-1}
\end{align}
where the second line follows from proposition \ref{prop_fusion_sklvspi}.
\end{proof}
The last formula can be used to quickly obtain the fusion rules with the bigger projectives. Thus,
\begin{equation*}
	\B{2i}{k\lc -2} \xf \dP{\lc-1}\xf \dP{1} \simeq \B{2i}{k\lc-2}\xf \dP{\lc} \simeq \overset{i-1}{\underset{p = 0}{\bigoplus}}\left(\ds{(k+2p)\lc} \xf \dP{\lc}\right) \oplus \dP{(k-1)\lc},
\end{equation*}
\begin{equation*}
	\B{2i}{kl -2} \xf \dP{\lc+1} \simeq \overset{i-1}{\underset{p = 0}{\bigoplus}}\left(\ds{(k+2p)\lc} \xf \dP{\lc+1}\right) \oplus \dP{(k-1)\lc+1},
\end{equation*}
\begin{equation*}
	\B{2i}{k\lc -2} \xf \dP{\lc+2} \simeq \overset{i-1}{\underset{p = 0}{\bigoplus}}\left(\ds{(k+2p)\lc} \xf \dP{\lc+2}\right) \oplus \dP{(k-1)\lc+2}.
\end{equation*}
Continuing in this manner eventually gives
\begin{equation*}
	\B{2i}{k\lc -2} \xf \dP{2\lc-1} \simeq \overset{i-1}{\underset{p = 0}{\bigoplus}}\left(\ds{(k+2p)\lc} \xf \dP{2\lc-1}\right) \oplus \dP{(k-2)\lc-1} \oplus \dP{k \lc -1}.
\end{equation*}
Note that if $k=2$, $\dP{(k-2)\lc-1}\simeq 0$. Repeating these arguments, the proof for the general formula is straightforward.
\begin{Prop}\label{prop:fusionBvsP}
For all $k>1,r>0$, $i>0$, $0\leq j<\lc$,
\begin{equation}
	\B{2i}{k\lc-2} \xf \dP{r\lc -1 +j} \simeq \overset{i-1}{\underset{p = 0}{\bigoplus}}\left(\ds{(k+2p)\lc} \xf \dP{r\lc-1+j}\right) \oplus \overset{k+r-2}{\underset{\underset{step=2}{p = \max\left( k-r,r-k+2\right)}}{\bigoplus}} \dP{p\lc-1+j}.
\end{equation}
\end{Prop}
The same method can be used to obtain the formulas for the fusion of $\B{2i}{k\lc-t}$. Fusing the formula in the preceding proposition with $\dP{1}$ yields
\begin{align}
\B{2i}{k\lc-2} \xf \dP{r\lc -1 +j} \xf \dP{1}  & \simeq \overset{i-1}{\underset{p = 0}{\bigoplus}}\left(\dP{(k+2p)\lc-1} \xf \dP{r \lc-1 +j}\right) \oplus \B{2i}{k\lc-3} \xf \dP{r \lc -1 -j} \\
& \simeq \overset{i-1}{\underset{p = 0}{\bigoplus}}\left(\left(\ds{(k+2p)\lc+1} \oplus \dP{(k+2p)\lc-1}\right) \xf \dP{r\lc-1+j}\right)\\
& \qquad \oplus \overset{k+r-2}{\underset{\underset{step=2}{p = \max\left( k-r,r-k+2\right)}}{\bigoplus}} \dP{p\lc-1+j}\xf \dP{1}.
\end{align}
Comparing the two lines yields
\begin{equation}
	\B{2i}{k\lc-3} \xf \dP{r\lc -1 +j} \simeq \overset{i-1}{\underset{p = 0}{\bigoplus}}\left(\ds{(k+2p)\lc+1} \xf \dP{r\lc-1+j}\right) \oplus \overset{k+r-2}{\underset{\underset{step=2}{p = \max\left( k-r,r-k+2\right)}}{\bigoplus}} \left(\dP{p\lc-1+j} \xf \dP{1}\right).
\end{equation}
Once again, this operation can be repeated and gives the following general formula.
\begin{Prop}
For all $k>1,r,i>0$, $0<t<\lc$, $0\leq j <\lc$,
\begin{equation}
	\B{2i}{k\lc-1-t} \xf \dP{r\lc -1 +j} \simeq \overset{i-1}{\underset{p = 0}{\bigoplus}}\left(\ds{(k+2p)\lc-1+t} \xf \dP{r\lc-1+j}\right) \oplus \overset{k+r-2}{\underset{\underset{step=2}{p = \max\left( k-r,r-k+2\right)}}{\bigoplus}}\left( \dP{p\lc-1+j}\xf \dP{t-1}\right).
\end{equation} 
\end{Prop}

Note that, in this section, the case $\B{2i}{\lc -2}$ has been avoided. In this case, there is a short-exact sequence
\begin{equation*}
	0 \longrightarrow \dl{\lc-2} \longrightarrow \B{2i}{\lc-2} \longrightarrow \T{2i-1}{\lc} \longrightarrow 0,
\end{equation*}
and since $\dl{\lc -2} \xf \dP{j} \simeq 0 $, for all $j\geq \lc-1 $, $$\B{2i}{\lc-2} \xf \dP{j} \simeq \T{2i-1}{\lc} \xf \dP{j}.$$ This case will be treated in section \ref{sec:fusionTvsP}
\end{subsection}
\begin{subsection}{Fusion of $\B{2i}{k}$ and standard modules}\label{sec:fusionBvsS}

We now want to compute the fusion of $\B{2i}{k}$ with the standard modules $\ds{q}$ that are not projective. The first step is to find a formula for $\B{2i}{k\lc -2}\xf \ds{\lc}$.

Using the projective cover of $\ds{\lc}$ (see section \ref{sec:rappel.indec}) and the right-exactness of fusion, one can obtain the exact sequence
\begin{equation}\label{eq:fusionBvsS.1}
	\B{2i}{k\lc -2} \xf \dP{\lc-2} \overset{f}{\longrightarrow} \B{2i}{k \lc -2} \xf \dP{\lc} \longrightarrow \B{2i}{k \lc -2} \xf \ds{\lc} \longrightarrow 0.
\end{equation}
Using propositions \ref{prop:fusionBvssmallP}, \ref{prop:fusionBvsP}, and \ref{prop_fusion_sqlvspkl} gives
\begin{equation*}
	\B{2i}{k\lc -2} \xf \dP{\lc-2} \simeq \overset{i-1}{\underset{p = 0}{\bigoplus}}\dP{(k+2p)\lc-1} \xf \dP{\lc-3} \oplus \B{2i}{(k-1)\lc},
\end{equation*}
and
\begin{equation*}
	\B{2i}{k\lc -2} \xf \dP{\lc} \simeq \overset{i-1}{\underset{p = 0}{\bigoplus}}\left(\dP{(k+2p)\lc-1} \xf \left(\dP{\lc-3} \oplus \dP{\lc-1}\right)\right) \oplus \overset{i}{\underset{p = 0}{\bigoplus}}\dP{(k-1+2p)\lc}.
\end{equation*}
Therefore
\begin{equation*}
 \left(\B{2i}{k\lc -2} \xf \dP{\lc}\right)/ \left( \B{2i}{k\lc -2} \xf \dP{\lc-2} \right) \simeq \overset{i-1}{\underset{p = 0}{\bigoplus}}\dP{(k+2p)\lc-1} \xf \dP{\lc-1} \oplus \B{2(i+1)}{(k-1)\lc-2}
\end{equation*}
where we used proposition \ref{prop:indec.prop}. If it can be proved that $f$ is injective, this will give a formula for $\B{2i}{k \lc -2} \xf \ds{\lc}$. To do this, we will prove that the dimension of $\B{2i}{k \lc -2} \xf \ds{\lc}$ is that of $\left(\B{2i}{k\lc -2} \xf \dP{\lc}\right)/ \left( \B{2i}{k\lc -2} \xf \dP{\lc-2} \right)$, and this will be done by induction on $i$.

Note first that by proposition \ref{prop:fusionIvsSl}
\begin{equation}
\B{2 \times 0}{k\lc -2} \xf \ds{\lc} \simeq \dl{k\lc-2} \xf \ds{\lc} \simeq \B{2}{(k-1)\lc-2} \simeq \left(\B{0}{k\lc -2} \xf \dP{\lc}\right)/ \left( \B{0}{k\lc -2} \xf \dP{\lc-2} \right).
\end{equation}
This gives the case $i=0$ for all $k>1$. Assume now that 
\begin{equation*}
 	\B{2i}{k \lc -2} \xf \ds{\lc} \simeq \overset{i-1}{\underset{p = 0}{\bigoplus}}\dP{(k+2p)\lc-1} \xf \dP{\lc-1} \oplus \B{2(i+1)}{(k-1)\lc-2}
\end{equation*}
for a certain  $i$ and all $k>1$. To proceed with the induction, we will use the short exact sequence
\begin{equation}
	0 \longrightarrow \B{2i}{k \lc -2} \longrightarrow \B{2(i+1)}{k\lc -2} \longrightarrow \ds{(k+2i)\lc}\longrightarrow 0.
\end{equation}
It can be seen by inspecting the Loewy diagram of the $B$ modules and proved using techniques developed in \cite{BelRiSy}.
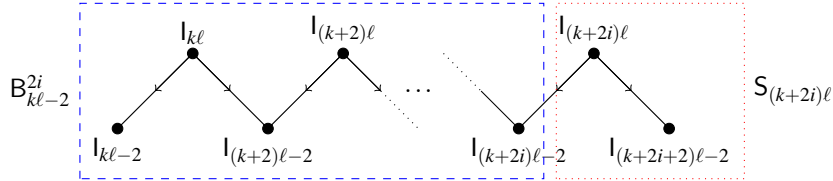
\begin{equation*}
\begin{tikzpicture}[scale=1/3]
\filldraw	
	   (0,0) circle (6 pt) node[anchor = north] {$\dl{k \lc -2}$}
 	-- (3,3) circle (6 pt) node[anchor = south] {$\dl{k \lc }$}
 	-- (6,0) circle (6 pt) node[anchor = north] {$\dl{(k+2)\lc-2}$}
	-- (9,3) circle (6 pt) node[anchor = south] {$\dl{(k+2)\lc}$};
\draw (9,3) -- (10.5,1.5);
\draw[dotted] (10.5,1.5) -- (12,0);
\node at (12,1.5) {$\hdots$};
\draw[dotted] (13,3) -- (14.5,1.5);
\draw (14.5,1.5) -- (16,0);
\filldraw
	   (16,0) circle (6 pt) node[anchor = north] {$\dl{(k+2i)\lc -2}$}
	-- (19,3) circle (6 pt) node[anchor = south] {$\dl{(k+2i)\lc}$}
	-- (22,0) circle (6 pt) node[anchor = north] {$\dl{ (k+2i+2)\lc-2}$};
\draw[->] (3,3) -- (1.5,1.5);
\draw[->] (3,3) -- (4.5,1.5);
\draw[->] (9,3) -- (7.5,1.5);
\draw[->] (9,3) -- (10.5,1.5);
\draw[->] (19,3) -- (17.5,1.5);
\draw[->] (19,3) -- (20.5,1.5);
\draw[dashed, blue] (-1.5,-2) -- (-1.5,5) -- (17,5) -- (17,-2) -- (-1.5,-2);
\node[anchor=east] at (-1.5,1.5) {$\B{2i}{k\lc-2}$};
\draw[dotted, red] (17.5,-2) -- (17.5,5) -- (25,5) -- (25,-2) -- (17.5,-2);
\node[anchor =west] at (25,1.5) {$\ds{(k+2i)\lc}$};
\end{tikzpicture}
\end{equation*}
The right-exactness of fusion yields the exact sequence
\begin{equation}
	\B{2i}{k \lc -2} \xf \ds{\lc}\overset{g}{\longrightarrow} \B{2(i+1)}{k \lc -2}\xf \ds{\lc} \longrightarrow \ds{(k+2i)\lc}\xf \ds{\lc}\longrightarrow 0.
\end{equation}
It follows that 
\begin{align*}
	\dim \left(  \B{2(i+1)}{k \lc -2}\xf \ds{\lc} \right) & \leq \dim\left(\B{2i}{k \lc -2} \xf \ds{\lc}\right) + \dim\left( \ds{(k+2i)\lc}\xf \ds{\lc} \right) \\
	& = \dim \left( \overset{i-1}{\underset{p = 0}{\bigoplus}}\dP{(k+2p)\lc-1} \xf \dP{\lc-1} \oplus \B{2(i+1)}{(k-1)\lc-2}\right) \\
	& \qquad + \dim \left( \dP{(k+2i)\lc-1}\xf \dP{\lc-1} \oplus \ds{(k+2i +1)\lc}\right) \\
	& = \dim \left(\overset{i+1-1}{\underset{p = 0}{\bigoplus}}\dP{(k+2p)\lc-1} \xf \dP{\lc-1} \oplus \B{2(i+2)}{(k-1)\lc-2}  \right)
\end{align*}

where the equality occurs if and only if $g$ is injective. The exact sequence \eqref{eq:fusionBvsS.1} gives
\begin{align*}
	\dim \left(\B{2(i+1)}{k \lc -2}\xf \ds{\lc} \right)& = \dim \left( \B{2(i+1)}{k \lc -2} \xf \dP{\lc} \right) - \dim \im f \\
	& \geq \dim \left( \B{2(i+1)}{k \lc -2} \xf \dP{\lc} \right) - \dim \left( \B{2i}{k\lc -2} \xf \dP{\lc-2} \right) \\
	& =  \dim \left( \overset{i+1-1}{\underset{p = 0}{\bigoplus}}\dP{(k+2p)\lc-1} \xf \dP{\lc-1} \oplus \B{2(i+2)}{(k-1)\lc-2}\right).\\
\end{align*}
It follows that $\ker f \simeq 0$, and the following result is thus proved.
\begin{Prop}\label{prop:fusionBvsSl}
For all $i\geq 0$, $k>1$,
\begin{equation}
	\B{2i}{k \lc -2} \xf \ds{\lc} \simeq \overset{i-1}{\underset{p = 0}{\bigoplus}}\dP{(k+2p)\lc-1} \xf \dP{\lc-1} \oplus \B{2(i+1)}{(k-1)\lc-2}.
\end{equation}
\end{Prop}
Fusion rules for bigger standard modules will not be needed to compute the fusion of irreducible modules but we include them for the sake of completeness. 
\begin{Prop}
For all $0<r <k$,
\begin{equation}\label{eq:BvsSql}
\B{2i}{k \lc -2} \xf \ds{r \lc} \simeq \overset{i-1}{\underset{p = 0}{\bigoplus}}\dP{(k+2p)\lc-1} \xf \dP{r\lc-1} \oplus \B{2(i+r)}{(k-r)\lc-2}.
\end{equation}
\end{Prop}
\begin{proof}
We proceed by induction on $r$, the case $r=1$ being given by the previous proposition. Assume the result for some $r<k-1$. Using propositions \ref{prop:fusionBvsP}, \ref{prop_fusion_sqlvspkl}, and \ref{prop:fusionproj}, we start by noting that
\begin{align}
	\B{2i}{k \lc -2} \xf \dP{r\lc -1} \xf \dP{\lc-1} & \simeq \overset{i-1}{\underset{p = 0}{\bigoplus}}\left(\ds{(k+2p)\lc} \xf \dP{r \lc-1} \xf \dP{\lc-1} \right)\oplus \overset{k+r-2}{\underset{\underset{step=2}{p = k-r}}{\bigoplus}}\left(\dP{p \lc -1} \xf \dP{\lc-1} \right) \notag\\
	& \simeq \overset{i-1}{\underset{p = 0}{\bigoplus}}\left(\dP{(k+2p)\lc -1} \xf \dP{r\lc -1} \xf \dP{\lc-2} \oplus \dP{(k+2p+1)\lc -1} \xf \dP{r \lc -1} \right) \notag\\
	& \qquad \oplus \overset{r-1}{\underset{p = 0}{\bigoplus}}\left(\dP{(k-r+2p)\lc -1} \xf \dP{\lc-1} \right),
\end{align}
and
\begin{equation*}
\dP{(k+2p+1)\lc-1} \xf \dP{r \lc -1} \simeq \dP{(k+2p)\lc-1} \xf \dP{(r-1)\lc-1} \oplus \dP{(k+r+2p)\lc -1}\xf \dP{\lc-1}.
\end{equation*}
Next, we fuse the left side of equation \eqref{eq:BvsSql} with $\ds{\lc}$ and use propositions \ref{prop:fusionBvsSl}, \ref{prop_fusion_sqlvspkl}, and \ref{prop:fusionstandvssl} to obtain
\begin{align}
 \B{2i}{k \lc -2} \xf \left(\ds{r \lc} \xf \ds{\lc}\right)& \simeq \B{2i}{k \lc -2} \xf \left( \dP{r\lc -1} \xf \dP{\lc-1} \oplus \ds{(r+1)\lc}\right), 
\end{align}
while fusing the right side of equation \eqref{eq:BvsSql} with $\ds{\lc}$ gives
\begin{align}
\overset{i-1}{\underset{p = 0}{\bigoplus}}\dP{(k+2p)\lc-1} & \xf \dP{r\lc-1} \xf \ds{\lc} \oplus \B{2(i+r)}{(k-r)\lc-2} \xf \ds{\lc} \notag\\
& \simeq\overset{i-1}{\underset{p = 0}{\bigoplus}}\left(\dP{(k+2p)\lc-1} \xf \left(\dP{r \lc -1} \xf \dP{\lc-2} \oplus \dP{(r-1)\lc-1} \oplus \dP{(r+1)\lc -1}\right)\right) \notag\\
& \oplus  \overset{i+r-1}{\underset{p = 0}{\bigoplus}}\left(\dP{(k-r+2p)\lc -1}\xf \dP{\lc-1}\right) \oplus \B{2(i+1+r)}{(k-r-1)\lc-2}.
\end{align}
Comparing the two results and using the previous observations gives the conclusion for $r+1$.
\end{proof}
Note that in all of these calculations, we carefully avoided the case $k=1$ (and $r=k$). There is a short exact sequence
\begin{equation}
0 \longrightarrow \dl{\lc-2} \longrightarrow \B{2i}{\lc-2} \longrightarrow \T{2i -1}{\lc} \longrightarrow 0,
\end{equation}
which can be seen by inspecting the Loewy diagram of $\B{2i}{\lc-2}$ and proved using techniques developed in \cite{BelRiSy}.
\begin{equation}
\begin{tikzpicture}[scale=1/3]
\filldraw	
	   (0,0) circle (6 pt) node[anchor = north] {$\dl{\lc-2}$}
 	-- (3,3) circle (6 pt) node[anchor = south] {$\dl{\lc}$}
 	-- (6,0) circle (6 pt) node[anchor = north] {$\dl{2\lc-2}$}
	-- (9,3) circle (6 pt) node[anchor = south] {$\dl{2\lc}$};
\draw (9,3) -- (10.5,1.5);
\draw[dotted] (10.5,1.5) -- (12,0);
\node at (12,1.5) {$\hdots$};
\draw[dotted] (13,3) -- (14.5,1.5);
\draw (14.5,1.5) -- (16,0);
\filldraw
	   (16,0) circle (6 pt) node[anchor = north] {$\dl{(2i-1)\lc-2}$}
	-- (19,3) circle (6 pt) node[anchor = south] {$\dl{(2i-1)\lc}$}
	-- (22,0) circle (6 pt) node[anchor = north] {$\dl{2i\lc -2}$};
\draw[->] (3,3) -- (1.5,1.5);
\draw[->] (3,3) -- (4.5,1.5);
\draw[->] (9,3) -- (7.5,1.5);
\draw[->] (9,3) -- (10.5,1.5);
\draw[->] (19,3) -- (17.5,1.5);
\draw[->] (19,3) -- (20.5,1.5);
\draw[dashed,blue]
		(2,6) -- (24,6)
	--  (24,-3) -- (2,-3)
	--	(1.5,6);
\node[anchor = west]  at (24,1.5) {$\T{2i-1}{\lc}$};
\end{tikzpicture}
\end{equation}
Since it was already noted in proposition \ref{prop:fusionIWeird} that $\dl{\lc-2} \xf \ds{q\lc} \simeq 0$ for all $q\geq 1$, 
\begin{equation}\label{eq:fusionBSWeird}
\B{2i}{\lc-2}\xf \ds{q\lc} \simeq \T{2i -1}{\lc} \xf \ds{q\lc}.
\end{equation}
Therefore, to compute $\dl{2\lc-2} \xf \ds{2\lc}$, we will have to compute
\begin{equation}
\left(\dl{2\lc-2} \xf \ds{\lc}\right) \xf \ds{\lc} \simeq \B{2}{\lc-2} \xf \ds{\lc} \simeq \T{1}{\lc} \xf \ds{\lc}.
\end{equation}
The fusion rules for $\T{2i-1}{\lc}$ will thus be needed to compute the fusion rules of the irreducible modules.
\end{subsection}
\begin{subsection}{Fusion of $\T{2i+1}{k}$ and projective modules}\label{sec:fusionTvsP}
The formulas for the induction of $\T{2i+1}{k}$ are \cite{BelRiSy}:
\begin{equation}
	\Ind{\T{2i+1}{n,k}} \simeq \T{2i+1}{n+1,k-1}  \oplus \T{2i+1}{n+1,k+1}\oplus \left. \begin{cases}
	\T{2i+1}{n+1,k}, & \text{ in } \dtl{n}\\
	0, & \text{ in } \tl{n}
	\end{cases}\right\},
\end{equation}
where 
\begin{equation}
	\T{2i+1}{n,k} \simeq \overset{i}{\underset{p=0}{\bigoplus}} \dP{n,k+2p\lc}
\end{equation}
if $k$ is critical and
\begin{equation}
	\T{2i+1}{n,-1} \simeq \overset{i}{\underset{p=1}{\bigoplus}} \dP{n,2p\lc-1}.
\end{equation}
Using the parity of the relevant modules gives the following fusion rules.
\begin{Prop}\label{prop:fusionTvs01}
For all $k$,$i$, in the dilute family
\begin{equation}
	\T{2i+1}{n,k}\xf\dP{1,0} \simeq \T{2i+1}{n+1,k},
\end{equation}
while in both families
\begin{equation}
	\T{2i+1}{n,k} \xf \dP{1,1} \simeq \T{2i+1}{n+1,k-1} \oplus \T{2i+1}{n+1,k+1}.
\end{equation} 
\end{Prop}
Once again, fusing these modules with $\dP{1,0}$ simply increases the parameter $n$. In the regular case, proposition \ref{prop:magickOthersP} gives $$\T{2i+1}{n,k} \xf \dP{2,0} \simeq \T{2i+1}{n+2,k}, $$ as long as $\lc \neq 2$. As before, the proofs will be independent of $n$ so we simply omit this parameter and assume $n$ to be big enough for the modules to exist. 

We start by studying the modules $\T{2i+1}{k\lc}$. Note that
\begin{equation*}
	\T{2i+1}{k\lc} \xf \dP{1} \simeq \overset{i}{\underset{p=0}{\bigoplus}} \dP{(k+2p)\lc-1} \oplus \T{2i+1}{k\lc +1}.
\end{equation*}
Fusing this expression with $\dP{1}$ yields
\begin{align*}
\T{2i+1}{k\lc} \xf \dP{1}\xf \dP{1} &\simeq \T{2i+1}{k\lc} \xf \left(\dP{0} \oplus \dP{2} \right)\\
	& \simeq \overset{i}{\underset{p=0}{\bigoplus}} \dP{(k+2p)\lc-1} \xf \dP{1} \oplus \T{2i+1}{k\lc} \oplus \T{2i+1}{k\lc+2}
\end{align*}
Comparing the first and second lines and using proposition \ref{prop:fusionTvs01} give the fusion rule
\begin{equation*}
	\T{2i+1}{k\lc}\xf\dP{2} \simeq \overset{i}{\underset{p=0}{\bigoplus}}\left( \dP{(k+2p)\lc-1} \xf \dP{1}\right) \oplus \T{2i+1}{k\lc+2}.
\end{equation*}
It is a simple exercise to repeat this argument and obtain the fusion rules for the other small projectives.
\begin{Prop}\label{prop:fusionTvssmallP}
For all $i,k \geq 0$, $0\leq j<\lc-1$,
\begin{equation}
\T{2i+1}{k\lc}\xf\dP{j} \simeq \overset{i}{\underset{p=0}{\bigoplus}}\left( \dP{(k+2p)\lc-1} \xf \dP{j-1} \right)\oplus \T{2i+1}{k\lc+j}.
\end{equation}
\end{Prop}
In particular,
\begin{equation*}
\T{2i+1}{k\lc} \xf \dP{\lc-2} \simeq \overset{i}{\underset{p=0}{\bigoplus}}\left( \dP{(k+2p)\lc-1} \xf \dP{\lc-3}\right) \oplus \T{2i+1}{(k+1)\lc-2}.
\end{equation*}
Fusing this expression with $\dP{1}$ gives
\begin{align*}
\T{2i+1}{k\lc} \xf \dP{\lc-2} \xf \dP{1} & \simeq \T{2i+1}{k\lc}\xf \left(\dP{\lc-3} \oplus \dP{\lc-1}\right) \\
	& \simeq \overset{i}{\underset{p=0}{\bigoplus}}\left( \dP{(k+2p)\lc-1} \xf \left(\dP{\lc-4} \oplus \dP{\lc-2} \right)\right) \oplus \T{2i+1}{(k+1)\lc-3} \\
	& \qquad \oplus \overset{i}{\underset{p=0}{\bigoplus}}\left( \dP{(k+2p+1)\lc-1}\right).
\end{align*}
Comparing the first and second lines gives
\begin{align}\label{eq:fusionTvsP.1}
	\T{2i+1}{k\lc} \xf \dP{\lc-1} & \simeq \overset{i}{\underset{p=0}{\bigoplus}}\left( \dP{(k+2p)\lc-1}\xf \dP{\lc-2} \oplus \dP{(k+1+2p)\lc-1}\right) \notag \\
	& \simeq \overset{i}{\underset{p=0}{\bigoplus}}\left( \ds{(k+2p)\lc} \xf \dP{\lc-1}\right) ,
\end{align}
where the known fusion rules for standard modules (proposition \ref{prop_fusion_sklvspi}) were used in the second line. Fusing this expression with $\dP{1}$ gives
\begin{equation*}
\T{2i+1}{k\lc} \xf \dP{\lc-1} \xf \dP{1} \simeq \T{2i+1}{k\lc} \xf \dP{\lc} \simeq \overset{i}{\underset{p=0}{\bigoplus}}\left( \ds{(k+2p)\lc} \xf \dP{\lc}\right) .
\end{equation*}
Fusing the latter expression again with $\dP{1}$ gives
\begin{align*}
	\T{2i+1}{k\lc} \xf \dP{\lc} \xf \dP{1} & \simeq \T{2i+1}{k\lc} \xf \left(2\dP{\lc-1} \oplus \dP{\lc+1}\right)\\
		& \simeq \overset{i}{\underset{p=0}{\bigoplus}}\left( \ds{(k+2p)\lc} \xf\left( 2\dP{\lc-1} \oplus \dP{\lc+1}\right)\right).
\end{align*}
Comparing the two lines yields the fusion rule
\begin{equation*}
	\T{2i+1}{k\lc} \xf \dP{\lc+1} \simeq \overset{i}{\underset{p=0}{\bigoplus}}\left( \ds{(k+2p)\lc} \xf \dP{\lc+1}\right). 
\end{equation*}
The same arguments prove the following proposition.
\begin{Prop}\label{prop:fusionTvsP}
For all $i, k  \geq 0$, $r\geq \lc-1$,
\begin{equation}\label{eq:fusionTklvsP}
	\T{2i+1}{k\lc} \xf \dP{r} \simeq \overset{i}{\underset{p=0}{\bigoplus}}\left( \ds{(k+2p)\lc} \xf \dP{r}\right),
\end{equation}
\end{Prop}
The fusion rules for $\T{2i+1}{k\lc +i}$ can be obtained from these formulas. We start by fusing \eqref{eq:fusionTklvsP} with $\dP{1}$.
\begin{align*}
	\T{2i+1}{k\lc} \xf \dP{r} \xf \dP{1} & \simeq \left(\overset{i}{\underset{p=0}{\bigoplus}} \dP{(k+2p)\lc-1} \oplus \T{2i+1}{k\lc +1} \right)\xf \dP{r} \\
	& \simeq \overset{i}{\underset{p=0}{\bigoplus}}\left(\left( \ds{(k+2p)\lc+1} \oplus \dP{(k+2p)\lc-1}\right) \xf \dP{r}\right).
\end{align*}
Comparing the two lines yields
\begin{equation}
\T{2i+1}{k\lc+1} \xf \dP{r} \simeq \overset{i}{\underset{p=0}{\bigoplus}}\left( \ds{(k+2p)\lc+1} \xf \dP{r}\right).
\end{equation}
This argument can be repeated to obtain the following proposition.
\begin{Prop}
For all $i,k \geq 0 $, $0<j<\lc$, $r\geq \lc-1$,
\begin{equation}
\T{2i+1}{k\lc-1+j} \xf \dP{r} \simeq \overset{i}{\underset{p=0}{\bigoplus}}\left( \ds{(k+2p)\lc-1+j} \xf \dP{r}\right).
\end{equation}
\end{Prop}
\end{subsection}
\begin{subsection}{Fusion of $\T{2i+1}{k}$ and standard modules}\label{sec:fusionTvsS}

We want to compute fusions of $\T{2i+1}{k}$ with non-projective standard modules. Proceeding as in the previous sections,  we start by computing $\T{2i+1}{k\lc}\xf \ds{\lc}$, where $k\neq 0$. 

There is a short-exact sequence
\begin{equation*}
0 \longrightarrow \dP{\lc-2} \longrightarrow \dP{\lc} \longrightarrow \ds{\lc} \longrightarrow 0,
\end{equation*}
which gives the exact sequence
\begin{equation}\label{eq:sesTvsS.1}
\T{2i+1}{k\lc}\xf\dP{\lc-2} \overset{f}{\longrightarrow} \T{2i+1}{k\lc}\xf\dP{\lc} \longrightarrow \T{2i+1}{kl}\xf\ds{\lc} \longrightarrow 0
\end{equation}
by using the right-exactness of fusion. Propositions \ref{prop:fusionTvssmallP} and equation \eqref{eq:fusionTvsP.1} give
\begin{equation*}
\T{2i+1}{k\lc} \xf \dP{\lc-2} \simeq \overset{i}{\underset{p=0}{\bigoplus}}\left( \dP{(k+2p)\lc-1} \xf \dP{\lc-3}\right) \oplus \T{2i+1}{(k+1)\lc-2},
\end{equation*}
\begin{equation*}
\T{2i+1}{k\lc}\xf\dP{\lc} \simeq \overset{i}{\underset{p=0}{\bigoplus}}\left( \dP{(k+2p)\lc-1}\xf \left(\dP{\lc-3}\oplus \dP{\lc-1}\right) \oplus \dP{(k+1+2p)\lc}\right).
\end{equation*}
Therefore
\begin{equation}
\left(\T{2i+1}{k\lc}\xf\dP{\lc}\right) / \left( \T{2i+1}{k\lc}\xf\dP{\lc-2}\right) \simeq \overset{i}{\underset{p=0}{\bigoplus}}\left( \dP{(k+2p)\lc-1} \xf \dP{\lc-1}\right) \oplus \T{2i+1}{(k+1)\lc},
\end{equation}
where proposition \ref{prop:indec.prop}, which gives $ \Big(\overset{i}{\underset{p=0}{\bigoplus}}  \dP{(k+1+2p)\lc} \Big)/ \T{2i+1}{(k+1)\lc-2} \simeq \T{2i+1}{(k+1)\lc}$, was used. The goal is now to prove that
\begin{equation}\label{eq:fusionTvsSl}
\T{2i+1}{k\lc}\xf\ds{l} \simeq \overset{i}{\underset{p=0}{\bigoplus}}\left( \dP{(k+2p)\lc-1} \xf \dP{\lc-1}\right) \oplus \T{2i+1}{(k+1)\lc},
\end{equation}
which is equivalent to $f$ being injective. Note that for $i=0$, this is just the fusion of two standard modules, and proposition \ref{prop:fusionstandvssl}, or \ref{prop:fusion_ds2vsds2} if $k=1$, $\lc =2$ in the regular family, agrees with \eqref{eq:fusionTvsSl}. We thus proceed by induction on $i$. Assume that \eqref{eq:fusionTvsSl} stands for $i$ and use the short exact sequence
\begin{equation*}
0 \longrightarrow \T{2i+1}{k\lc} \longrightarrow \T{2i+3}{k\lc} \longrightarrow \ds{(k+2i+2)\lc} \longrightarrow 0,
\end{equation*}
to obtain the exact sequence
\begin{equation}
\T{2i+1}{k\lc}\xf \ds{\lc} \longrightarrow \T{2i+3}{k\lc}\xf \ds{\lc} \longrightarrow \ds{(k+2i+2)\lc}\xf \ds{\lc} \longrightarrow 0.
\end{equation}
It gives the inequality
\begin{align*}
	\dim \left(\T{2i+3}{k\lc}\xf \ds{\lc} \right) & \leq \dim \left(\T{2i+1}{k\lc}\xf \ds{\lc} \right) + \dim \left( \ds{(k+2i+2)\lc}\xf \ds{\lc} \right)\\
	&= \dim \left( \overset{i}{\underset{p=0}{\bigoplus}}\left( \dP{(k+2p)\lc-1} \xf \dP{\lc-1}\right) \oplus \T{2i+1}{(k+1)\lc}\right)\\
	& \qquad + \dim \left(\left( \dP{(k+2i+2)\lc-1} \xf \dP{\lc-1}\right) \oplus \ds{(k+2i + 3)\lc} \right)\\
	& = \dim \left(\overset{i+1}{\underset{p=0}{\bigoplus}}\left( \dP{(k+2p)\lc-1} \xf \dP{\lc-1}\right) \oplus \T{2i+3}{(k+1)\lc} \right)
\end{align*}
 
However, the exact sequence \eqref{eq:sesTvsS.1} also give the inequality
\begin{align*}
\dim \left( \T{2i+3}{k\lc}\xf \ds{\lc}\right) & = \dim \left(\T{2i+3}{k\lc}\xf\dP{\lc} \right) - \dim \im f \\
& \geq \dim \left( \overset{i+1}{\underset{p=0}{\bigoplus}}\left( \dP{(k+2p)\lc-1} \xf \dP{\lc-1}\right) \oplus \T{2i+3}{(k+1)\lc} \right)
\end{align*}

Comparing the two bounds shows that $\dim\left(\text{\rm im}f\right) = \dim\left(\T{2i+3}{k\lc}\xf\dP{\lc-2}\right)$ and thus that $f$ is injective. Formula \eqref{eq:fusionTvsSl} must therefore stand for $i+1$, proving the following proposition.
\begin{Prop}\label{prop:fusionTvsSl}
For $i\geq 0$, $k>0$,
\begin{equation}
\T{2i+1}{k\lc}\xf\ds{\lc} \simeq \overset{i}{\underset{p=0}{\bigoplus}}\left( \dP{(k+2p)\lc-1} \xf \dP{\lc-1}\right) \oplus \T{2i+1}{(k+1)\lc}.
\end{equation}
\end{Prop}
Fusions with the bigger standard modules and the other $\T{2i+1}{k\lc +i}$ will not be needed but are presented for the sake of completeness.
\begin{Prop}
For all $i\geq0$, $k,r>0$,
\begin{equation}\label{eq:fusionTvsSql}
\T{2i+1}{k \lc} \xf \ds{r \lc} \simeq \overset{i}{\underset{p=0}{\bigoplus}}\left( \dP{(k+2p)\lc-1} \xf \dP{r\lc-1}\right) \oplus \T{2i+1}{(k+r)\lc}.
\end{equation}
\end{Prop}
\begin{proof}
We proceed by induction on $r$. The case $r=1$ being contained in proposition \ref{prop:fusionTvsSl} , suppose that the result holds for a certain $r>1$. Then, we start by noticing that by propositions \ref{prop:fusionTvsP}, and \ref{prop_fusion_sklvspi}, 
\begin{align*}
	\T{2i+1}{k \lc} \xf \dP{r \lc -1} \xf \dP{\lc -1} &\simeq \overset{i}{\underset{p=0}{\bigoplus}}\left( \ds{(k+2p)\lc} \xf \dP{r\lc-1} \xf \dP{\lc -1}\right)\\
	& \simeq \overset{i}{\underset{p=0}{\bigoplus}}\left(\dP{(k+2p)\lc-1}\xf \dP{r\lc -1}\xf \dP{l-2} \oplus \dP{(k+2p+1)\lc -1} \xf \dP{r\lc -1} \right),
\end{align*}
and by proposition \ref{prop:fusionproj},
\begin{equation}
\dP{(k+2p+1)\lc -1} \xf \dP{r\lc -1} \simeq \dP{(k+2p)\lc -1} \xf \dP{(r-1)\lc -1} \oplus \dP{(k+2p +r)\lc -1}\xf \dP{\lc-1}.
\end{equation}
Then, fuse the left side of \eqref{eq:fusionTvsSql} with $\ds{\lc}$ and use propositions \ref{prop:fusionTvsP}, and \ref{prop_fusion_sklvspi}, with equation \eqref{eq:fusionBSWeird} to obtain
\begin{equation}
	\T{2i+1}{k \lc } \xf \left(\ds{r\lc} \xf \ds{\lc}\right)  \simeq \T{2i+1}{k \lc } \xf\left(\dP{r \lc -1}\xf\dP{\lc -1} \oplus \ds{(r+1)\lc} \right), 
\end{equation}
while fusing its right side with $\ds{\lc}$ and using propositions \ref{prop_fusion_sqlvspkl}, and \ref{prop:fusionTvsSl} gives
\begin{align}
\overset{i}{\underset{p=0}{\bigoplus}}& \left( \dP{(k+2p)\lc-1}  \xf \left(\dP{r\lc-1}\xf \ds{\lc}\right)\right) \oplus \T{2i+1}{(k+r)\lc}\xf \ds{\lc} \notag \\
 & \qquad \simeq \overset{i}{\underset{p=0}{\bigoplus}}\left( \dP{(k+2p)\lc-1} \xf \left(\dP{r\lc-1} \xf \dP{\lc-2} \oplus \dP{(r+1)\lc -1} \oplus \dP{(r-1)\lc -1}\right)\right)\notag\\
	& \qquad \simeq \overset{i}{\underset{p=0}{\bigoplus}}\left(\dP{(k+2p+r)\lc -1}\xf \dP{\lc-1}\right) \oplus \T{2i+1}{(k+r+1)\lc}.
\end{align} 
Comparing the two sides and using the preceding observations gives the conclusion for $i+1$.
\end{proof}
\end{subsection}
\begin{subsection}{The fusion of irreducible and standard modules, second part}
We now have the tools needed to compute the fusion of an irreducible module and a non-projective standard module.
\begin{Prop}\label{prop:fusionIlvsSql}
For $k>1$,$r>0$, and in the regular family if $\lc \neq 2$,
\begin{equation}\label{eq:fusionIvsS.1}
\dl{k \lc-2} \xf \ds{r\lc} \simeq \begin{cases}
\B{2r}{(k-r)\lc -2} & \text{ if } k>q\\
\T{2k-3}{(2+r-k)\lc} & \text{ if } k\leq q
\end{cases}.
\end{equation}
In the regular family, if $\lc = 2$,
\begin{equation}
\dl{k \lc} \xf \ds{r\lc} \simeq \begin{cases}
\B{2r}{(k-r)\lc} & \text{ if } k>r\\
\T{2k+1}{(1+r-k)\lc} & \text{ if } k\leq r
\end{cases}.
\end{equation}
\end{Prop}
\begin{proof}
We proceed by induction on $r$. Proposition \ref{prop:fusionIvsSl} already gives the case $r=1$, so suppose that the result holds for some $1 \leq r<k-1$. Fuse the left side of equation \eqref{eq:fusionIvsS.1} with $\ds{\lc}$ and use propositions \ref{prop:fusionstandvssl}, and \ref{prop:fusionIvsP} to obtain
\begin{align}
	\dl{k \lc -2} \xf \left(\ds{r \lc} \xf \ds{ \lc}\right) &\simeq \dl{k \lc -2} \xf \left(\dP{r\lc-1} \xf \dP{\lc-1} \oplus \ds{(r+1)\lc} \right)\notag\\
	& \simeq \overset{r-1}{\underset{p = 0}{\bigoplus}}\dP{(k+2p-r)\lc-1} \xf \dP{\lc -1} \oplus  \dl{k \lc -2} \xf \ds{(r+1)\lc}.
\end{align}
Then, fuse the right side of equation \eqref{eq:fusionIvsS.1} with $\ds{\lc}$ and use propositions \ref{prop:fusionBvsSl}, and \ref{prop:fusionTvsSl} to obtain
\begin{equation}
\B{2q}{(k-r)\lc -2} \xf \ds{\lc} \simeq \overset{r-1}{\underset{p = 0}{\bigoplus}}\dP{(k+2p-r)\lc-1} \xf \dP{\lc -1} \oplus \left. \begin{cases}
\B{2(r+1)}{(k-1-r)\lc-2}, & \text{ if }  r< k-1\\
\T{2(k-2)+1}{2\lc}, & \text{ if } r=k-1
\end{cases}\right\rbrace .
\end{equation}
Comparing the two results gives the conclusion for $r+1$. In particular, this gives the conclusion for all $r \leq k$. 

 Suppose now that the result holds for some $r\geq k$. Fuse the left side of equation \eqref{eq:fusionIvsS.1} with $\ds{\lc}$ and use propositions \ref{prop:fusionstandvssl}, and \ref{prop:fusionIvsP} to obtain
 \begin{align}
 	\dl{k \lc -2} \xf \left( \ds{r \lc} \xf \ds{ \lc} \right)& \simeq \dl{k \lc -2} \xf \left(\dP{r\lc-1} \xf \dP{\lc-1} \oplus \ds{(r+1)\lc} \right) \notag \\
 	& \simeq  \overset{k-2}{\underset{p=0}{\bigoplus}}\left( \dP{(2+r-k+2p)\lc-1} \xf \dP{\lc-1}\right) \oplus  \dl{k \lc -2} \xf \ds{(r+1)\lc}.
 \end{align}
Fusing the right side of equation \eqref{eq:fusionIvsS.1} with $\ds{\lc}$ and using propositions \ref{prop:fusionTvsSl} instead gives
\begin{equation}
\T{2k-3}{(2+r-k)\lc} \xf \ds{\lc} \simeq \overset{k-2}{\underset{p=0}{\bigoplus}}\left( \dP{(2+r-k+2p)\lc-1} \xf \dP{\lc-1}\right) \oplus \T{2k-3}{(2-k+r+1)\lc}.
\end{equation}
Comparing the two results then give the conclusion for $r+1$.

In the regular family, the case where $\lc=2$ is slightly different because then $\B{2i}{0} \simeq \T{2i-1}{2}$. Nevertheless, the arguments are nearly identical.
\end{proof}
\begin{Prop}\label{prop:fusionIvsstand}
For $k>1$, $r\geq 1$, $0<i,j<\lc$,
\begin{equation}\label{eq:fusionIvsstand}
\dl{k \lc-1 -i} \xf \ds{r\lc-1+j} \simeq \begin{cases}
\B{2r}{(k-r)\lc -1-j} \xf \dP{i-1} & \text{ if } k>r\\
\T{2k-3}{(2+r-k)\lc+(j-1)} \xf \dP{i-1} & \text{ if } k\leq r
\end{cases}.
\end{equation}
\end{Prop}
\begin{proof}
 The proof mimics those of previous sections so we will only give a rough outline. Proceed by induction on $i,j$, using proposition \ref{prop:fusionIlvsSql} for the case $i=j=1$. To induce on $i$, fuse both sides of equation \eqref{eq:fusionIvsstand} with $\dP{1}$, use propositions \ref{prop:fusionIvssmallP}, \ref{prop:fusionBvssmallP}, and \ref{prop:fusionTvssmallP} and compare the two results. Then, induce on $j$ by doing the same thing but with propositions \ref{prop_fusion_sklvspi}, \ref{prop:fusionBvssmallP}, and \ref{prop:fusionTvssmallP}, instead.
\end{proof}
\end{subsection}
\begin{subsection}{Fusion of two irreducible modules, first part}\label{sec:fusionIvsIp1}
Now that the fusion of standard modules with irreducible ones are know, the fusion of two irreducible modules can be directly computed.
\begin{Prop}\label{prop:fusion.IvsI1}
For $k\geq r >1$, and in the regular family, $\lc \neq 2$,
\begin{equation}
\dl{k\lc -2} \xf \dl{r \lc -2} \simeq \begin{cases}
\overset{k+r-2}{\underset{\underset{step=2}{p=k-r+2}}{\bigoplus}}\dl{p \lc -2} & \text{ if } r < k \\
\B{1}{0} \oplus \overset{2k-2}{\underset{\underset{step=2}{p=4}}{\bigoplus}}\dl{p \lc -2} & \text{ if } r=k
\end{cases}.
\end{equation}
\end{Prop}
\begin{proof}
Start with the exact sequence
\begin{equation}
\ds{r \lc} \longrightarrow \ds{r \lc -2} \longrightarrow \dl{r \lc -2} \longrightarrow 0,
\end{equation}
which becomes
\begin{equation}
\left.\begin{cases}
\B{2 r}{(k-r)\lc -2} & \text{ if } r<k\\
\T{2k-3}{2\lc} & \text{ if } r=k
\end{cases}\right\} \overset{g}{\longrightarrow} \B{2(r-1)}{(k-r)\lc} \longrightarrow \dl{r \lc -2}\xf \dl{k \lc -2} \longrightarrow 0
\end{equation}
by using the right-exactness of fusion together with proposition \ref{prop:fusionIvsstand}. Then, build the following exact commuting diagram:
\begin{equation}
\begin{tikzpicture}[scale=1/3]
	\node[anchor = west] at (3,0) {$\left.\begin{cases}
\B{2 r}{(k-r)\lc -2} & \text{ if } r<k\\
\T{2k-3}{2\lc} & \text{ if } r=k
\end{cases}\right\}$};
	\draw[->] (14,0) -- (17,0);
		\node[anchor = south] at (15.5,0) {$g$};
		\node[anchor=west] at (17,0) {$\B{2(r-1)}{(k-r)\lc}$};
	\draw[->] (21,0) -- (24,0);
		\node[anchor=south] at (22.5,0) {$\bar{g}$};
		\node[anchor=west] at (24,0) {$\dl{r \lc -2}\xf \dl{k \lc -2} \longrightarrow 0$};
		\node at (0,-5) {$0$};
	\draw[->] (1,-5) -- (7,-5);
		\node[anchor=west] at (7,-5) {$\overset{k+r-2}{\underset{\underset{step=2}{p=k-r}}{\bigoplus}}\dl{p \lc }$};
	\draw[->] (11,-5) -- (17,-5);
		\node[anchor=north] at (14,-5) {$\alpha$};
		\node[anchor=west] at (17,-5) {$\B{2(r-1)}{(k-r)\lc}$};
	\draw[->] (21,-5) -- (24,-5);
		\node[anchor=north] at (22.5,-5) {$\bar{\alpha}$};
		\node[anchor=west] at (24,-5) {$\overset{k+r-2}{\underset{\underset{step=2}{p=k-r+2}}{\bigoplus}}\dl{p \lc -2} \longrightarrow 0$};
	\draw[->] (9,-1.5)--(9,-3);
		\node[anchor=east] at (9,-2) {$\gamma $};
	\draw[->] (19,-1) -- (19,-4);
		\node[anchor=west] at (19,-2.5) {$\id $};
	\draw[->] (27,-1)--(27,-3);
		\node[anchor=west]at (27,-2) {$f $};
	\draw[->] (19,3) -- (19,1);
		\node[anchor=south] at (19,3) {$0$};
	\draw[->] (27,3) -- (27,1);
		\node[anchor=south] at (27,3) {$\ker f $};
	\draw[->] (19,-6) -- (19,-8);
		\node[anchor=north] at (19,-8) {$0$};
\end{tikzpicture}.
\end{equation}
Here, $f$ exists by universality of the Cokernel of $g$ because
\begin{equation}
\Hom{\B{2 r}{(k-r)\lc -2},\overset{k+r-2}{\underset{\underset{step=2}{p=k-r+2}}{\bigoplus}}\dl{p \lc -2}} \simeq \Hom{\T{2k -3}{2\lc},\overset{2k-2}{\underset{\underset{step=2}{p=2}}{\bigoplus}}\dl{p \lc -2}} \simeq 0,
\end{equation} 
and thus $\bar{\alpha}g = 0$, which also give the existence of $\gamma$ by universality of $\ker \bar\alpha$.
 The snake lemma then gives $\Coker{f} \simeq 0$ and $\ker f \simeq \Coker{\gamma}$. Our goal is now to prove that 
\begin{equation}\label{eq:IvsI.Goal}
 \Hom{\overset{k+r-2}{\underset{\underset{step=2}{p=k-r}}{\bigoplus}}\dl{p \lc },\dl{r \lc -2}\xf \dl{k \lc -2} } \simeq 0, 
\end{equation}
because that would imply that $\ker f = 0 $, and thus that $f$ is an isomorphism. But, if there is a non-zero morphism from some $\dl{p\lc}$ to $\dl{r \lc -2}\xf \dl{k \lc -2}$, it has to be injective since $\dl{p\lc}$ is irreducible, and there must thus be a morphism from $\dl{r \lc -2}\xf \dl{k \lc -2}$ to $\dP{p\lc}$, the injective hull of $\dl{p\lc}$ (when $p\neq 0$). We are therefore trying to compute $$\Hom{\dl{r \lc -2}\xf \dl{k \lc -2}, \dP{p \lc}},$$ for $k-r \leq p \leq k+r-2$.
 
   Now, recall that $\dl{r \lc -2} \xf \dP{\lc-1} \simeq \dP{(r-1)\lc -1}$ which implies that
\begin{equation}\label{eq:fusionII}
\dl{k \lc -2} \xf \dl{r\lc -2} \xf \dP{\lc-1} \simeq \dl{k\lc-2} \xf \dP{(r-1)\lc -1} \simeq\overset{k+r-2}{\underset{\underset{step=2}{s=k-r+2}}{\bigoplus}}\dl{s\lc -2} \xf \dP{\lc -1}.
\end{equation}
Using this observation with the definition of the fusion quotient (see section \ref{sec:fusion_quotient}) and proposition \ref{prop:fusionq.proj} give
\begin{align}
	\Hom{\dl{k \lc -2} \xf \dl{r\lc -2} \xf \dP{\lc-1},\dP{p\lc-(\lc-1)}} & \simeq \Hom{\overset{k +r -2}{\underset{\underset{step=2}{s=k-r+2}}{\bigoplus}}\dl{s \lc -2} \xf \dP{\lc -1},\dP{p\lc - (\lc -1)}} \notag\\
	& \simeq \Hom{\overset{k +r -2}{\underset{\underset{step=2}{s=k-r+2}}{\bigoplus}}\dl{s \lc -2} ,\dP{p\lc - (\lc-1)}\xf \dP{\lc -1}}\notag\\
	& \simeq 0,\label{eq:fusionIvsI.1}
\end{align}
where the last line is obtained in the following way. Start by using proposition \ref{prop:fusionproj} to obtain
\begin{equation}
\dP{p\lc - (\lc-1)}\xf \dP{\lc -1} \simeq \dP{p\lc} \oplus \dP{(p-2)\lc} 
\oplus \overset{\lc-3}{\underset{\underset{step=2}{\sigma = (\lc -1) \mod 2}}{\bigoplus}} \dP{(p-1)\lc -1 + \sigma}
\end{equation}
if $\lc \neq 2$, and
\begin{equation}
\dP{p \times 2 - (2-1)}\xf \dP{2 -1} \simeq \dP{p\times 2},
\end{equation}
when $\lc =2$. Then, notice that the projective modules $\dP{s \lc-2}$, the only projective module containing $\dl{s\lc-2}$ as a submodule never appears in these fusions for any $p \in [k-r, k-r+2, \hdots, k+r-2 ] $.

However, using the definition of the fusion quotient (see section \ref{sec:fusion_quotient}) and proposition \ref{prop:fusionq.proj} also give
\begin{align}
	\Hom{\dl{k \lc -2} \xf \dl{r\lc -2} \xf \dP{\lc-1},\dP{p\lc-(\lc-1)}} & \simeq \Hom{\dl{k \lc -2} \xf \dl{r\lc -2}, \dP{p\lc - (\lc-1)} \div_{f} \dP{\lc-1}} \notag\\
	& \simeq \Hom{\dl{k \lc -2} \xf \dl{r\lc -2}, \dP{p\lc - (\lc-1)} \xf \dP{\lc-1}}. \label{eq:fusionIvsI.2}
\end{align}
It follows that $$\Hom{\dl{k \lc -2} \xf \dl{r\lc -2}, \dP{p\lc - (\lc-1)} \xf \dP{\lc-1}} \simeq 0,$$ and in particular $$\Hom{\dl{k \lc -2} \xf \dl{r\lc -2}, \dP{p\lc}}\simeq 0$$ for all $p \in [k-r, k-r+2, \hdots, k+r-2 ] $. Equation \eqref{eq:IvsI.Goal} is thus proved, and the conclusion when $r \neq k$ is obtained.

When $r=k$, the proof above does not work because then the injective hull of $\dl{(k-r)\lc}\simeq\dl{0}$ is $\B{1}{0} $ instead of $\dP{0}$. But the Loewy diagram of  $\T{2k-3}{2 \lc}$, figure \ref{fig:LdiagT2lc}, shows that $\dl{0}$ is not one of its quotient, and thus $\dl{0} \subset \ker f$.
\begin{figure}[h!]
\caption{The Loewy diagram of $\T{2k-3}{2 \lc}$.}\label{fig:LdiagT2lc}
\begin{tikzpicture}[scale=1/3]
\filldraw	
 	   (3,3) circle (6 pt) node[anchor = south] {$\dl{2 \lc}$}
 	-- (6,0) circle (6 pt) node[anchor = north] {$\dl{4\lc -2}$}
	-- (9,3) circle (6 pt) node[anchor = south] {$\dl{4\lc}$};
\draw (9,3) -- (10.5,1.5);
\draw[dotted] (10.5,1.5) -- (12,0);
\node at (12,1.5) {$\hdots$};
\draw[dotted] (13,3) -- (14.5,1.5);
\draw (14.5,1.5) -- (16,0);
\filldraw
	   (16,0) circle (6 pt) node[anchor = north] {$\dl{2(k-1)\lc -2}$}
	-- (19,3) circle (6 pt) node[anchor = south] {$\dl{2(k-1)\lc}$}
	-- (22,0) circle (6 pt) node[anchor = north] {$\dl{2k\lc-2 }$};
\draw[->] (3,3) -- (4.5,1.5);
\draw[->] (9,3) -- (7.5,1.5);
\draw[->] (9,3) -- (10.5,1.5);
\draw[->] (19,3) -- (17.5,1.5);
\draw[->] (19,3) -- (20.5,1.5);
\end{tikzpicture}
\end{figure}
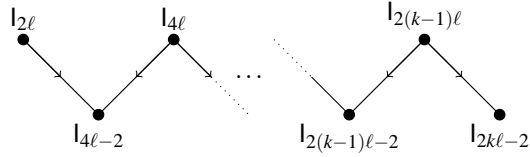
 The same argument as for the case $k\neq r$ can then be used to rule out the appearance of the other irreducible modules, and it follows that $\dl{0} \simeq \ker f$. However, proposition \ref{prop:extensions} shows that the only irreducible module which can be extended by $\dl{0}$ is $\dl{2\lc -2}$, giving
\begin{equation}
\dl{k \lc -2} \xf \dl{k \lc -2} \simeq M \oplus \overset{2k-2}{\underset{\underset{step=2}{p=4}}{\bigoplus}}\dl{p \lc -2},
\end{equation}
where $M$ satisfy the short exact sequence
\begin{equation*}
0 \longrightarrow \dl{0} \longrightarrow M \longrightarrow \dl{2 \lc -2} \longrightarrow 0.
\end{equation*}
Note that this sequence cannot split since $\Hom{\B{2(k-1)}{0},\dl{0}} \simeq 0$. Comparing this sequence with the definition of the $B$ modules then gives $$M \simeq \B{1}{0}.$$
\end{proof}
Using proposition \ref{prop:fusionIvssmallP}, this can be used to compute the fusion of the other irreducibles. However to do so requires the fusion of $\B{1}{0}$ with projective modules.
\end{subsection}
\begin{subsection}{Fusion of $\B{1}{n,k}$ and projective modules}
We start by giving the behaviour of $\B{1}{n,k}$ under induction \cite{BelRiSy}.
\begin{Prop}
For all $n \geq k^{1}$, and $k$ not critical,
\begin{equation}
\Ind{\B{1}{n,k}} \simeq \B{1}{n+1,k-1}  \oplus \B{1}{n+1,k+1} \oplus \left. \begin{cases}
\B{1}{n+1,k}, & \text{ in } \dtl{n}\\
0, & \text{ in } \tl{n}
\end{cases}\right\}.
\end{equation}
where $\B{1}{n,k \pm 1} \simeq \dP{n,k \pm 1}$ when $k$ is critical.
\end{Prop}
Using proposition \ref{prop:fusion_parity} together with the parity of the modules yields the following fusion rules.
\begin{Prop}
For all non-critical $k$, in the dilute Temperley-Lieb family
\begin{equation}
\B{1}{n,k}\xf\dP{1,0} \simeq \B{1}{n+1,k},
\end{equation}
while in both families
\begin{equation}
\B{1}{n,k} \xf \dP{1,1} \simeq \B{1}{n+1,k-1} \oplus \B{1}{n+1,k+1}.
\end{equation}
\end{Prop}
As usual, fusing $\B{1}{n,k}$ with $\dP{1,0}$ simply increases the parameter $n$. We will thus omit this parameter and always assume that it is big enough for the modules to exist.

We now compute the fusion rules for $\B{1}{k l}$, $k\geq 0$. The preceding proposition gives
\begin{equation*}
\B{1}{k \lc} \xf \dP{1} \simeq \dP{k \lc -1} \oplus \B{1}{k \lc +1},
\end{equation*}
where it is understood that $\dP{k \lc -1} \simeq 0$ if $k=0$. Fusing this result with $\dP{1}$ yields
\begin{equation*}
\begin{tabular}{r l}
$\B{1}{k\lc} \xf \dP{1}\xf\dP{1} $&$\simeq \B{1}{k\lc} \xf \left(\dP{0} \oplus \dP{2} \right)$\\
&$\simeq \dP{k \lc -1} \xf \dP{1} \oplus \B{1}{k \lc} \oplus \B{1}{k \lc +2}$.
\end{tabular}
\end{equation*} 
Comparing the two lines gives the fusion rule
\begin{equation*}
\B{1}{k\lc} \xf \dP{2} \simeq \dP{k \lc -1}\xf \dP{1} \oplus \B{1}{k \lc +2}.
\end{equation*}
Repeating the argument yields
\begin{equation*}
\B{1}{k\lc} \xf \dP{3} \simeq \dP{k \lc -1} \xf \dP{2} \oplus \B{1}{k \lc +3},
\end{equation*}
\begin{equation*}
\B{1}{k\lc} \xf \dP{4} \simeq \dP{k \lc -1} \xf \dP{3} \oplus \B{1}{k\lc +4}.
\end{equation*}
This arguments can be repeated as needed to obtain the following fusion rules.
\begin{Prop}\label{prop:fusion.B1vsSmallP}
For $0< i \leq \lc-1$, $k\geq 0$,
\begin{equation}
\B{1}{k\lc} \xf \dP{i} \simeq \dP{k \lc -1} \xf \dP{i-1} \oplus \B{1}{k\lc +i},
\end{equation}
where it is understood that $\dP{-1} \simeq 0$.
\end{Prop}
If $k\neq 0$, this proposition gives
\begin{equation*}
\B{1}{k\lc} \xf \dP{\lc -1} \simeq \dP{k\lc -1} \xf \dP{\lc -2} \oplus \dP{(k+1)\lc-1} \simeq \ds{k \lc} \xf \dP{\lc -1},
\end{equation*}
where we used proposition \ref{prop_fusion_sklvspi}. Fusing this expression with $\dP{1}$ gives
\begin{equation*}
\B{1}{k \lc} \xf \dP{\lc -1} \xf \dP{1} \simeq \B{1}{k \lc} \xf \dP{\lc} \simeq \ds{k \lc} \xf \dP{\lc}.
\end{equation*}
Fusing this with $\dP{1}$ again gives
\begin{equation*}
\begin{tabular}{r l}
$\B{1}{k \lc} \xf \dP{\lc}\xf \dP{1} $&$\simeq \B{1}{k\lc} \xf \left(2\dP{\lc -1} \oplus \dP{\lc +1}\right)$\\
& $\simeq \ds{k \lc} \xf \left(2\dP{\lc -1} \oplus \dP{\lc +1} \right)$.
\end{tabular}
\end{equation*}
Comparing the two lines gives the fusion rule
\begin{equation*}
\B{1}{k \lc} \xf \dP{\lc+1} \simeq \ds{k \lc } \xf \dP{\lc+1}.
\end{equation*}
It is simple enough to repeat this argument and obtain the general formula.
\begin{Prop}\label{prop:fusion.B1vsP}
For $k>0$,$r\geq \lc-1$,
\begin{equation}
\B{1}{k \lc} \xf \dP{r} \simeq \ds{k \lc} \xf \dP{r}.
\end{equation}
\end{Prop}
For $k=0$, recall  the short exact sequence
\begin{equation}
0 \longrightarrow \dl{0} \longrightarrow \B{1}{0} \longrightarrow \dl{2\lc -2} \longrightarrow 0.
\end{equation}
Since $\dl{0} \xf \dP{r} \simeq 0$ for all $r\geq \lc -1$ (see propositions \ref{prop:fusionIWeird} and \ref{prop:fusionIvssmallP}), the following result is obtained.
\begin{Prop}
For all $r\geq 1$, $0\leq j< \lc $,
\begin{equation}
\B{1}{0} \xf \dP{r\lc -1 +j} \simeq \dl{2\lc -2} \xf \dP{r \lc -1 +j} \simeq \dP{r \lc -1 +j}.
\end{equation}
\end{Prop}
More general results could be easily obtained but we will stop here since we have all we need to finish the computation of fusions of irreducible modules.
\end{subsection}
\begin{subsection}{Fusion of two irreducible modules, second part}\label{sec:fusionIvsIfinal}
Proposition \ref{prop:fusion.IvsI1} gives
\begin{equation*}
\dl{k\lc -2} \xf \dl{r \lc -2} \simeq \begin{cases}
\overset{k+r-2}{\underset{\underset{step=2}{p=k-r+2}}{\bigoplus}}\dl{p \lc -2} & \text{ if } r < k \\
\B{1}{0} \oplus \overset{2k-2}{\underset{\underset{step=2}{p=4}}{\bigoplus}}\dl{p \lc -2} & \text{ if } r=k
\end{cases},
\end{equation*}
and proposition \ref{prop:fusionIvssmallP} gives
\begin{equation*}
\dl{k\lc -2} \xf \dP{i} \simeq \dl{k\lc -2-i},
\end{equation*}
for all $0\leq i <\lc -1$. To obtain $\dl{k \lc -2-i} \xf \dl{r \lc -2 -j}$ we must therefore compute $\dl{p \lc -2} \xf \dP{i}\xf \dP{j}$ and $\B{1}{0}\xf \dP{i} \xf \dP{j}$. Using propositions \ref{prop:fusionproj}, and \ref{prop:fusionIvsP},
\begin{align*}
\dl{p \lc -2} \xf \dP{i} \xf \dP{j} & \simeq  \dl{p \lc -2} \xf \Big(\overset{\min (i+j,2\lc-(i+j)-4)}{\underset{\underset{step=2}{\sigma = |i-j|}}{\bigoplus}}\dP{\sigma} \oplus \overset{i+j -\lc+1}{\underset{\underset{step=2}{\sigma =(i+j+\lc+ 1)\text{mod} 2 }}{\bigoplus}}\dP{\lc -1 +\sigma} \Big)\\
	& \simeq \overset{\min(i+j,2\lc-(i+j)-4)}{\underset{\underset{step=2}{\sigma = |i-j|}}{\bigoplus}}\dl{p \lc -2 - \sigma} \oplus \overset{i+j -\lc+1}{\underset{\underset{step=2}{\sigma =(i+j+\lc+1)\text{mod} 2 }}{\bigoplus}}\dP{(p-1)\lc -1 +\sigma}.
\end{align*}

Similarly, using proposition \ref{prop:fusionproj} with propositions \ref{prop:fusion.B1vsSmallP} and \ref{prop:fusion.B1vsP} gives
\begin{equation}
\B{1}{0} \xf  \dP{i} \xf \dP{j} \simeq \overset{\min(i+j,2\lc-(i+j)-4)}{\underset{\underset{step=2}{\sigma = |i-j|}}{\bigoplus}}\B{1}{\sigma} \oplus \overset{i+j -\lc+1}{\underset{\underset{step=2}{\sigma =(i+j+\lc+1)\text{mod} 2 }}{\bigoplus}}\dP{\lc -1 +\sigma}.
\end{equation}
These give the final result.
\begin{Thm}
For $1<r< k$, $0<i,j<\lc-1$,
\begin{equation}
\dl{k\lc -2-i} \xf \dl{r \lc -2-j} \simeq \overset{k+r-2}{\underset{\underset{step=2}{p=k-r+2}}{\bigoplus}}\Big( \overset{\min(i+j,2\lc-(i+j)-4)}{\underset{\underset{step=2}{\sigma = |i-j|}}{\bigoplus}}\dl{p \lc -2 - \sigma} \oplus \overset{i+j -\lc+1}{\underset{\underset{step=2}{\sigma =(i+j+\lc+1)\text{mod} 2 }}{\bigoplus}}\dP{(p-1)\lc -1 +\sigma} \Big),
\end{equation}
\begin{multline}
\dl{k \lc -2 -i} \xf \dl{k \lc -2 -j} \simeq \overset{\min(i+j,2\lc-(i+j)-4)}{\underset{\underset{step=2}{\sigma = |i-j|}}{\bigoplus}}\B{1}{\sigma} \oplus \overset{i+j -\lc+1}{\underset{\underset{step=2}{\sigma =(i+j+\lc+1)\text{mod} 2 }}{\bigoplus}}\dP{\lc -1 +\sigma} \\
\oplus \overset{2k-2}{\underset{\underset{step=2}{p=4}}{\bigoplus}}\Big( \overset{\min(i+j,2\lc-(i+j)-4)}{\underset{\underset{step=2}{\sigma = |i-j|}}{\bigoplus}}\dl{p \lc -2 - \sigma} \oplus \overset{i+j -\lc+1}{\underset{\underset{step=2}{\sigma =(i+j+\lc+1)\text{mod} 2 }}{\bigoplus}}\dP{(p-1)\lc -1 +\sigma} \Big).
\end{multline}
\end{Thm}
We still need to compute the fusion rules for the irreducibles with $k=1$. Recall that it was established in proposition \ref{prop:fusionIWeird} that 
\begin{equation*}
\dl{\lc -2} \xf \dl{r} \simeq 0
\end{equation*}
for all $r\geq \lc$. Using the short-exact sequence
\begin{equation*}
0 \longrightarrow \dl{\lc} \longrightarrow \dP{\lc -2} \longrightarrow \dl{\lc -2} \longrightarrow 0,
\end{equation*}
with the right exactness of fusion, it follows that
\begin{equation}
\dl{\lc -2} \xf \dl{\lc -2} \simeq \dP{\lc -2} \xf \dl{\lc -2} \simeq \dl{0},
\end{equation}
and thus that
\begin{align}
\dl{\lc -2 -i} \xf \dl{\lc -2-j}  & \simeq \dl{\lc-2}\xf\dl{\lc -2} \xf \left( \dP{i} \xf \dP{j} \right)\notag \\
 & \simeq \dl{0}\xf \left(\overset{\min (i+j,2\lc-(i+j)-4)}{\underset{\underset{step=2}{\sigma = |i-j|}}{\bigoplus}}\dP{\sigma} \oplus \overset{i+j -\lc+1}{\underset{\underset{step=2}{\sigma =(i+j+\lc+1)\text{mod} 2 }}{\bigoplus}}\dP{\lc -1 +\sigma} \right)\notag\\
 & \simeq \overset{\min (i+j,2\lc-(i+j)-4)}{\underset{\underset{step=2}{\sigma = |i-j|}}{\bigoplus}}\dl{\sigma},
\end{align}
where the last line is obtained by proposition \ref{prop:fusionIvssmallP}. The following theorem is then obtained by changing the indices.
\begin{Thm}
For all $0\leq i,j <\lc-1$,
\begin{equation}\label{eq:fusionI.Kacs}
\dl{i}\xf\dl{j} \simeq \overset{\min\left\{ i+j, 2\lc -(i+j)-4\right\}}{\underset{\underset{step=2}{p=|i-j|}}{\bigoplus}}\dl{p}.
\end{equation}
\end{Thm}
It should be noted that for a minimal model $M(p',p)$ of the Virasoro algebra, the fusion rule between two primary fields is
\begin{equation}
\phi_{1,1+s} \xf \phi_{1,1+r} = \sum_{\underset{step=2}{l=|r-s|}}^{\text{Min}\left(r+s, 2p -(s+r) - 4) \right)} \phi_{1,1+l},
\end{equation}
which is identical to \eqref{eq:fusionI.Kacs} under the correspondence $\ell \to p$, $\dl{i} \to \phi_{1,1+i}$.
\end{subsection}
\end{section}
\begin{section}{Conclusion}
The main results of the paper are now reviewed. A definition of a fusion product on the Temperley-Lieb family as been proposed by Read and Saleur \cite{ReadSaleur1,ReadSaleur2}; in section \ref{sec:deffusion}, we generalize their definition to more general families of associative algebras, including the dilute Temperley-Lieb algebra. A straightforward consequence of this definition is that the fusion of pairs of projective modules are also projective. In the Temperley-Lieb algebras, when $q$ is not a root of unity, the projective modules $\dP{n,k}$ behaves under fusion like irreducible $su(2)$ representations under tensor product: $$\dP{n,k} \xf \dP{m,r} \simeq \overset{k+r}{\underset{\underset{step =2}{p = |k-r|}}{\bigoplus}} \dP{n+m,p}. $$
When $q$ is a root of unity, they behave like a polynomial ring, with a basis of Chebyshev polynomials of the second kind:
$$\dP{n,i} \to U_{i}(x), \qquad \dP{n,\kc} \to U_{\kc}(x), \qquad \dP{n,\kc +i} \to U_{\kc-i}(x) + U_{\kc+i}(x).$$

In section \ref{sec:fusion.standard}, we use this information to compute fusion products of standard modules $\ds{n,k}$ with projective modules and other standard modules. It is shown that these can once again be interpreted as a polynomial ring with a basis of Chebyshev polynomials, albeit with a different product. The correspondence is $$\ds{n,k} \to U_{k}(x),$$ and when taking a product, the result must first be re-written in terms of the polynomials representing projective modules $\dP{n,p}$, starting with the smallest $p$; the remaining polynomials are then identified with the standard modules.

In section \ref{sec:fusionquotients}, it is shown how to use fusion rules obtained previously to construct more complex ones. In particular, we compute the fusion product of an irreducible modules, and a standard modules. This shows the appearance of two other classes of indecomposable modules, the $B$'s and the $T$'s. After computing their fusion rules in section \ref{sec:fusionBvsP},\ref{sec:fusionBvsS},\ref{sec:fusionTvsP} and \ref{sec:fusionTvsS}, the fusion product of pairs of irreducible modules is computed in section \ref{sec:fusionIvsIp1}. Here, we use the adjoint of the fusion product, the fusion quotient, which simplifies the proofs greatly. Finally, in section \ref{sec:fusionIvsIfinal} we find a general formula for the fusion product of pairs of irreducible modules lying on the left of the first critical line.

It is then recognized that the irreducible modules $\dl{n,i}$, with $i\leq \ell -2$, behave under fusion like primary fields in the first line of the Kacs table of a Virasoro minimal model $M(p',p)$, with $p = \ell$. 

There are still many fusion rules between indecomposable modules which we have yet to compute. We chose to limit ourselves to the projective, standard and irreducible modules because they are very important in the representation theory of the Temperley-Lieb algebras, but it would be interesting to find out how the other, more exotic, modules behave under this fusion product, as they do appear in physical problems CIT. We believe that the arguments used here could be extended to obtain these fusions.

The appearance of the fusion quotient is a simple consequence of the definition of the fusion product. However, while it is conjectured that the fusion product corresponds to the fusion product on the Virasoro algebra in the limit, the meaning of this fusion quotient is unclear. Is there a corresponding functor on the Virasoro algebra?
\end{section}
\appendix

\begin{section}{Fusion of $\dP{2,0}$ in $\tl{n}$}\label{sec:magicp2}
We investigate here the fusion of $\dP{2,0}$ with various modules in the regular family $\tl{n}$ when $\lc \neq 2$.
\begin{Prop}\label{prop:magickSP2}
If $\beta \neq 0$, then $\ds{n,i} \xf \dP{2,0} \simeq \ds{n+2,i}.$
\end{Prop}
\begin{proof}
Pick 
\begin{equation*}
z = \begin{tikzpicture}[baseline={(current bounding box.center)}, scale=1/3]
		\draw[very thick] (0,-1/2) -- (0,15/2);
			\draw (0,0) .. controls (1,0) and (1,1) .. (0,1);
				\node[anchor = south] at (1/2,1) {$\vdots$};
			\draw (0,3) .. controls (1,3) and (1,4) .. (0,4);
			\draw (0,5) -- (3,5);
				\node[anchor = south] at (1/2,5) {$\vdots$};
			\draw (0,7) -- (3,7);
				\node[anchor = east] at (0,7) {\small{$1$}};
				\node[anchor = east] at (0,5) {\small{$i$}};
				\node[anchor = east] at (0,4) {\small{$i+1$}};
				\node[anchor = east] at (0,0) {\small{$n$}};
\end{tikzpicture}, \qquad x = \begin{tikzpicture}[baseline= {(current bounding box.center)},scale = 1/3]
	\draw[very thick] (0,-3/4) -- (0,3/4);
	\draw (0,-1/2) .. controls (1,-1/2) and (1,1/2) .. (0,1/2);
\end{tikzpicture}, 
\end{equation*}
where $z$ is a generator of $\ds{n,i}$ because $<z,z> = \beta^{\frac{n-i}{2}} \neq 0$, and $x$ is a generator of $\dP{2,0} = \mathbb{C}x$. Note that
\begin{equation}
\begin{tabular}{r l}
	$\ds{n,i} \xf \dP{2,0}$&$ \simeq \tl{n+2} \otimes \left(\ds{n,i} \otimes \dP{2,0}\right)$\\
	&$	\simeq \tl{n+2} \otimes \left(\tl{n}\otimes \tl{2} \right)(z \otimes x )$\\
 & $\simeq \tl{n+2} \otimes \left(z \otimes x \right)$.
\end{tabular}
\end{equation}
With the usual generator,
\begin{equation}
e_{j}=
	\begin{tikzpicture}[baseline={(current bounding box.center)}, scale=1/3]
		\draw[very thick] (0,-1/2) -- (0,15/2);
		\draw[very thick] (3,-1/2) -- (3,15/2);
		\draw (0,0) -- (3,0);
			\node[anchor = east] at (0,0) {\small{$n$}};
			\node[anchor = south] at (3/2,0) {$\vdots$};
		\draw (0,2) -- (3,2);
			\node[anchor = east] at (0,2) {\small{$j+2$}};
		\draw (0,3) .. controls (1,3) and (1,4) .. (0,4);
			\node[anchor = east] at (0,3) {\small{$j+1$}};
			\node[anchor = east] at (0,4) {\small{$j$}};
		\draw (3,3) .. controls (2,3) and (2,4) .. (3,4);		
		\draw (0,5) -- (3,5);
			\node[anchor = east] at (0,5) {\small{$j-1$}};
			\node[anchor = south] at (3/2,5) {$\vdots$};
		\draw (0,7) -- (3,7);
			\node[anchor = east] at (0,7) {\small{$1$}};
	\end{tikzpicture},
\end{equation}
we note that $$e_{i+1}e_{i+3}\hdots e_{n-1} \otimes (z \otimes x) = \beta^{\frac{n-i}{2} +1} (z\otimes x).$$ It thus follows that
\begin{equation}
\begin{tabular}{r l}
	$\ds{n,i} \xf \dP{2,0}$&$ \simeq \tl{n+2} \otimes \left(e_{i+1}e_{i+3}\hdots e_{n-1} \otimes e_{1}\right) (z \otimes x)$\\
	& $\simeq \tl{n+2} \left(e_{i+1}e_{i+3}\hdots e_{n-1} e_{n+1}  \right)\otimes (z \otimes x)$\\
	& $\simeq \text{Span}_{\mathbb{C}} \left\{ \;
	\begin{tikzpicture}[baseline={(current bounding box.center)}, scale=1/3]
		\draw[very thick] (0,-1/2) -- (0,15/2);
		\draw[very thick] (3,-1/2) -- (3,15/2);
				\node[anchor = west] at (0,4) {$u$};
			\draw (3,0) .. controls (2,0) and (2,1) .. (3,1);
				\node[anchor=west] at (3,0) {\small{$n+2$}};
				\node[anchor=west] at (3,1) {\small{$n+1$}};
				\node[anchor=south] at (5/2,1) {$\vdots$};
			\draw (3,3) .. controls (2,3) and (2,4) .. (3,4);
				\node[anchor=west] at (3,3) {\small{$i+2$}};
				\node[anchor=west] at (3,4) {\small{$i+1$}};
			\draw (3,5) -- (2,5);
				\node[anchor=west] at (3,5) {\small{$i$}};
				\node[anchor=south] at (5/2,5) {$\vdots$};
			\draw (3,7) -- (2,7);
				\node[anchor=west] at (3,7) {\small{$1$}};
	\end{tikzpicture} \otimes (z \otimes x)| \text{ where } u\in \ds{n+2,i} \right\}$\\
	& $\simeq \ds{n+2,i}$,
\end{tabular}
\end{equation}
where the last two lines are obtained by straightforward calculations.
\end{proof}
\begin{Prop}\label{prop:magickPP2}
If $\beta \neq 0$, then $\dP{n,i} \xf \dP{2,0} \simeq \dP{n+2,i}.$
\end{Prop}
\begin{proof}
If $i_{-}<0$(see definition of $i_{\pm}$ in section \ref{sec:rappel.indec}), then $\dP{n,i} \simeq \ds{n,i}$ and the result is given by proposition \ref{prop:magickSP2}. If $i_{-} \geq 0$, there is a short exact sequence
\begin{equation}
0 \longrightarrow \ds{n,i_{-}} \longrightarrow \dP{n,i} \longrightarrow \ds{n,i} \longrightarrow 0,
\end{equation}
which becomes 
\begin{equation}
\ds{n+2,i_{-}} \longrightarrow \dP{n,i}\xf\dP{2,0} \longrightarrow \ds{n+2,i} \longrightarrow 0,
\end{equation}
by fusing it with $\dP{2,0}$ and using proposition \ref{prop:magickSP2}. However, since the fusion of two projective modules is projective, it follows that $\dP{n,i}\xf\dP{2,0} $ is a projective module having $\ds{n+2,i}$ as a quotient, and whose dimension is at most $\dim \ds{n+2,i_{-}} + \dim \ds{n+2,i} = \dim \dP{n+2,i}$. Since $\dP{n+2,i}$ is the projective cover of $\ds{n+2,i}$, the conclusion follows.
\end{proof}
Similar arguments can be used to compute the action of $\dP{2,0}$ on other modules. We simply state the result.
\begin{Prop}\label{prop:magickOthersP}
If $n\geq i_{+}$, $$\dl{n,i} \xf \dP{2,0} \simeq \dl{n+2,i}, $$
and if $n\geq k$, $$\B{j}{n,k}\xf \dP{2,0} \simeq \B{2j}{n+2,k}, $$ $$\T{j}{n,k}\xf \dP{2,0} \simeq \T{j}{n+2,k}.$$
\end{Prop}
\end{section}
\begin{section}{Fusion quotient}\label{sec:fusion_quotient}
We present here a brief study of the operator adjoint to the fusion product, the fusion quotient. We begin with the definition then present the basic properties that follows from it. Finally, we give the fusion quotients of a few Temperley-Lieb modules to show that the two operations, while giving similar results, are not equivalent.
\begin{subsection}{Definition of the fusion quotient}
\begin{Prop}\label{f_frob}
Consider a family of algebras $(A_i)_{i \in \mathbb{N}}$ on which fusion is defined (see the beginning of section \ref{sec:deffusion}), $U$ a $A_i$-module, $V$ a $A_{j}$-module and $W$ a $A_{i+j}$-module. There is an isomorphism of vector spaces
\begin{equation}
\text{\rm Hom}_{A_{i+j}}\left(U \xf V, W\right) \simeq \text{\rm Hom}_{A_{i}}\left(U,\text{Hom}_{A_{i+j}}\left( A_{i}\xf V, W\right)\right)
\end{equation}
where $A_{i}\xf V$ is seen as a left $A_{i+j}$-module and a right $A_{i}$-module.
\end{Prop}
\begin{Def}
For $U$ a $A_{i}$-module and $V$ a $A_{i+j}$-module. The \emph{fusion quotient} of $V$ by $U$, denoted by $V \div_{f} U$, is the $A_{j}$-module
\begin{equation}
V \div_{f} U = \text{\rm Hom}_{A_{i+j}}\left(A_{j}\xf U, V\right)
\end{equation}
where the module structure is given by
\begin{equation}
(a g): b \otimes_{A_{j}\otimes A_{i}}\left(c \otimes_{\mathbb{C}} x\right) \mapsto g\left(b \otimes_{A_{j}\otimes A_{i}}\left(c a\otimes_{\mathbb{C}} x\right) \right),
\end{equation}
where $a,c \in A_{j}$, $b \in A_{i+j}$, $x \in U$,$g \in \text{Hom}_{A_{i+j}}\left(A_{j} \xf U,V \right)$.
\end{Def}
If the fusion product has additional properties, like linearity, associativity and commutativity the fusion quotient will inherit some of those.
\begin{Prop}\label{prop:fusionq.prop}
Let $Q$ and $\bar{Q}$ be a pair of $A_{i+j+k}$-modules, $U, \bar{U}$ two $A_{j}$-modules and $V$ a $A_{k}$-module,
\begin{equation}\label{fq_linear}
\left(Q \oplus \bar{Q} \right)\div_{f} \left(U \oplus \bar{U}\right) \simeq \left(Q \div_{f} U \right) \oplus \left(\bar{Q} \div_{f} U \right) \oplus\left(Q \div_{f} \bar{U} \right) \oplus\left(\bar{Q} \div_{f} \bar{U} \right).
\end{equation} 
If the fusion product on the family $\lbrace A_{i} \rbrace$ is associative, then
\begin{equation}\label{fq_assoc}
\left(Q \div_{f} U\right) \div_{f} V \simeq Q \div_{f} \left(V \xf U\right).
\end{equation}
If the fusion product is also commutative, then
\begin{equation}\label{fq_comm}
\left(Q \div_{f} U\right) \div_{f} V \simeq \left(Q \div_{f} V\right) \div_{f} U.
\end{equation}
\end{Prop}
\begin{proof}
The proof of \eqref{fq_linear} follows from the linearity of the fusion product and of the $\text{Hom}$ functor. If the fusion product on the family $\lbrace A_{i} \rbrace$ is associative, then
\begin{align}
\left(Q \div_{f} U\right) \div_{f} V & = \text{Hom}_{A_{i+k}}\left(A_{i}\xf V, Q \div_{f}U\right) \\
& \simeq \text{Hom}_{A_{i+k+j}}\left(\left(A_{i}\xf V\right)\xf U, Q \right)\\
& \simeq \text{Hom}_{A_{i+j+k}}\left(A_{i}\xf \left(V\xf U \right),Q\right)\\
& = Q \div_{f} \left(V \xf U \right).
\end{align}
The first and last lines are simply the definition of the fusion quotient, while the second is proposition \ref{f_frob} and the third is obtained by using the associativity of the fusion product. If the product is also commutative, it is clear that
\begin{equation}
\left(Q \div_{f} U\right) \div_{f} V  \simeq Q \div_{f} \left(V \xf U\right) \simeq Q \div_{f} \left(U \xf V\right) \simeq  \left(Q \div_{f} V\right) \div_{f} U 
\end{equation}
by using \eqref{fq_assoc}.
\end{proof}
The following proposition gives the behaviour of short exact sequences under the fusion quotient.
\begin{Prop}\label{prop:fusionq.functor}
Let $$0 \longrightarrow U \longrightarrow V \longrightarrow W \longrightarrow 0 $$ be a short exact sequence of $A_{i}$-modules and $Q$ be a $A_{j}$-module. If $i>j$, the sequence of $A_{i-j}$-modules
\begin{equation}
0 \longrightarrow U \div_{f} Q \longrightarrow V\div_{f} Q \overset{f}{\longrightarrow} W \div_{f} Q
\end{equation}
is exact. If $Q$ is projective, then $f$ is surjective. If $j>i$, the sequence of $A_{j-i}$-module
\begin{equation}
0 \longrightarrow Q\div_{f} W \longrightarrow Q\div_{f} V \longrightarrow Q \div_{f} U
\end{equation}
is exact.
\end{Prop}
\begin{proof}
For the case $i>j$ simply use the fact that $\Hom{P, - }$ is always left-exact for all module $P$. If moreover $Q$ is projective, proposition \ref{prop:fusion_resproj} shows that $A_{i-j} \xf Q$ is projective so that $\text{Hom}_{A_{i+j}}\left\{A_{i-j} \xf Q, - \right\}$ is also right-exact. For the other case, the right-exactness of the fusion product is used to obtain the exact sequence
\begin{equation}
A_{j-i} \xf U \longrightarrow A_{j-i} \xf V \longrightarrow A_{j-i} \xf W \longrightarrow 0.
\end{equation}
The final result is obtained by using the fact that $\Hom{ - ,P}$ is always left-exact and contravariant.
\end{proof}
Note also that the fusion quotient of an $A_{i+j}$-module  $U$ by $A_{j}$ has the structure of a $A_{i}\otimes_{\mathbb{C}}A_{j}$-module. It can be seen that this quotient is in fact isomorphic to the restriction of $U$ to the subalgebra $A_{i}\otimes_{\mathbb{C}}A_{j}$. The following proposition relates this structure to the quotient of $U$ by a $A_{j}$-module $V$.
\begin{Prop}\label{prop:fusionq.hom}
For $U$ a $A_{i+j}$-module and $V$ a $A_{j}$-module,
\begin{equation}
U\div_{f} V \simeq \text{\rm Hom}_{A_{j}}\left(V, U \div_{f} A_{i}\right)
\end{equation}
where the action of $A_{i}$ on $\text{\rm Hom}_{A_{j}}\left(V, U \div_{f} A_{i}\right)$ is given by
\begin{equation}
(a_i g): v_{j} \mapsto \left(b_{i+j} \otimes_{A_{j}\otimes A_{i}} \left(c_{j}\otimes_{\mathbb{C}} d_i\right) \mapsto g(v_{j})\left(b_{i+j} \otimes_{A_{j}\otimes A_{i}} \left(c_{j} \otimes_{\mathbb{C}} d_{i} a_{i}\right)\right) \right),
\end{equation}
where the indices on $a_i,b_{i+j},\hdots$ refers to which of $A_{i},A_{i+j},\hdots$ they belong.
\end{Prop}
\begin{proof}
The proof proceeds by construction. Define the vector space homomorphism $\phi:U\div_f V \to \text{Hom}_{A_{j}}\left(V, U \div_{f} A_{i}\right)$ by
\begin{equation}
\phi(g)= \left(v_j \mapsto\left(b_{i+j} \otimes_{A_{j}\otimes A_{i}} \left(c_{j} \otimes_{\mathbb{C}} d_{i}\right) \mapsto g(b_{i+j} \otimes_{A_{j}\otimes A_{i}} \left(c_{j}v_{j} \otimes_{\mathbb{C}} d_{i}\right)) \right)\right) 
\end{equation}
and another homomorphism $\psi: \text{Hom}_{A_{j}}\left(V, U \div_{f} A_{i}\right) \to U\div_f V $ by
\begin{equation}
\psi(g) = \left(b_{i+j}\otimes_{A_{j}\otimes A_{i}} (v_{j}\otimes_{\mathbb{C}} d_{i}) \mapsto g(v_{j})(b_{i+j}\otimes_{A_{j}\otimes A_{i}} (\id_{A_{j}}\otimes_{\mathbb{C}}d_{i}) ) \right).
\end{equation}
 It is straightforward to verify that these two morphisms are inverse of each other, and that the action of $A_{i}$ defined in the proposition makes them into $A_{i}$-module homomorphisms.
\end{proof}
\end{subsection}
\begin{subsection}{Fusion quotient in the Temperley-Lieb families}
We present here the fusion quotient of some modules in the $\tl{n}$ and $\dtl{n}$ families. 
\begin{Prop}
	Let $A_n$ be $\tl{n}$ or $\dtl{n}$. For any $A_{n+1}$-module $U$,
	$$U \div_f  A_{1} \simeq \Res{U}, $$
	where the restriction functor is $\Res{-} =  _{A_{n}}\left(A_{n+1}\right)_{A_{n+1}} \otimes_{A_{n+1}} -$.
\end{Prop}
\begin{proof}
The functor $\Res{-}$ is the adjoint of the functor $\Ind{-}$ defined in section \ref{sec:rappel.induction}. Since $ - \xf A_{1}$ is equivalent to $\Ind{-}$, their adjoints must also be equivalent.
\end{proof}
 This restriction functor as also been computed \cite{RiSY,BelSY,BelRiSy}.
\begin{Prop}
For $0 \leq i <\ell$, $0 < j<\ell $ such that $n+1 \geq \kc +i $,
\begin{align}
\Res{\dP{n+1,\kc+i}} \simeq & \left. \begin{cases}
2 \dP{n,\kc},& \text{ if } i=1\\
0, & \text{ if } i=0 \\
\dP{n,\kc+i-1}, & \text{otherwise}
\end{cases}\right\} \oplus \left. \begin{cases}
\dP{n,\kc+i}, &\text{ in } \dtl{n} \text{ and } n\geq \kc +i\\
\ds{n,\kc-i}, &\text{ in } \dtl{n} \text{ and } n < \kc +i\\
0, &\text{ in } \tl{n}
\end{cases}\right\} \notag\\
& \oplus \left. \begin{cases}
\dP{n,\kc-\ell} \oplus \dP{n,\kc+\ell}, & \text{ if } i = \ell -1 \text{ and } n \geq \kc + \ell\\
\dP{n,\kc-\ell}, & \text{ if } i = \ell -1 \text{ and } n < \kc + \ell\\
\ds{n,\kc -( i+1)}, & \text{ if } i \neq \ell -1 \text{ and } n < \kc + i + 1\\
\dP{n,\kc + i+1}, & \text{ if } i \neq \ell -1 \text{ and } n \geq \kc + i + 1
\end{cases}\right\},
\end{align}
\begin{align}
\Res{\ds{n+1, \kc +i}} \simeq & \left. \begin{cases}
	\dP{n,\kc}, & \text{ if } i=1\\
	\ds{n,\kc + i-1}, & \text{ otherwise }
\end{cases}\right\rbrace \oplus 
\left. \begin{cases}
\ds{n, \kc + i}, &\text{ in } \dtl{n} \text{ and } n\geq \kc +i\\
0, &\text{ in } \dtl{n} \text{ and } n < \kc +i\\
0, & \text{ in } \tl{n}
\end{cases}\right\rbrace \notag\\
& \oplus \left. \begin{cases}
	\dP{n,\kc + \ell}, & \text{ if } i = \lc -1 \text{ and } n \geq \kc + \ell \\
	\ds{n,\kc + i +1}, & \text{ if } i \neq \lc - 1 \text{ and } n \geq \kc + i+1 \\
	0, & \text{ if } n < \kc + i +1
\end{cases} \right\rbrace.
\end{align}
\end{Prop}
\begin{Coro}\label{coro:res.ind.proj}
If $k \leq n$,
\begin{equation}
\Res{\dP{n+2,k}} \simeq \Ind{\dP{n,k}}, \qquad \Res{\ds{n+2,k}} \simeq \Ind{\ds{n,k}}.
\end{equation}
\end{Coro} 
As for the fusion product, we now need to compute the fusion quotient of a standard module by $\dP{2,0}$ in $\tl{n}$.
\begin{Prop}
In the regular $\tl{n}$ family, if $\ell \neq 2$, and $n- 2m \geq q $, then 
\begin{equation}
\ds{n,q}\div_{f} \dP{2m,0} \simeq \ds{n-2m,q}.
\end{equation}
If $n-2m < q$, then $\ds{n,q}\div_{f} \dP{2m,0}  \simeq 0$.
\end{Prop}
\begin{proof}
Start with the case $m=1$. The first step is to prove that the two modules have the same dimension. For this, note that
\begin{equation}
\begin{tabular}{r l}
 $\ds{n,q}\div_{f} \dP{2,0}$ &$ \overset{1}{\simeq} \Hom{\tl{n-2} \xf \dP{2,0}, \ds{n,q}}$\\
 &$ \overset{2}{\simeq} \overset{n-2}{\underset{\underset{step=2}{i=n \mod 2}}{\bigoplus}}\left( \left(\dim\dl{n-2,i}\right) \Hom{\dP{n-2,i} \xf \dP{2,0}, \ds{n,q}}\right)$\\
 &$ \overset{3}{\simeq} \overset{n-2}{\underset{\underset{step=2}{i=n \mod 2}}{\bigoplus}}\left( \left(\dim\dl{n-2,i}\right) \Hom{\dP{n,i}, \ds{n,q}}\right)$\\
 &$ \overset{4}{\simeq} \mathbb{C} \left(\dim \dl{n-2,q} + \dim \dl{n-2,q_{1}}\right),$
 \end{tabular}
\end{equation}
where the isomorphism are morphism of vector spaces. Here, $1$ is simply the definition of the fusion quotient while $2$ is Wedderburn's theorem with linearity of Hom. The morphism $3$ is obtained by using proposition \ref{prop:fusionproj} while $4$ is obtained by inspecting the Loewy diagrams of the projective modules to find the morphism from the $\mathsf{P}$s to $\ds{n,q}$. It follows that $$\dim (\ds{n,q} \div_{f}\dP{2,0}) = \dim \dl{n-2,q} + \dim \dl{n-2,q_{1}} = \dim \ds{n-2,q}.$$ Note that one or both of these irreducible modules may not be defined, in which case we simply set their dimension to zero. In particular if both $q,q_{1} > n-2$, then $\ds{n,q} \div_{f}\dP{2,0} \simeq 0$.

To identify the action of $\tl{n-2}$ on $\ds{n,q} \div_{f} \dP{2,0}$, we proceed as follows. Note that \linebreak
$\tl{n-2}\xf \dP{2,0}$ is isomorphic as a left $\tl{n}$-module and as a right $\tl{n-2}$-module to $J$, the left ideal of $\tl{n}$ spanned by diagrams where the bottom two nodes on their right side are linked together, i.e. those of the form
\begin{equation*}
	\begin{tikzpicture}[scale=1/3]
		\draw (0,-1/2) -- (0,9/2);
		\draw (3,-1/2) -- (3,9/2);
			\draw (3,0) .. controls (2,0) and (2,1) .. (3,1);
		\node[anchor = west] at (0,2) {$u_i$};
		\node[anchor = east] at (3,3) {$v_i$};
		\node[anchor= west] at (3,2) {,};
	\end{tikzpicture}
\end{equation*}
where $u_i \in \ds{n,i}$, $v_i \in \ds{n-2,i}$ for some $0 \leq i \leq n-2$, and where the action of $\tl{n-2}$ on $J$ is obtained by adding two straight lines at the bottom of every diagram. To see this, verify that $\phi: a \mapsto az$, defines a bi-module isomorphism between the two, where 
$$z= \id_{\tl{n}} \otimes_{\tl{n-2}\xf \tl{2}} \left(\id_{\tl{n-2}} \otimes \:\begin{tikzpicture}[baseline={(current bounding box.center)},scale=1/3]
\draw (0,-1/2) -- (0,3/2);
\draw (0,0) .. controls (1,0) and (1,1) .. (0,1);
\end{tikzpicture} \right).$$
Next, notice that $g$ is an homomorphism from $\tl{n}$ to $\ds{n,q}$ if and only if there exists a unique $x$ in $\ds{n,q}$ such that $g \equiv g_{x} :a \mapsto a x$. Furthermore, since $J$ is isomorphic to $\tl{n-2} \xf \dP{2,0}$, it is a direct summand of $\tl{n}$, and thus every morphism from $J$ to $\ds{n,q}$ must be of the form $g_{x} \circ i$ for some $x$, where $i$ is the canonical injection. 

Now, consider the diagram
\begin{equation*}
	\begin{tikzpicture}[scale=1/3]
		\draw (0,-1/2) -- (0,9/2);
		\draw (3,-1/2) -- (3,9/2);
			\draw (0,0) .. controls (1,0) and (1,1) .. (0,1);
			\draw (3,0) .. controls (2,0) and (2,1) .. (3,1);
			\draw (0,2) -- (3,2);
				\node[anchor = south] at (3/2,2) {$\vdots$};
			\draw (0,4) -- (3,4);
				\node[anchor = east] at (0,2) {$e=$};
	\end{tikzpicture}
\end{equation*}
 in $\tl{n}$ and notice that for any $a \in J$, $a e \frac{1}{\beta} = a$. It follows that $$ \Hom{J, \ds{n,q}} \simeq  \left\lbrace (a\mapsto a x)| x \in \ds{n,q} \text{ such that } e x = \beta x \right\rbrace.$$  Note now that any link diagram in $\ds{n,q}$ where the two bottom nodes are linked together will define such a morphism. These span a vector space of dimension $\dim \ds{n-2,q}$. 
 
Using the action of $\tl{n-2}$ defined on $\ds{n,q}$ by adding two straight lines at the bottom of every diagram, it can be directly verified that for any $b \in \tl{n-2}$, $$(bg_{x} \circ i): a \mapsto abx = g_{bx} \circ i,$$ and thus, that $\Hom{J,\ds{n,q}}$ is isomorphic, as a left $\tl{n-2}$-module, to the submodule of $\ds{n,q}$ spanned by diagrams where the two bottom nodes are linked together. Comparing these link diagrams with a basis of $\ds{n-2,q}$ gives the conclusion.

The proof then proceed by induction on $m$. The case $m=1$ is proved so assume that the result stands for some $m$. Then
\begin{equation}
\left(\ds{n,q}\div_{f} \dP{2m,0} \right)\div_{f} \dP{2,0} \simeq \ds{n,q} \div_{f}\left(\dP{2m,0}\xf\dP{2,0}\right) \simeq \ds{n,q}\div_{f} \dP{2(m+1),0} \simeq \ds{n-2(m+1),q}, 
\end{equation}
where we simply used propositions \ref{prop:fusionproj} and \ref{prop:fusionq.prop}.
\end{proof}
\begin{Coro}\label{coro:fusionq.proj.reg}
In $\tl{n}$, if $\ell \neq 2$ and $n-2m \geq q$,
\begin{equation}
\dP{n,q}\div_{f} \dP{2m,0} \simeq \dP{n-2m,q}.
\end{equation}
\end{Coro}
\begin{proof}
If $q<\ell-1$ or if $q$ is critical, this is trivial. If $q >\ell-1$ is not critical, there is a short-exact sequence
$$0 \longrightarrow \ds{n,q_{-1}} \longrightarrow \dP{n,q} \longrightarrow \ds{n,q} \longrightarrow 0,$$
which gives the short-exact sequence of $\tl{n-m}$-modules $$0 \longrightarrow \ds{n-2m,q_{-1}} \longrightarrow \dP{n,q}\div_{f} \dP{2m,0} \longrightarrow \ds{n-2m,q} \longrightarrow 0,$$ by using proposition \ref{prop:fusionq.functor}. Since $$\Hom{\ds{n-2m,q}, \dP{n,q}\div_{f}\dP{2m,0}} \simeq \Hom{\ds{n,q},\dP{n,q}} \simeq \mathbb{C},$$ the only morphism from $\ds{n-2m,q}$ to $\dP{n,q}\div_{f} \dP{2m,0}$ must be the one which goes through $\ds{n-2m,q_{-1}}$, and thus this sequence does not split. Comparing this sequence with the definition of $\dP{n-2m,q}$ gives the conclusion.
\end{proof}
Note that a consequence of this is that $\dP{n,q}\div_{f} \dP{2m,0} \simeq \ds{n-2m,q_{-1}}$ if $q_{-1}\leq n < q$, and $\dP{n,q}\div_{f} \dP{2m,0} \simeq 0$ if $q_{-1}>n$.
\begin{Prop}
If $U$ is a $\dtl{n+m}$-module, $V$ a $\tl{m}$-module, both with well-defined parity, then $U\div_{f} V$ is even if they are both of the same parity and odd otherwise.
\end{Prop}
\begin{proof}
It was argued in a comment preceding proposition \ref{prop:fusion_parity} that for $W,V$, two modules with well-defined parities, $W\xf V$ is even if they are both of the same parity and odd otherwise. In particular, take $W = \edtl{n}$ , the even ideal of $\dtl{n}$. Then $W \xf V$ is even (odd), if and only if $V$ is even (odd). But, by definition, 
\begin{equation*}
\Hom{W\xf  V, U} \simeq \Hom{W,U \div_{f} V}.
\end{equation*}
The right side of this equality is non-zero if and only if $U \div_{f} V$ is even, while the left side vanishes unless $U$ is of the same parity as $W \xf V$. It follows that $U \div_{f} V$ is even if and only if $U$ is of the same parity as $V$.
\end{proof}
\begin{Prop}\label{prop:fusionq.proj}
Unless $\ell =2$ in the regular family, for all $0\leq i \leq n$ and $0\leq j \leq m$,
\begin{equation}\label{eq:fusionq.proj}
\dP{m+2n,j} \div_{f} \dP{n,i} \simeq \dP{m,j} \xf \dP{n,i},
\end{equation}
\begin{equation}
\ds{m+2n,j} \div_{f} \dP{n,i} \simeq \ds{m,j} \xf \dP{n,i}.
\end{equation}
If $\ell =2$ in the regular family, the statement is still true for $i=n=1$.
\end{Prop}
\begin{proof}
We do the proof for the first equality as that of the second is identical. Using the restriction of $\dP{m+2n,j}$ with the preceding proposition gives the conclusion for $i=1$ in both families, and $i=0$ in the dilute family. The case $i=0$ in the regular family is contained in corollary \ref{coro:fusionq.proj.reg}. We thus proceed by induction on $i$. If the result stands for $i$, then
\begin{align*}
	\left(\dP{m+2n,j}\div_{f} \dP{n-1,i}\right) \div_{f} \dP{1,1} & \overset{1}{\simeq} \dP{m+2n,j} \div_{f} \left(\dP{n-1,i} \xf \dP{1,1} \right)\\
	& \overset{2}{\simeq} \dP{m+2n,j} \div_{f} \left(\dP{n,i-1} \oplus \dP{n,i+1} \right),
\end{align*}
and
\begin{align*}
	\left(\dP{m+2n,j}\div_{f} \dP{n-1,i}\right) \div_{f} \dP{1,1} & \overset{3}{\simeq} \left(\dP{m+2,j} \xf \dP{n-1,i} \right)\div_{f} \dP{1,1}\\
	&\overset{4}{\simeq}  \underset{\lambda \in \Lambda}{\bigoplus} \dP{m+n+1,\lambda} \div_{f} \dP{1,1}\\
	&\overset{5}{\simeq} \underset{\lambda \in \Lambda}{\bigoplus} \dP{m+n-1,\lambda} \xf \dP{1,1}\\
	&\overset{6}{\simeq} \left(\dP{m,j}\xf \dP{n-1,i} \right)\xf \dP{1,1}\\
	&\overset{7}{\simeq} \dP{m,j} \xf \left(\dP{n,i-1} \oplus \dP{n,i+1}\right),
\end{align*}
where we assumed, for simplicity, that $i, i\pm 1$ were not critical. These cases are simple generalizations of the same arguments. The isomorphism $1$ is simply proposition \ref{prop:fusionq.prop}, while $2$ is proposition \ref{prop:fusionproj}. The isomorphism $3$ is obtained by applying $- \div_{f} \dP{1,1}$ on the right side of \eqref{eq:fusionq.proj}, and $4$ is obtained by applying proposition \ref{prop:fusionproj}, where all the index appearing in the projective modules were grouped in the family $\Lambda$. Noting that $\lambda \leq j+i \leq m+n$, for all $\lambda \in \Lambda$, proposition \ref{prop:fusionq.proj} with $i=1$ can be used, obtaining $5$. Finally, use again proposition \ref{prop:fusionproj} to obtain $6$, and use the associativity of the fusion product with, again proposition \ref{prop:fusionproj}, to obtain $7$. Comparing $2$ and $7$ and using the induction hypothesis gives the conclusion.
\end{proof}
What happens when we take a quotient of the form $\dP{n+2m,j} \div_{f} \dP{m,i}$, but $j>n$? It can be seen that
\begin{multline}
\dP{n+2m,j}\div_{f} \dP{m,i} \simeq \left(\dP{n+2m+(j-n),j} \div_{f} \dP{j-n,0}  \right)\div_{f} \dP{m,i} \\
\simeq \left( \dP{j+2m,j} \div_{f} \dP{m,i}\right) \div_{f} \dP{(j-n),0} \simeq \left(\dP{j,j} \xf \dP{m,i} \right)\div_{f} \dP{j-n,0},
\end{multline}
where we simply used propositions \ref{prop:fusionq.prop} and \ref{prop:fusionq.proj}. There is thus the following ``recipe": start by computing $\dP{n,j} \xf \dP{m,i}$, by applying proposition \ref{prop:fusionproj}, ignoring the fact that $\dP{n,j}$ is not well-defined. Then, use the fact that, by definition
\begin{equation*}
\dP{n,k} \equiv \begin{cases}
\ds{n,k_{-}}, & \text{when } k_{-}\leq n< j\\
0, & \text{when } k_{-}>n
\end{cases}.
\end{equation*}
For instance, in $\ell =5$,
$$\dP{10,9} \div_{f} \dP{4,4} \simeq \underbrace{\dP{6,9}}_{0} \oplus \underbrace{\dP{6,11}}_{0} \oplus \underbrace{\dP{6,13}}_{\ds{6,5}} \simeq \ds{6,5}.$$

More complex fusion quotients could be computed by using arguments similar to those we used to compute fusion products. However, the focus of this paper is on the fusion product, we only give one fairly simple case to show that the two operations are distinct.
\begin{Prop}
For $n\geq \ell$,
\begin{equation}
\dP{m+2n, j} \div_{f} \ds{n,\ell} \simeq \begin{cases}
\dP{m,j-1} \xf \dP{n ,\ell -1} \oplus \T{2}{m+n, \ell -2- j}, & 0\leq j < \ell -1\\
\dP{m,j}\xf \ds{n,\ell}, & \text{otherwise}
\end{cases}.
\end{equation}
\end{Prop}
\end{subsection}
\end{section}
	
\end{document}